\documentclass[12pt]{article} 
\renewcommand{\baselinestretch}{1.2}

\usepackage{epsf}

\newcommand{\bfig}{\begin{figure}}
\newcommand{\efig}{\end{figure}}
\newcommand{\bea}{\begin{eqnarray}}
\newcommand{\eea}{\end{eqnarray}}
\newcommand{\be}{\begin{equation}}
\newcommand{\ee}{\end{equation}}

\newcommand{\eps}{\epsilon}
\newcommand{\ldB}{\lambda_{dB}}
\newcommand{\bi}{\begin{itemize}}
\newcommand{\ei}{\end{itemize}}

\title{Bose-Einstein condensates in atomic gases: simple
theoretical results}
\author{Yvan Castin}
\date{\small 25 October 2000}

\begin{document}
\maketitle
\bibliographystyle{alpha}
\tableofcontents

\section{Introduction}
\markright{Introduction}

\subsection{1925: Einstein's prediction for the ideal Bose gas}
Einstein considered $N$ non-interacting bosonic and non-relativistic
particles in a cubic box of volume $L^3$
with periodic boundary conditions.
In the {\it thermodynamic limit}, defined as
\be
N,L \rightarrow \infty
\ \ \ \mbox{with}\ \ \ \frac{N}{L^3}=\rho=\mbox{constant},
\ee
a phase transition occurs at a temperature $T_c$ defined by:
\be
\rho \lambda_{dB}^3(T_c) = \zeta(3/2) = 2.612...
\label{eq:idealbox}
\ee
where we have defined the thermal de Broglie wavelength of the
gas as function of the temperature $T$:
\be
\ldB(T)=\left(\frac{2\pi\hbar^2}{m k_B T}\right)^{1/2}
\ee
and where
$\zeta(\alpha)=\sum_{k=1}^\infty 1/k^\alpha$ is the
Riemann Zeta function. 

The order parameter of this phase transition is the fraction $N_0/N$ of particles
in the ground state of the box, that is in the plane wave with momentum $\vec{p}=\vec{0}$.
For temperatures lower than $T_c$ this fraction $N_0/N$ remains finite
at the thermodynamic limit, whereas it tends to zero when $T>T_c$:
\bea
T>T_c &\hspace{2cm}& \frac{N_0}{N} \rightarrow 0 \\
T<T_c &\hspace{2cm}& \frac{N_0}{N}
        \rightarrow 1-\left(\frac{T}{T_c} \right)^{3/2}.
\label{eq:condfracidealbox}
\eea
For $T<T_c$ the system has formed a Bose-Einstein condensate in $\vec{p}=\vec{0}$.
The number $N_0$ of particles in the condensate is on the order of $N$, that is macroscopic.
As we will see, the {\it macroscopic population of a single quantum state} is the key feature
of a Bose-Einstein condensate, and
gives rise to interesting properties, e.g.\ coherence (as for the laser).

\subsection{Experimental proof?}
The major problem encountered experimentally to verify Einstein's predictions
is that at densities and temperatures required by Eq.(\ref{eq:idealbox})
at thermodynamic equilibrium almost all materials are in the solid state.

An exception is He$^4$ which is a fluid at $T=0$. However He$^4$ is a
strongly interacting system. In He$^4$ in sharp contrast with the
prediction for the ideal gas Eq.(\ref{eq:condfracidealbox}), $N_0/N < 10\%$
even at zero temperature \cite{Helium}.
\footnote{Amusingly the ideal gas prediction
Eq.(\ref{eq:idealbox})
does not give a too wrong result for the transition temperature in helium.
Note that the condensate fraction $N_0/N$ should not be confused
with the superfluid fraction: at $T=0$ the superfluid fraction is equal
to unity.}

The solution which victoriously led to Bose-Einstein condensation in atomic gases is to bring the
system to extremely low densities (much lower than in a normal gas) and
to cool it rapidly enough so that it has no time to recombine and solidify.
The price to pay for an ultralow density is the necessity to cool at extremely low temperatures.
Typically one has in the experiments with condensates:
\bea
\rho & < & 10^{15} \mbox{atoms}/\mbox{cm}^3  \\
T & < & 1 \mu K .
\eea
The critical temperatures range from $20$ nK to the $\mu$K range.

Bose-Einstein condensation was achieved for the first time in atomic gases
in 1995. The group of Eric Cornell and Carl Wieman at JILA was first,
with ${}^{87}$Rb atoms \cite{JILA_first}. They were closely followed
by the group of Wolfgang Ketterle at MIT with ${}^{23}$Na atoms \cite{MIT_first}
and the group of Randy Hulet at Rice University with ${}^{7}$Li atoms \cite{Rice_first}.
Nowadays there are many condensates mainly with rubidium or sodium atoms.
No other alkali atoms than the ones of year 1995 has been condensed. Atomic hydrogen has been condensed
in 1998 at MIT in the group of Dan Kleppner \cite{Hydrogen}; the experiments on hydrogen
were actually the first ones to start and played a fundamental pioneering role in
developing many of the experimental techniques having led the alkalis atoms
to success, such as magnetic trapping and evaporative
cooling of atoms.

In our lectures we do not consider the experimental techniques used to obtained and
to study Bose-Einstein condensates as they are treated in the lectures of Wolgang Ketterle
and of Eric Cornell at this school.

\subsection{Why interesting?}
\subsubsection{Simple systems for the theory}
An important theoretical frame
for Bose-Einstein condensation in interacting systems was developed
in the 50's by Beliaev, Bogoliubov, Gross, Pitaevskii in the context of
superfluid helium.
This theory however
is supposed to work better if applied to Bose condensed gases where the
interactions are much weaker.

The interactions in ultracold atomic gases can be described by a single
parameter $a$, the so-called {\it scattering length},
as interactions take place
between atoms with very low relative kinetic energy. The gaseous condensates
are {\sl dilute} systems as the mean interparticle separation is much larger
than the scattering length $a$:
\be
\rho |a|^3 \ll 1.
\ee
This provides a small parameter to the theory and, as we shall see, simple
mean field approaches can be used with success to describe most
of the properties of the atomic condensates.

\subsubsection{New features}
Atomic gases offer some new interesting features with respect to superfluid helium 4:
\begin{itemize}
\item {\sl Spatial inhomogeneity}:
This feature can be used as a tool to detect the presence of a 
Bose-Einstein condensate inside the trap: in an inhomogeneous gas Bose-Einstein
condensation occurs not only in momentum
space but also in position space!

\item {\sl Finite size effects}:
The number of atoms in condensates of alkali gases is usually $N_0<10^7$.
The hydrogen condensate obtained at MIT by Kleppner is larger $N_0 \simeq 10^9$.
Interesting finite size effects, that is effects which disappear
at the thermodynamic limit, such as Bose-Einstein condensates with effective attractive
interactions ($a<0$),  can be studied in relatively small condensates.

It is also interesting to consider small condensates where some interesting
quantum aspects concerning coherence properties of the condensates,
such as collapses and revivals of the relative phase between
two condensates \cite{mesure_franges2}, 
could perhaps be measured \cite{resur}.

\item {\sl Tunability}:
Condensates in atomic gases can be manipulated and studied using
the powerful techniques of atomic physics (see the lectures of Wolfgang Ketterle and
Eric Cornell). Almost all the parameters can be controlled at will, including the
interaction strength $a$ between the particles. The atoms can be imaged not only in position
space, but also in momentum space, allowing one to see the momentum distribution
of atoms in the condensate! One can also tailor the shape and intensity of the trapping
potential containing the condensate.

\end{itemize}

\section{The ideal Bose gas in a trap}
\markright{Ideal Bose gas}
\label{Cap:ideal}
Let us consider a gas of non-interacting bosonic particles trapped in a potential $U(\vec{r}\,)$
at thermal equilibrium. As the particles do not interact thermal equilibrium has to be provided
by coupling to an external reservoir. In the grand-canonical ensemble the state of the gas
is described by the equilibrium $N$-body density matrix
\be
\hat{\rho} = \frac{1}{\Xi} \exp [ -\beta
                \left( \hat{H} - \mu \hat{N} \right)]
\ee
where $\Xi$ is a normalization factor,
$\hat{H}$ is the Hamiltonian containing the kinetic energy and trapping potential energy
of all the particles,
$\hat{N}$ is the operator giving the total number of particles,
$\beta = 1/k_B T$ where $T$ is the temperature, and $\mu$ is the chemical
potential. One more conveniently introduces the fugacity:
\be
{\mathrm z} = \exp [ \beta \mu ].
\ee

\subsection{Bose-Einstein condensation in a harmonic trap}
Let us consider the case of a harmonic trapping potential $U(\vec{r}\,)$:
\be
U(\vec{r}\,)=\frac{1}{2} m (\omega_x^2 x^2 + \omega_y^2 y^2 + \omega_z^2 z^2).
\ee
We wish to determine the properties of the trapped gas at thermal equilibrium; the
calculations can be done
in the basis of harmonic levels or
in position space. 

\subsubsection{In the basis of harmonic levels}
Let us consider the single particle eigenstates of the harmonic potential
with eigenvalues $\eps_{\vec{l}}$ labeled by the vector:
\be
\vec{l} = (l_x,l_y,l_z) \hspace{1cm} l_\alpha=0,1,2,3... \; (\alpha=x,y,z).
\ee
One has:
\be
\eps_{\vec{l}}=l_x \hbar \omega_x + l_y \hbar \omega_y + l_z \hbar \omega_z 
\ee
where the zero-point energy $(\hbar/2 )(\omega_x + \omega_y + \omega_z)$ has
been absorbed for convenience in the definition of the chemical potential.
Let us consider the case
of an isotropic potential for which all the $\omega_\alpha$'s are equal to $\omega$,
so that $\eps_{\vec{l}}=l\hbar \omega $ with $l\equiv l_x+l_y+l_z$.

The mean occupation number of each single particle eigenstate in the trap is given by the Bose
distribution:
\be
n_{\vec{l}} = \frac{1}{\exp[\beta (\eps_{\vec{l}}-\mu)]-1}
          = \left[ \frac{1}{\mathrm z} \exp (\beta l \hbar \omega) - 1 \right]^{-1}.
\label{eq:bon}
\ee
Since $n_{\vec{l}}$ has to remain positive (for $l$=0,1,2 ...), the range of variation
of the fugacity $\mathrm z$ is given by
\be
0 < {\mathrm z} < 1.
\ee
The average total number of particles $N$ is obtained by summing over
all the occupation numbers: $N=\sum_{\vec{l}}  n_{\vec{l}}$, a relation that
can be used in principle to eliminate $\mathrm z$ in terms of $N$. It is useful to
keep in mind that for a fixed temperature $T$, $N$ is an increasing function of $\mathrm z$.

In the limit ${\mathrm z} \rightarrow 0$ one recovers Boltzmann statistics:
$n_{\vec{l}} \propto \exp (-\beta \eps_{\vec{l}}) $. We are interested here
in the opposite, quantum degenerate limit where 
the occupation number of the ground state $l=0$ of the trap, given by
\be
N_0 = n_{\vec{0}} = \frac{\mathrm z}{1-{\mathrm z}},
\ee
diverges when ${\mathrm z} \rightarrow 1$, which indicates the presence of a Bose-Einstein condensate
in the ground state of the trap.

We wish to watch the formation of the condensate when $\mathrm z$ is getting closer to one, that
is when one gradually increases the total number of particles $N$.
The essence of Bose-Einstein condensation is actually the phenomenon of {\sl saturation}
of the population of the excited levels in the trap, a direct consequence of the Bose
distribution function.
Consider indeed the sum of the occupation numbers of the single particle excited states in
the trap:
\be
N'=\sum_{\vec{l} \neq \vec{0}}  n_{\vec{l}}\;.
\ee
The key point is that for a given temperature $T$, $N'$ is bounded from
above:
\be
N'=\sum_{\vec{l} \neq \vec{0}} 
   \left[ \frac{1}{\mathrm z} \exp (\beta l \hbar \omega) - 1 \right]^{-1} <
\sum_{\vec{l} \neq \vec{0}}
   \left[ \exp (\beta l \hbar \omega) - 1 \right]^{-1} \equiv N'_{\mbox{\scriptsize max}}\;.
\ee
Note that we can safely set ${\mathrm z}=1$ since the above sum excludes the term $l=0$.

If the temperature $T$ is fixed and we start adding particles to the system,
particles will be forced to pile up in the ground state of the trap 
when $N>N'_{\mbox{\scriptsize max}}$,
where they will form a condensate. Let us now estimate the \lq\lq critical" value
of particle number $N'_{\mbox{\scriptsize max}}$. 

We will restrict to the interesting regime
$k_B T \gg \hbar\omega$: in this regime Bose statistics allows one to accumulate most
of the particles in a single quantum state of the trap while having the system in contact
with a thermostat at a temperature much higher than the quantum of oscillation $\hbar\omega$,
a very counter-intuitive result for someone used to Boltzmann statistics! 
On the contrary the regime $k_B T \ll \hbar \omega$ would lead to a large
occupation number of the ground state of the trap even for Boltzmann statistics.

A first way to calculate $N'_{\mbox{\scriptsize max}}$ is to realize that the generic
term of the sum varies slowly with $l$ as $k_B T \gg \hbar \omega$ so that one can replace
the discrete sum
$\sum_{\vec{l} \neq \vec{0}}$ by an integral $\int_{l_\alpha \geq 0} d^3\vec{l}$.
As we are in the case of a three-dimensional harmonic trap there is no divergence of the integral
around $\vec{l}=\vec{0}$.

We will rather use a second method, which allows one to calculate also the first
correction to the leading term in $k_B T/\hbar\omega$. We use the series expansion
\be
\frac{1}{e^x-1}=\frac{e^{-x}}{1-e^{-x}}=\sum_{k=1}^\infty e^{-kx} 
\label{eq:see}
\ee
which leads to the following expression for $N'_{\mbox{\scriptsize max}}$, 
if one exchanges the summations
over $\vec{l}$ and $k$:
\be
N'_{\mbox{\scriptsize max}} = \sum_{k=1}^\infty \sum_{\vec{l} \neq \vec{0}}
        \exp [ -k \beta \hbar \omega \sum_\alpha l_\alpha] 
        = \sum_{k=1}^\infty  \left[ \left(
        \frac{1}{1-\exp [-\beta \hbar \omega k ] } \right)^3 -1 \right].
\ee
We now expand the expression inside the brackets for small $x$:
\be
\left[ \left( \frac{1}{1-\exp [-x ] } \right)^3 -1 \right] =
\frac{1}{x^3} + \frac{3}{2 x^2} + ...
\label{eq:serie}
\ee
and we sum term by term to obtain
\be
N'_{\mbox{\scriptsize max}}=\left( \frac{k_BT }{\hbar \omega} \right)^3 \zeta (3) + 
    \frac{3}{2} \left( \frac{k_BT }{\hbar \omega} \right)^2 \zeta (2) + ...
\label{eq:nmaxex}
\ee
Note that the exchange of summation over $k$ and summation over the order of expansion 
in Eq.(\ref{eq:serie}) is no longer allowed for the next term $1/x$, which would lead 
to a logarithmic divergence (that one can cut \lq\lq by hand" at $k\simeq k_B T/\hbar\omega$).

One then finds to leading order for the fraction of population in the single particle
ground state:
\be
\frac{N_0}{N} \simeq \frac{N-N'_{\mbox{\scriptsize max}}}{N} \simeq 1- \zeta (3)
        \left( \frac{k_BT }{\hbar \omega} \right)^3 \frac{1}{N} =
        1- \left(\frac{T}{T_c^0}\right)^3
\label{eq:cfrac}
\ee
where the critical temperature $T_c^0$ is defined by:
\be
\zeta (3) \left( \frac{k_BT_c^0 }{\hbar \omega} \right)^3 =N
\label{eq:tc0}
\ee
and $\zeta(3) = 1.202...$. Note that the universal law (\ref{eq:cfrac})
differs from the one obtained in the homogeneous case 
(\ref{eq:condfracidealbox}) usually considered in the
literature.

The present calculation is easily extended to the case of an anisotropic harmonic
trap. To leading order one finds
\be
N'_{\mbox{\scriptsize max}}\simeq \left( \frac{k_BT }{\hbar \bar{\omega}} \right)^3 \zeta (3)
\ee
where $\bar{\omega}= (\omega_x \omega_y \omega_z)^{1/3}$ is the geometric mean
of the trap frequencies.
One can also calculate $N'_{\mbox{\scriptsize max}}$ in two-dimensional and one-dimensional
models. One also finds in these cases a finite value for $N'_{\mbox{\scriptsize max}}$:
the saturation of population in the single particle
excited states applies as well and one can form a condensate, a situation very
different from the thermodynamical limit in the homogeneous 1D and 2D cases.

\subsubsection{Comparison with the exact calculation}
\begin{figure}[htb]
\epsfysize=8cm \centerline{\epsfbox{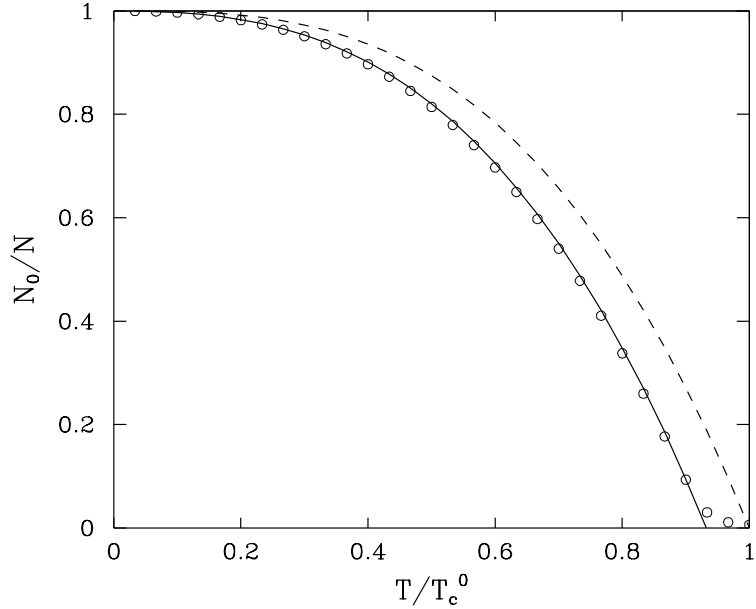}}
\caption{\small Condensate fraction versus temperature for an ideal 
Bose gas in a spherically
symmetric trap with $N=1000$ particles. The circles correspond to the exact quantum
calculation. The solid line 
corresponds to the prediction 
$N_0/N \simeq  1-N'_{\mbox{\scriptsize max}}/N$ with $N'_{\mbox{\scriptsize max}}$
given by the two terms in the expansion Eq.(\ref{eq:nmaxex}). 
The dashed line corresponds to the prediction 
$N_0/N \simeq  1-N'_{\mbox{\scriptsize max}}/N$ with $N'_{\mbox{\scriptsize max}}$
given by the leading term in Eq.(\ref{eq:nmaxex}). This figure was taken from 
\cite{StringariRev}.
}\label{fig:compare}
\end{figure}

One can see in figure \ref{fig:compare} that the first two terms in the
expansion Eq.(\ref{eq:nmaxex}), combined with the approximation
$N_0/N \simeq  1-N'_{\mbox{\scriptsize max}}/N$,
give a very good approximation to the exact condensate fraction 
for $N=1000$ particles only.

\subsubsection{In position space}
A very important object in the description of the state of the gas is the so-called
one-body density matrix.  We can define it as follows.

Consider a one-body observable
\be
{\cal X}=\sum_{i=1}^{\hat{N}} X(i)
\ee
where $X(i)$ is the observable for particle number $i$
and where $\hat{N}$ is the operator giving the total number of
particles.
The one-body density matrix $\hat{\rho}_1$ is defined by the requirement 
that for any $X$:
\be
\langle {\cal X} \rangle \equiv \mbox{Tr} [\hat{\rho}_1 X(1)].
\label{eq:rho1}
\ee
For the particular case of $X$ equal to the identity
it follows ${\cal X}=\hat{N}$ and
$\langle {\cal X} \rangle = \mbox{Tr} [\hat{\rho}_1]=\langle \hat{N}\rangle$
so that our one-body density matrix is normalized to the mean number
of particles in the system.

An equivalent definition of $\hat{\rho}_1$ in the second quantized
formalism is simply
\be
\langle \vec{r}\,' | \hat{\rho}_1 |\vec{r}\, \rangle =
\langle \hat{\psi}^\dagger(\vec{r}\,)  \hat{\psi}(\vec{r}\,') \rangle
\ee
where $\hat{\psi}(\vec{r}\,)$ is the atomic field operator, annihilating an atom in $\vec{r}$.

At thermal equilibrium in the grand-canonical ensemble, the one-body density matrix
of the ideal Bose gas is given by
\be
\hat{\rho}_1=\frac{1}{{\mathrm z}^{-1} \exp (\beta \hat{h}_1) -1}
\label{eq:rho1ideal}
\ee
where the single-particle Hamiltonian in the case of a spherically symmetric
harmonic trap is
\be
\hat{h}_1 = \frac{\vec{p}\,^2}{2m} + \frac{1}{2} m \omega^2 \vec{r}\,^2
                - \frac{3}{2} \hbar \omega.
\label{eq:H0}
\ee
Here again we have subtracted the zero-point energy for convenience.
The Bose formula Eq.(\ref{eq:bon}) corresponds to the diagonal element of
$\hat{\rho}_1$ in the eigenbasis of the harmonic oscillator (the off-diagonal
elements of course vanish).
In position space the diagonal term
\be
\langle \vec{r}\, | \hat{\rho}_1 | \vec{r}\, \rangle = \rho(\vec{r}\,)
\ee
gives the mean spatial density of the gas.

In order to calculate the density we use the series expansion Eq.(\ref{eq:see})
to rewrite $\hat{\rho}_1$ as follows:
\be
\hat{\rho}_1 = \sum_{k=1}^{\infty}{\mathrm z}^k e^{-\beta k \hat{h}_1}.
\label{eq:rho1sum}
\ee
This writing takes advantage of the fact that the matrix elements
\be
\langle \vec{r}\, |e^{-\beta k \hat{h}_1} | \vec{r}\,' \rangle 
\ee
are known for an harmonic oscillator potential \cite{Landau}. One then
obtains explicitly:
\be
\rho(\vec{r}\,) = \left( \frac{m \omega}{\pi \hbar} \right)^{3/2}
  \sum_{k=1}^{\infty} {\mathrm z}^k
  \left( 1 - \exp (-2 \beta k \hbar \omega) \right)^{-3/2}
  \exp \left[ - \frac{m\omega r^2}{\hbar}
  \tanh\left(\frac{\beta k \hbar \omega}{2}\right) \right]
\label{eq:Landau}
\ee
One can identify the contribution of the condensate
to this sum when ${\mathrm z}\rightarrow 1^{-}$. 
When the summation index $k$ is large, what determines the convergence of the series is indeed
the factor ${\mathrm z}^k$. Replacing the other factors in the summand by their asymptotic
value for $k\rightarrow +\infty$ we identify the diverging part when ${\mathrm z}=1$:
\be
\left( \frac{m \omega}{\pi \hbar} \right)^{3/2}
  \sum_{k=1}^{\infty} {\mathrm z}^k
  \exp \left[ - \frac{m\omega r^2}{\hbar} \right] =
  \frac{{\mathrm z}}{1-{\mathrm z}} |\phi_{0,0,0} (\vec{r}\,)|^2 =
  N_0 |\phi_{0,0,0} (\vec{r}\,)|^2
\ee
where $\phi_{0,0,0} (\vec{r}\,)$ is the ground state wave function of the
harmonic oscillator.

\begin{figure}[htb]
\epsfysize=8cm \centerline{\epsfbox{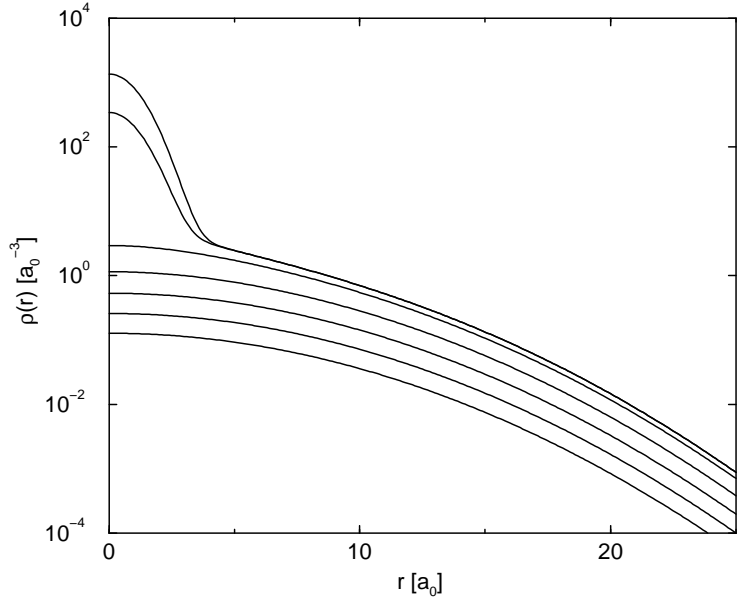}}
\caption{\small Spatial density for an ideal Bose gas at thermal equilibrium
in a harmonic trap of frequency $\omega$. The temperature is
fixed to $k_B T=20 \hbar\omega$ and the number of particles ranges
from $N=500$ to $N=32 000$ between the lowest curve and the upper
curve, with a geometrical reason equal to 2. The unit of
length for the figure is $a_0=(\hbar/2m\omega)^{1/2}$, that is the
spatial radius of the ground state of the trap.} \label{fig:densite}
\end{figure}

Numerically we have calculated the total density $\rho(\vec{r}\,)$ for
a fixed temperature $k_B T= 20 \hbar\omega$ and for increasing number of
particles (see figure \ref{fig:densite}). Here the maximal number
of particles one can put in the excited states of the trap
is $N'_{\mbox{\scriptsize max}} \simeq \zeta(3) (k_B T/\hbar\omega)^3 \simeq 10^4$.
When $N\ll N'_{\mbox{\scriptsize max}}$ the effect of an increase of $N$ is mainly
to multiply the density by some global factor (the curves in logarithmic scale in
figure \ref{fig:densite} are parallel one to the other). When $N$ is becoming
larger than $N'_{\mbox{\scriptsize max}}$ a peak in density grows around $r=0$, indicating
the formation of the condensate, whereas the far wings of the density distribution 
saturate, which reflects the saturation of the population of the excited
levels of the trap.

\subsubsection{Relation to Einstein's condition $\rho \lambda_{dB}^3 = \zeta(3/2)$}

In the limit $k_B T \gg \hbar\omega$ we can actually calculate the value
$\rho'_{\mbox{\scriptsize max}}(\vec{r}\,)$ to which the density $\rho'(\vec{r}\,)$  of
particles in the excited states of the trap saturates when ${\mathrm z}\rightarrow 1$.
We simply use the expansion Eq.(\ref{eq:Landau}), subtracting from 
the total density $\rho(\vec{r}\,)$ the contribution of the condensate
$N_0 |\phi_{0,0,0} (\vec{r}\,)|^2$. The resulting series is converging
even for ${\mathrm z}=1$
so that we can take safely the
semiclassical limit $k_B T \gg \hbar \omega$ term by term in the sum:
\be
\rho'(\vec{r}\,) \simeq \frac{1}{\ldB^3} \sum_{k=1}^\infty
        \frac{{\mathrm z}^k}{k^{3/2}}
        \exp \left( -\frac{1}{2} k \beta m \omega^2 r^2 \right)
        = \frac{1}{\ldB^3} g_{3/2}
        \left[{\mathrm z}  \exp \left( -\beta \frac{1}{2} m \omega^2 r^2 \right)\right]
\label{eq:rhosc}
\ee
where 
\be
g_\alpha(x) = \sum_{k=1}^\infty \frac{x^k}{k^\alpha}.
\label{eq:ga}
\ee
We term this approximation semiclassical as (i) 
one can imagine that the classical limit $\hbar\rightarrow 0$
is taken in each term $k$ of the sum, giving the usual Gaussian distribution for the
density of a classical harmonic oscillator at temperature $k_B T/k$, but (ii) the
distribution still reflects the quantum Bose statistics.

If now we set ${\mathrm z}=1$ in (\ref{eq:rhosc}) to express the fact that a
condensate is formed
we obtain
\be
\rho'_{\mbox{\scriptsize max}}(\vec{r} = \vec{0}) \simeq \frac{1}{\ldB^3} g_{3/2}(1) =
                   \frac{\zeta(3/2)}{\ldB^3}.
\ee
We therefore recover Einstein's condition provided one replaces the density
$\rho$ of the homogeneous case by the density at the center of the trap.

\subsection{Bose-Einstein condensation in a more general trap}
We now extend the idea of the previous semiclassical limit to more
general non-harmonic potentials. This allows to find the condition
for Bose-Einstein condensation in presence of a non-harmonic potential.
This will prove useful in presence 
of interactions between the particles
where the non-harmonicity is provided by the mean field potential.

\subsubsection{The Wigner distribution}
The idea is to find a representation of the one-body density matrix having
a simple (non pathological) behavior when $\hbar \rightarrow 0$.
Let us take as an example a single harmonic oscillator.
The density matrix is then of the form:
\be
\hat{\sigma}=\frac{1}{Z} e^{-\beta \hat{H}_{\mbox{\scriptsize ho}}}
\ee
where $\hat{H}_{\mbox{\scriptsize ho}}$ is the harmonic oscillator Hamiltonian. 
As shown in \cite{Landau}
all the matrix elements of $\hat{\sigma}$ can be calculated exactly:
\be
\langle \vec{r}\, |\hat{\sigma}  |\vec{r}\,' \rangle =
\frac{1}{(2 \pi)^{3/2} (\Delta r)^3}
\exp \left[ -\frac{[(\vec{r}+\vec{r}\,'\,)/2]^2}{2 (\Delta r)^2} \right]
\exp \left[ -\frac{(\vec{r}-\vec{r}\,'\,)^2}{2 \xi^2} \right]
\ee
The relevant length scales are the spatial width of the cloud
$\Delta r$:
\be
(\Delta r)^2 = \frac{\hbar}{2 m \omega}
        \mbox{cotanh} \left( \frac{\hbar \omega}{2 k_B T} \right)
\ee
and the coherence length $\xi$:
\be
\xi^2 = \frac{2 \hbar}{m \omega}
        \tanh \left( \frac{\hbar \omega}{2 k_B T} \right).
\ee
If we now take the classical limit $\hbar \rightarrow 0$
(in more physical terms the limit $\hbar \omega \ll k_B T$) then:
\bea
(\Delta r)^2 &\rightarrow& \frac{k_B T}{m \omega^2} \\
\xi^2 &\sim& \frac{\hbar^2}{m k_B T}= \frac{\ldB^2}{2 \pi}.
\eea
In the limit $\hbar \rightarrow 0$ the $\hbar$ dependence of $\xi$
causes $\langle \vec{r}\, |\hat{\sigma}  |\vec{r}\,' \rangle \rightarrow 0$ for
fixed values of $\vec{r},\vec{r}\,'$
unless $\vec{r} = \vec{r'}$: the limit is singular.

To avoid this problem one can use the Wigner representation of the
density matrix, introduced also in the lectures of Zurek and Paz:
\be
W[\hat{\sigma}] (\vec{r},\vec{p}\,) =
\int \frac{d^3 \vec{u}}{h^3} \left. \langle \vec{r}- \frac{\vec{u}}{2}\right|
        \hat{\sigma} \left|\vec{r} + \frac{\vec{u}}{2}\rangle\right.
        e^{i \vec{p} \cdot \vec{u} /\hbar}.
\label{eq:wigner}
\ee
The Wigner distribution is the quantum analog of the classical phase space
distribution. In particular one can check that the Wigner distribution is normalized
to unity and that
\bea
\int d^3 \vec{r}\;  W(\vec{r},\vec{p}\,) &=&
        \langle \vec{p}\, |\hat{\sigma}  |\vec{p}\, \rangle \\
\int d^3 \vec{p}\;  W(\vec{r},\vec{p}) &=&
        \langle \vec{r}\, |\hat{\sigma}  |\vec{r}\, \rangle.
\eea
An important caveat is that $W$ is not necessarily positive.

For the harmonic oscillator at thermal equilibrium
the integral over $\vec{u}$ in Eq.(\ref{eq:wigner}) is Gaussian and can be
performed exactly:
\be
W(\vec{r},\vec{p}\,) = \frac{1}{(2 \pi \Delta r \Delta p)^3}
        \exp (- \frac{r^2}{2 (\Delta r)^2})
        \exp (- \frac{p^2}{2 (\Delta p)^2})
\ee
where $\Delta p\equiv {\hbar}/{\xi}$.
If we take now the limit $\hbar \rightarrow 0$:
\bea
(\Delta r)^2 &\rightarrow& \frac{k_B T}{m \omega^2} \\
(\Delta p)^2 &\rightarrow& m k_B T
\eea
so that $W(\vec{r},\vec{p}\,)$ tends to the classical phase space density.

\subsubsection{Critical temperature in the semiclassical limit}
Let us turn back to our problem of trapped atoms in a non-harmonic trap
where  the single particle Hamiltonian is given by 
\be
\hat{h}_1 = \frac{\vec{p}\,^2}{2m} + U(\vec{r}\,)
\ee
and the one-body
density matrix
is given by Eq.(\ref{eq:rho1sum}). For $\hbar \rightarrow 0$ we have:
\be
W[e^{-\beta k \hat{h}_1}](\vec{r},\vec{p}) \simeq \frac{1}{h^3}
        \exp \left[-k \beta \left(\frac{p^2}{2m}+ U(\vec{r}\,)\right)\right].
\ee
As we did before we put apart the contribution of the condensate.
One then gets for the one-body density matrix of the non-condensed fraction
of the gas in the semiclassical limit:
\bea
W[\hat{\rho}'_1]_{sc} &=& \frac{1}{h^3} \sum_{k=1}^{+\infty} 
        {\mathrm z}^k\exp \left[-k \beta \left(\frac{p^2}{2m}+ U(\vec{r}\,)\right)\right]
\label{eq:wsc} \\
&=& \frac{1}{h^3}\left\{\frac{1}{{\mathrm z}}
        \exp \left[\beta \left(\frac{p^2}{2m}+U(\vec{r}\,)\right)\right] -1 \right\}^{-1}.
\eea

We are now interested in the spatial density of the non-condensed particles
in the semiclassical limit. By integrating Eq.(\ref{eq:wsc}) over $p$ we obtain:
\be
\rho'_{sc}(\vec{r}\,) = \frac{1}{\ldB^3} g_{3/2} ({\mathrm z} e^{-\beta U(\vec{r}\,)}) 
\label{eq:secforrho}
\ee
where $g_{\alpha}$ is defined in Eq.(\ref{eq:ga}).
The condition for Bose-Einstein condensation
is ${\mathrm z} \rightarrow e^{\beta U_{\mbox{\scriptsize min}}}$ where
$U_{\mbox{\scriptsize min}}=\mbox{min}_{\vec{r}}\, U(\vec{r}\,)$ is the minimal
value of the trapping potential, achieved in the point $\vec{r}_{\mbox{\scriptsize min}}$.
For ${\mathrm z} = e^{\beta U_{\mbox{\scriptsize min}}}$ the semiclassical approximation for
the non-condensed density gives in this point:
\be
\rho'_{sc}(\vec{r}_{\mbox{\scriptsize min}}) = \frac{1}{\ldB^3} g_{3/2}(1)
\ee
or
\be
\rho \ldB^3= 2.612...
\ee
Again Einstein's formula is recovered with $\rho$ being the maximal
density of the non-condensed cloud, that is the non-condensed density at the center of the trap.

The semiclassical calculation that we have just presented was initially put forward
in \cite{Bagnato}.
We do not discuss in details the validity of this semi-classical approximation.
Intuitively a necessary condition is $k_B T \gg \Delta E$ where $\Delta E$ is
the maximal level spacing of the single particle Hamiltonian among
the states thermally populated. Some situations, where the trapping potential is not
just a single well, may actually require more care.
The case of Bose-Einstein  condensation in a periodic
potential is an interesting example that we leave as an exercise to the reader.

\subsection{Is the ideal Bose gas model sufficient: experimental verdict}
\label{subsec:verdict}

\subsubsection{Condensed fraction as a function of temperature}
\begin{figure}[htb]
\epsfysize=8cm \centerline{\epsfbox{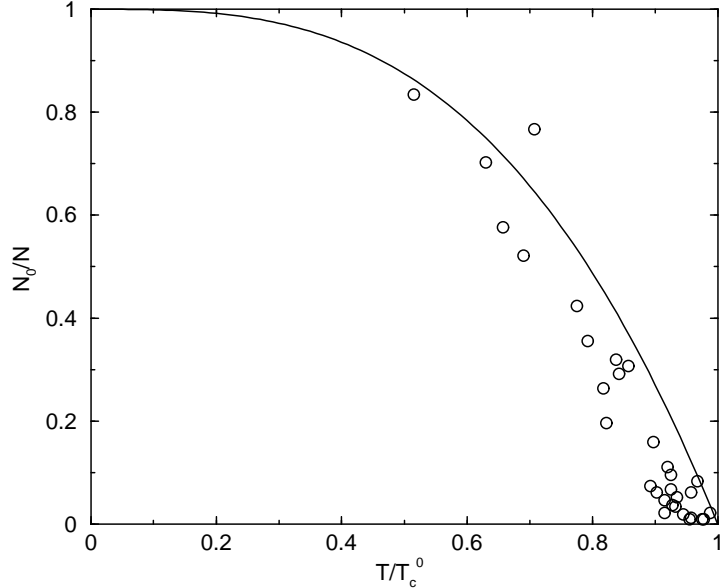}}
\caption{\small Condensate fraction $N_0/N$ as function of $T/T_c^0$ where $T_c^0$
is the leading order ideal Bose gas prediction Eq.(\ref{eq:tc0}). Circles
are the experimental results of \cite{n0_sur_n} while the dashed line
is Eq.(\ref{eq:cfrac}).} \label{fig:n0_sur_n}
\end{figure}

The groups at MIT and JILA have measured the condensate fraction
$N_0/N$ as function of temperature for a typical number of particles $N = 10^5$ or larger.
We reproduce here the results of JILA \cite{n0_sur_n} (see figure
\ref{fig:n0_sur_n}).
This figure shows that the leading order prediction of the ideal Bose gas 
Eq.(\ref{eq:cfrac}) is quite good, even if there is a clear indication from the
experimental data that the actual transition temperature is lower than $T_c^0$.
This deviation may be due to finite size effects and interaction effects
but the large experimental error has not allowed yet a fully quantitative comparison
to theory.

\subsubsection{Energy of the gas as function of temperature and
number of particles}
In the experiments one produces first a Bose condensed gas at thermal equilibrium.
Then one switches off suddenly the trapping potential. The cloud then expands
ballistically, and after a time long enough that the expansion velocity has
reached a steady state value one measures the kinetic energy of the expanding cloud.

Suppose that the trap is
switched off at $t=0$. For $t=0^-$ the total energy of the gas can be written as
\be
E_{\mbox{\scriptsize tot}}(0^-) 
= E_{\mbox{\scriptsize kin}} + E_{\mbox{\scriptsize trap}} + E_{\mbox{\scriptsize int}},
\ee
that is as the sum of kinetic energy, trapping potential energy and interaction
energy.
At time $t=0^+$ there is no trapping potential anymore
so that the total energy of the gas reduces to
\be
E_{\mbox{\scriptsize tot}}(0^+) = E_{\mbox{\scriptsize kin}} + E_{\mbox{\scriptsize int}}.
\ee
In the limit $t\rightarrow +\infty$ the gas expands, the density and therefore the interaction
energy drop, and all the energy $E_{\mbox{\scriptsize tot}}(0^+)$ is converted into kinetic energy, 
which is measured.

\begin{figure}[htb]
\epsfysize=8cm \centerline{\epsfbox{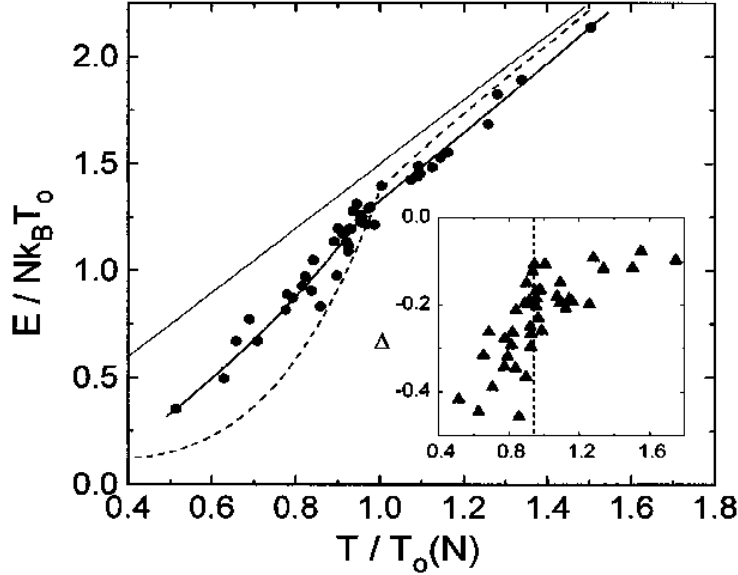}}
\caption{\small Expansion energy of the gas $E_{\mbox{\scriptsize tot}}(0^+)$ per particle and
in units of $k_B T_c^0$ as function of the temperature in units of $T_c^0$.
The disks correspond to the experimental results of \cite{n0_sur_n}.
The straight solid line is the prediction of Boltzmann statistics. The dashed curve exhibiting
a change of slope is the ideal Bose gas prediction. The curved solid line is a piecewise
polynomial fit to the data. The inset is a magnification
showing the change of slope of the energy as function of $T$ close to $T=T_c^0$.
The figure is taken from \cite{n0_sur_n}.}
\label{fig:chal}
\end{figure}

In figure \ref{fig:chal} we show the results of JILA for $E_{\mbox{\scriptsize tot}}(0^+)$
for temperatures around $T_c^0$ \cite{n0_sur_n} together with the ideal Bose gas prediction.
The main feature of the ideal Bose gas prediction is a change in the slope of the energy
as function of temperature when $T$ crosses $T_c$. One observes indeed a change of slope
in the experimental results (see the magnified inset)! 

For $T>T_c$ the ideal
Bose gas model is in good agreement with the experiment. For $T < T_c$ we observe however
that the experiment significantly deviates from the ideal Bose gas. 

\begin{figure}[htb]
\epsfysize=8cm \centerline{\epsfbox{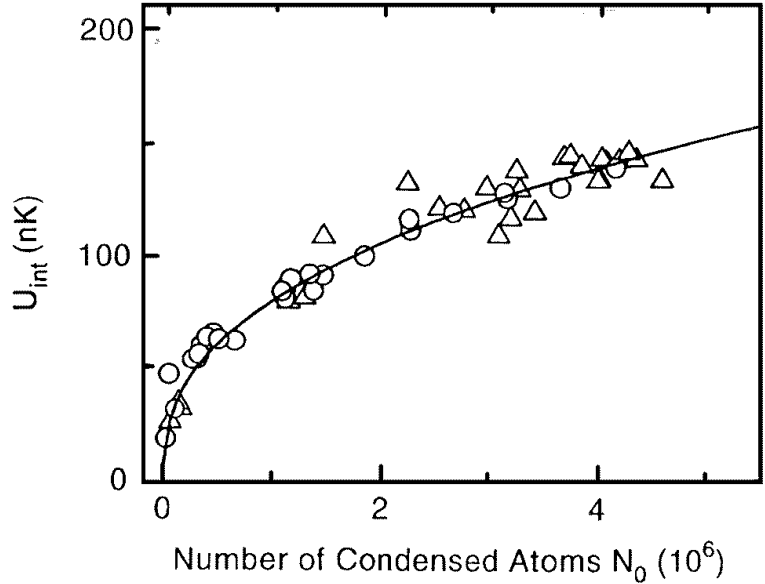}}
\caption{\small Expansion energy of the condensate per particle in the condensate, divided by
$k_B$, as a function of the number of particles in the condensate.
The experiment is performed at temperatures $T\ll T_c$. The triangles correspond
to cases where the non-condensed cloud was not visible experimentally. The disks
correspond to case where the non-condensed cloud could be seen. The figure
is taken from \cite{MIT_expans}. The solid line is a fit of the interacting
Bose gas prediction of \S\ref{Cap:GPE}.}\label{fig:mit_ener}
\end{figure}

What happens at even lower values of $T/T_c^0$? We show in figure \ref{fig:mit_ener} 
the expansion energy of the condensate per particle 
in the regime of an almost pure condensate
\cite{MIT_expans}.
This energy then depends almost only on the number of condensate particles $N_0$, in 
a non-linear fashion. This is in complete violation with the ideal Bose gas
model, which predicts an energy per particle in the condensate independent of $N_0$.
More precisely the ideal Bose gas prediction would be $\hbar(\omega_x + \omega_y +\omega_z)/4$
where the $\omega_\alpha$'s are the trap frequencies. In units of $k_B$ this would be
in the 10 nK range, an order of magnitude smaller than the measured values.

\subsubsection{Density profile of the condensate}
\begin{figure}[htb]
\epsfysize=8cm \centerline{\epsfbox{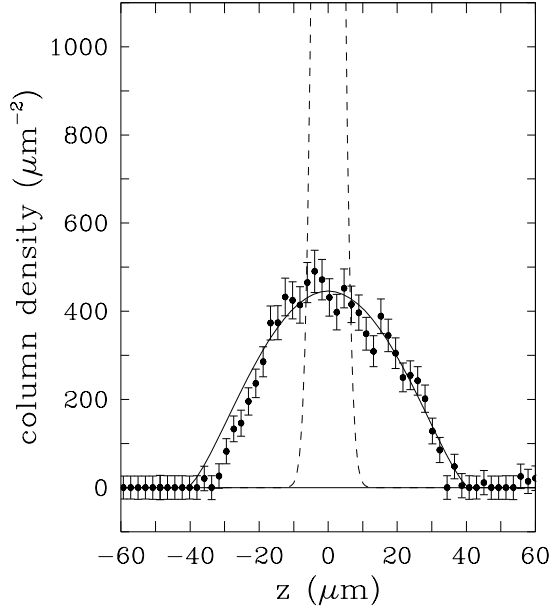}}
\caption{\small Column density profile (see text) of a condensate along the weak axis $z$
of a cigar-shaped trap. The experimental results of \cite{Hau} (dots) are very different
from the ideal Bose gas prediction (dashed line). The solid line corresponds to 
the theoretical prediction  of \S\ref{Cap:GPE}.}\label{fig:Hau}
\end{figure}

The group of Lene Hau at Rowland Institute has measured 
the density profile of the condensate in a cigar-shaped trap, along the weakly
confining axis $z$ of the trap. As imaging with a light beam is used 
the actual density obtained in the experiment is the density integrated
along the direction $y$ of propagation of the laser beam, plotted in figure
\ref{fig:Hau} for $x=0$ as function of $z$ \cite{Hau}. 
The measured profile is very different from and much broader than the Gaussian
density profile of the ground state wavefunction of the harmonic oscillator.

\subsubsection{Response frequencies of the condensate}
By modulating the harmonic frequencies of the trapping 
potential one can excite breathing
modes of the condensate. For example the group at MIT modulated the trap frequency
along the slow axis $z$ of a cigar-shaped trap and observed at $T\ll T_c$
subsequent breathing of the 
condensate at a frequency $1.569(4)\omega_z$. This frequency is not an integer
multiple of $\omega_z$ and can therefore not be obtained in the ideal Bose
gas model.

In conclusion the ideal Bose gas model may be acceptable as long as no significant
condensate has been formed. If a condensate is formed interaction effects 
become important, and dominant at $T\ll T_c$. This serves as a motivation to the 
next sections of this lecture, which will deal with the interacting Bose gas
problem.
\section{A model for the atomic interactions}
\markright{Model for interactions}
\label{Cap:model}
The previous section \ref{Cap:ideal} has shown that the ideal Bose gas model
is insufficient to explain the experimental results when a condensate is
formed. In this section we choose the model potential to be used in this lecture
to take into account the atomic interactions.
The reader interested in a more careful discussion of real interaction potentials
is referred to \cite{Jean_varenna}.

\subsection{Reminder of scattering theory}
We consider two particles of mass $m$ 
interacting in free space via the potential $V(\vec{r_1}-\vec{r_2})$
depending on the positions $\vec{r_1},\vec{r_2}$ only through the relative vector $\vec{r_1}-\vec{r_2}$.
The center of mass of the two particles
is then decoupled from their relative motion, and the evolution of the relative motion
is governed by the Hamiltonian:
\be
H_{\mbox{\scriptsize rel}} = {\vec{p}\,^2\over 2\mu} + V(\vec{r}\,)
\ee
where $\vec{r}=\vec{r_1}-\vec{r_2}$ is the vector of
coordinates of the relative motion, 
$\vec{p}=(\vec{p_1}-\vec{p_2})/2$ is the relative momentum and $\mu=m/2$ is the reduced mass.
We assume in what follows that the potential $V(\vec{r}\,)$ is vanishing in the limit
$r\rightarrow\infty$.

\subsubsection{General results of scattering theory}
The scattering states $\psi(\vec{r}\,)$ 
of the relative motion of the two particles are the eigenstates of $H_{\mbox{\scriptsize rel}}$
with positive energy $E$. Writing $E=\hbar^2k^2/2\mu$ and multiplying the eigenvalue equation
by $2\mu/\hbar^2$ we obtain
\be
(\Delta + k^2) \psi(\vec{r}\,) = {2\mu\over\hbar^2} V(\vec{r}\,)\psi(\vec{r}\,).
\label{eq:eve}
\ee
One has also to specify boundary conditions on $\psi$ to get the full description
of a scattering state. This is achieved by means of an integral formulation
of the eigenvalue equation.

\begin{itemize}
\item {\sl Integral equation}
\end{itemize}
To obtain the integral formulation of the scattering problem
we write the right hand side of the eigenvalue equation Eq.(\ref{eq:eve})
as a continuous sum of Dirac distributions:
\be
(\Delta + k^2) \psi(\vec{r}\,)
= \int d^3\vec{r}\,'{2\mu\over\hbar^2} V(\vec{r}\,')\psi(\vec{r}\,')\delta(\vec{r}-\vec{r}\,').
\ee
We then find a solution of this equation with a single Dirac distribution on the right hand side:
\be
(\Delta_{\vec{r}} + k^2) \psi_G(\vec{r}\,) = \delta(\vec{r}-\vec{r}\,')
\ee
having the form of an outgoing spherical wave for $r\rightarrow\infty$:
\be
\psi_G(\vec{r}\,) = -{1\over 4\pi} {e^{ik|\vec{r}-\vec{r}\,'|}\over |\vec{r}-\vec{r}\,'|}.
\ee
This is actually a Green's function of the operator $\Delta + k^2$.
The scattering state of the full problem can then be written as
\be
 \psi(\vec{r}\,) =  \psi_0(\vec{r}\,) -{2\mu\over 4\pi\hbar^2} \int d^3\vec{r}\,' 
{e^{ik|\vec{r}-\vec{r}\,'|}\over |\vec{r}-\vec{r}\,'|}
V(\vec{r}\,') \psi(\vec{r}\,')\,.
\label{eq:integrale}
\ee
The first term $\psi_0$ is the incoming free wave of the collision, solving
$(\Delta + k^2) \psi_0=0$; we simply assume here that the incoming wave is
a plane wave of wavevector $\vec{k}$:
\be
\psi_0({\vec{r}\,}) = \exp[i{\vec{k}}\cdot{\vec{r}}\,].
\ee
The remaining part of $\psi$ is then simply the scattered wave.

\begin{itemize}
\item {\sl Born expansion}
\end{itemize}
When the interaction potential is weak one sometimes expands the scattering state
$\psi$ in powers of $V$. In the integral formulation Eq.(\ref{eq:integrale})
of the eigenvalue equation
this corresponds to successive iterations of the integral, the approximation for
$\psi$ at order $n+1$ in $V$ being obtained by replacing $\psi$ by its approximation
at order $n$ in the right-hand side of the integral equation. E.g. to zeroth order
in $V$, $\psi=\psi_0$, and to first order in $V$ we get the so-called Born approximation:
\be
\psi_{\mbox{\scriptsize Born}}(\vec{r}\,) = \psi_0(\vec{r}\,) -{2\mu\over 4\pi\hbar^2} 
\int d^3\vec{r}\,'\;
{e^{ik|\vec{r}-\vec{r}\,'|}\over |\vec{r}-\vec{r}\,'|}
V(\vec{r}\,')\psi_0(\vec{r}\,').
\ee

\subsubsection{Low energy limit for scattering by a finite range potential}
Some results can be obtained in a simple way when the potential $V$ has a finite range $b$,
that is when it vanishes when $r>b$.

\begin{itemize}
\item {\sl asymptotic behavior for large $r$}
\end{itemize}
As the integration over the variable $\vec{r}\,'$ is limited to a range of radius $b$ one can
expand the distance from $\vec{r}$ to $\vec{r}\,'$ in powers of $r$ when $r\gg b$:
\be
|\vec{r}-\vec{r}\,'|= r -\vec{r}\,'\cdot\vec{n} + O\left({1\over r}\right)
\ee
where $\vec{n}=\vec{r}/r$ is the direction of scattering. The neglected term, scaling as
$b^2/r$, has a negligible contribution to the phase $\exp[ik|\vec{r}-\vec{r}\,'|]$ when
$r\gg kb^2$. One then enters the asymptotic regime for $\psi$:
\be
\psi(\vec{r}\,) = \psi_0(\vec{r}\,) +{e^{ikr}\over r} f_{\vec{k}}(\vec{n})+
O\left({1\over r^2}\right)
\ee
where the factor $f_{\vec{k}}$, the so-called scattering amplitude, does not depend
on the distance $r$:
\be
 f_{\vec{k}}(\vec{n}) =  -{2\mu\over 4\pi\hbar^2} \int d^3\vec{r}\,'\;
 e^{-ik\vec{n}\cdot\vec{r}\,'}
V(\vec{r}\,')\psi(\vec{r}\,').
\label{eq:scat_amp}
\ee

If the mean distance between the particles in the gas, on the order of $\rho^{-1/3}$, where
$\rho$ is the density, lies in the asymptotic regime for $\psi$ (that is $\rho^{-1/3}\gg b,kb^2$)
the effect of binary interactions on the macroscopic properties of the gas will be sensitive to
the scattering amplitude $f_{\vec{k}}$, and no longer to the details of the scattering
potential. This is the key property that we shall use later in this low
density regime to replace the exact interaction potential by a model potential having approximately the same scattering
amplitude.

\begin{itemize}
\item {\sl limit of low energy collisions}
\end{itemize}
Another simplification comes from the fact that collisions take place at low energy in the Bose condensed
gases: as $\hbar^2 k^2/2\mu$ is on the order of $k_BT$ in the thermal gas, 
$k$ becomes small at low temperature.

If $kb\ll 1$ the phase factor $\exp[-ik\vec{n}\cdot\vec{r}\,']$ becomes close to one in the integral
Eq.(\ref{eq:scat_amp})
giving the scattering amplitude. The scattering amplitude $f_{\vec{k}}$ then no longer depends on
the scattering direction $\vec{n}$, the asymptotic part of the scattered wave becomes
spherically symmetric (even if the scattering potential is not!): one then says that
scattering takes place
in the $s$-wave only.

Going to the mathematical limit $k\rightarrow 0$ we get for the scattering amplitude:
\be
f_{\vec{k}}(\vec{n}) \rightarrow -a.
\ee
The quantity $a$ is the so-called scattering amplitude; it will be the only parameter of our theory
describing the interactions between the particles, and our model potential will be adjusted to
have the same scattering length as the exact potential.
When $k$ is going to zero, the scattering state converges to the zero energy scattering state,
behaving for large $r$ as
\be
\psi_{E=0}(\vec{r}\,) = 1-{a\over r} +O\left({1\over r^2}\right).
\ee
A numerical calculation of this zero energy scattering state is an efficient way of
calculating $a$ for a given potential $V$. Note that there is of course no
connection between $a$ and
$b$, except for particular potentials like the hard sphere potential.

\subsubsection{Power law potentials}
In real life the interaction potential between atoms is not of finite range, as it contains
the Van der Waals tail scaling as $1/r^6$ for large $r$
\footnote{or even as $1/r^7$ if $r$ is larger than the optical
wavelength.}. It is fortunately possible to
show for the class of power-law potentials, scaling as $1/r^n$, that several of our
conclusions, obtained in the finite range case, hold provided that $n>3$.
E.g.\ in the limit of small $k$'s only the $s$-wave scattering survives, and $f_{\vec{k}}$
has a well defined limit for $k\rightarrow 0$, allowing one to define the scattering
length.

\subsection{The model potential used in this lecture}

\subsubsection{Why not keep the exact interaction potential ?}\label{subsubsec:why_not}
\begin{figure}[htb]
\epsfysize=8cm \centerline{\epsfbox{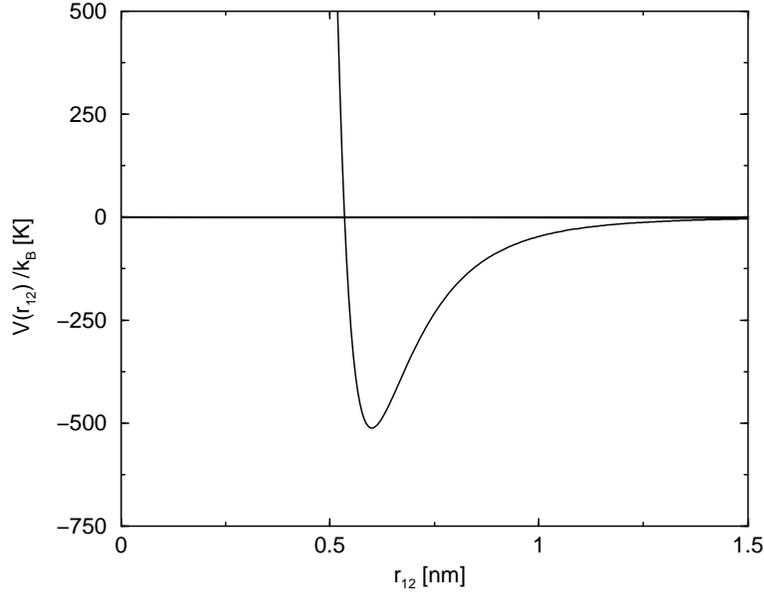}}
\caption{\small Typical shape of the interaction potential between two atoms, as function of 
the interatomic distance $r_{12}$. The numbers are indicative and correspond
to cesium.}\label{fig:cs}
\end{figure}

For alkali atoms the exact interaction potential has a repulsive hard core, is very deep (as deep
as $10^3$ Kelvins times $k_B$ 
for ${}^{133}$Cs), has a minimum at a  distance $r_{12}$ on the order
of 6 \AA (for cesium), 
and contains many bound states corresponding to molecular states of two alkali
atoms (see figure \ref{fig:cs}).

There are several disadvantages to use the exact interaction potential in a theoretical treatment 
of Bose-Einstein condensation:
\begin{enumerate}
\item $V$ is difficult to calculate precisely, and a small error on $V$ may result in a large
error on the scattering length $a$. In practice $a$ is measured experimentally, and this is
the most relevant information on $V$ in the low density, low temperature limit.
\item the presence of bound states of $V$ with a binding energy much smaller than the temperature
of the gas (there are 9 
orders of magnitude between the potential depth $10^3$K and the gas temperature
$\simeq 1\mu$K) clearly indicates that the Bose condensed gases are in a metastable state;
at the experimental temperatures and densities the complete thermal equilibrium of the 
system
would be a solid. Direct thermal equilibrium theory, such as the thermal $N$-body
density matrix $\exp[-\beta H]$, cannot therefore be used with $V$. This is why
even in the exact
Quantum Monte Carlo calculations performed for alkali gases \cite{Werner1} $V$ is
replaced by a hard sphere potential. Such a complication was absent for liquid helium,
where the well-known exact $V$ can be used \cite{Ceperley}.
\item $V$ can not be treated in the Born approximation, because 
it is very strongly repulsive at short distances and has many bound states: even
if the scattering length was zero, one would have to resum the whole Born series to obtain the
correct result
[We recall that for a potential as gentle as a square well of 
radius $b$, the Born approximation applies when
the zero-point energy for confinement within a domain of radius $b$, 
$\hbar^2 /2\mu b^2$, 
is much larger than the potential depth, which implies that no bound state is present in
the well.] As a consequence naive mean field approximations, which neglect 
the correlations between
particles due to interactions, implicitly relying on the Born approximation, cannot be used with the exact $V$.
\end{enumerate}

The key idea is therefore to replace the exact interaction potential by a model potential
(i) having the same scattering properties at low energy, that is the same scattering length,
and (ii) which should be treatable in the Born approximation, so that naive mean field approaches
apply.

The model potential satisfying these requirements with the minimal number of parameters (one!)
is the zero-range pseudo-potential initially introduced by Enrico Fermi
\cite{Fermi, Huang} and having the following
action on any two-body wavefunction:
\be
\langle{\vec{r_1}},{\vec{r_2}}|V|\psi_{1,2}\rangle \equiv
g\delta(\vec{r_1}-\vec{r_2}) \left[{\partial\over\partial{r_{12}}}\left(r_{12}
\psi_{1,2}(\vec{r_1},\vec{r_2})\right)\right]_{r_{12}=0}.
\label{eq:ppf}
\ee
The factor $g$ is the so-called coupling constant
\be
g={4\pi \hbar^2\over m}a
\ee
where $a$ is the scattering length of the exact potential. The pseudo-potential involves a Dirac
distribution and a regularizing operator.

\begin{itemize}
\item {\sl Effect of regularization}
\end{itemize}
When the wavefunction $\psi_{1,2}$ is regular close to $\vec{r_1}=\vec{r_2}$, one can check that
the regularizing operator has no effect, so that the pseudo-potential can be viewed as a mere
contact potential $g\delta(\vec{r_1}-\vec{r_2})$.

When the wavefunction  $\psi_{1,2}$ has a $1/r_{12}$ divergence:
\be
\psi_{1,2}(\vec{r_1},\vec{r_2}) = {A(\vec{r_1}+\vec{r_2})\over r_{12}} +\mbox{regular}
\ee
where $A$ is the function of the center of mass coordinates only
the regularizing operator removes the diverging part:
\be
{\partial\over\partial{r_{12}}}\left(r_{12}{A(\vec{r_1}+\vec{r_2})\over r_{12}}\right) =0.
\ee
In this way we have extended the Hilbert space of the state vectors of the particles with
wave functions diverging as $1/r_{12}$; note that these wavefunctions remain square
integrable, as the element of volume scales as $r_{12}^2$ in 3D. As we shall see this $1/r_{12}$
divergence is a consequence of the zero-range of the pseudo-potential.

\subsubsection{Scattering states of the pseudo-potential}
Turning back to the relative motion of two particles we now derive the scattering
states of the pseudo-potential from the integral equation Eq.(\ref{eq:integrale}). As the pseudo-potential
involves a Dirac $\delta(\vec{r}\,')$ the integral over $\vec{r}\,'$ can be performed explicitly:
\be
\psi(\vec{r}\,) = e^{i\vec{k}\cdot\vec{r}} -a {e^{ikr}\over r} 
\left[{\partial\over\partial{\vec{r}\,'}}\left(r'\psi(\vec{r}\,')\right)\right]_{r'=0}.
\label{eq:pour_C}
\ee
As the factor 
\be
C=\left[{\partial\over\partial{\vec{r}\,'}}\left(r'\psi(\vec{r}\,')\right)\right]_{r'=0}
\ee
does not depend on $\vec{r}$ we find that $\psi$ has the standard asymptotic behavior of a scattering
state in $r$ but {\sl everywhere} in space, not only for large $r$. This is due to the zero-range
of the pseudo-potential.
To calculate $C$, we multiply Eq.(\ref{eq:pour_C}) by $r$, we take the derivative with respect
to $r$ and set $r$ to zero. On the left hand side we recover the constant $C$ by
definition. We finally obtain:
\be
C= 1 -aCik
\ee
so that $C=1/(1+ika)$ and the scattering states of the pseudo-potential are exactly given by
\be
\psi_{\vec{k}}(\vec{r}\,) = e^{i\vec{k}\cdot\vec{r}} -{a\over 1+ika} {e^{ikr}\over r}.
\label{eq:scatt_pp}
\ee

The corresponding scattering amplitude,
\be
f_k = -{a\over 1+ika}
\ee
does not depend on the direction of scattering, so that the pseudo-potential scatters 
only in the $s$-wave, whatever the modulus $k$ is. The scattering length of the pseudo-potential,
$-f_{k=0}=a$, coincides with the one of the exact potential. 

Finally we note that the total cross-section for scattering of identical bosons 
by the pseudo-potential is given by a Lorentzian in $k$,
\be
\sigma = 8\pi |f_{\vec{k}}(\vec{n})|^2 =  {8\pi a^2\over 1+k^2 a^2},
\ee
and that the pseudo-potential obeys the optical theorem.

\subsubsection{Bound states of the pseudo-potential}
As a mathematical curiosity we now point out that not only the scattering states but also the bound
states of the pseudo-potential can be calculated. A first way of obtaining the bound states
is a direct solution of Schr\"odinger's equation. A more amusing way is to use the
following closure relation:
\be
\int {d^3\vec{k}\over (2\pi)^3} |\psi_{\vec{k}}\rangle \langle|\psi_{\vec{k}}| =
1 - P_{\mbox{\scriptsize bound}}
\ee
where $|\psi_{\vec{k}}\rangle$ is the scattering state given in Eq.(\ref{eq:scatt_pp}) and
$ P_{\mbox{\scriptsize bound}}$ is the projector on the bound states of the pseudo-potential.

In calculating the matrix elements of this closure relation between perfectly localized state vectors
$|\vec{r}\,\rangle$ and $|\vec{r}\,'\rangle$ and using spherical coordinates for the integration over $\vec{k}$ 
one ultimately faces the following type of integrals:
\be
I=\int_{-\infty}^{+\infty} dk\; {e^{ik(r+r')}\over 1+ika}.
\label{eq:tbc}
\ee
We calculate $I$ using the residues formula, by extending the integration variable $k$
to the complex plane and closing the contour of integration by a circle of infinite radius,
which has to be in the upper half of the complex plane as $r+r'>0$. 
As the integrand in $I$ has a pole in $k=i/a$, we find that $I$ vanishes for $a<0$, as the
pole is then in the lower half of the complex plane. For $a>0$ the pole gives a non-zero
contribution to the integral:
\be
I={2\pi\over a} e^{-(r+r')/a}.
\ee
Finally we find that $ P_{\mbox{\scriptsize bound}}=0$ for $a<0$, corresponding to the absence
of bound states, and $ P_{\mbox{\scriptsize bound}}=|\psi_{\mbox{\scriptsize bound}}\rangle \langle 
\psi_{\mbox{\scriptsize bound}}|$ for $a>0$, corresponding to the existence of a single
bound state:
\be
\psi_{\mbox{\scriptsize bound}}(\vec{r}\,) = {1\over\sqrt{2\pi a}} {e^{-r/a}\over r}.
\ee
From Schr\"odinger's equation, we find for the energy of the bound state:
\be
E_{\mbox{\scriptsize bound}} = - {\hbar^2\over ma^2}.
\ee

The existence of a bound state for $a>0$ and its absence for $a<0$ is a paradoxical
situation. As we shall see in the mean field approximation, the case $a>0$ corresponds to effective repulsive 
interactions between the atoms, whereas the case $a<0$ corresponds to effective
attractive interactions. In the purely 1D case, the situation is more intuitive,
the potential $g_{1D} \delta(x)$ having a bound state only in the effective attractive
case $g_{1D}<0$. This paradox in 3D comes from the non-intuitive effect of the regularizing
operator (an operation not required in 1D), which makes the pseudo-potential different
from a delta potential; actually one can shown in 3D that a delta potential viewed as
a limit of square well potentials with decreasing width $b$ and constant area does not
scattered in the limit $b\rightarrow 0$.

\subsection{Perturbative vs non-perturbative regimes for the pseudo-potential}
\subsubsection{Regime of the Born approximation}\label{subsubsec:rba}
As we will use mean field approximations requiring that the scattering potential is treatable
in the Born approximation, we identify the regime of validity of the Born approximation
for the pseudo-potential.

As we have seen in the previous subsection the integral equation for the scattering states
of the pseudo-potential can be reduced to the equation for $C$:
\be
C = 1-ika C,
\label{eq:sur_C}
\ee
the scattering state being given by
\be
\psi_{\vec{k}}(\vec{r}\,) = e^{i\vec{k}\cdot\vec{r}} -aC {e^{ikr}\over r}.
\ee
The Born expansion will then reduces to iterations of Eq.(\ref{eq:sur_C}).
To zeroth order in the interaction potential, we obtain $C_0=0$ so that $\psi_{\vec{k}}$ reduces to the
incoming wave.
To first order, we get the Born approximation 
\be
C_1 = 1 -ika C_0 = 1.
\ee
To second order and third order we obtain
\bea
C_2 &=&1 -ika C_1 = 1-ika \\
C_3 &=&1 -ika C_2 = 1-ika +(ika)^2
\eea
so that the Born expansion is a geometrical series expansion of the exact result
$C=1/(1+ika)$ in powers of $ika$.

The validity condition of the Born approximation is that the first order result is a small correction
to the zeroth order result. For the scattering amplitude  this requires
\be
k|a| \ll 1.
\ee
For the scattering state this requires 
\be
r\gg a. 
\ee
If one takes for $r$ the typical distance $\rho^{-1/3}$ between the particles
in the gas, where $\rho$ is the density, this leads to 
\be
\rho^{1/3} |a| \ll 1.
\label{eq:sgp}
\ee

\begin{itemize}
\item {\sl Are the conditions for the Born approximation satisfied in the experiments ?}
\end{itemize}

To estimate the order of magnitude of $k$ we average $k^2$ over a Maxwell-Boltzmann distribution 
of atoms with a temperature $T=1\mu$K, typically larger than the critical temperature for alkali
gases; the average gives a root mean square for $k$ equal to
\be
\Delta k = \left({3mk_BT\over2\hbar^2}\right)^{1/2}.
\label{eq:dk}
\ee
For ${}^{23}$Na atoms used at MIT, with a scattering length of 
$50\, a_{\mbox{\scriptsize Bohr}}$, where the Bohr
radius is $a_{\mbox{\scriptsize Bohr}}=0.53$ \AA,
we obtain $\Delta k\, a =2\times 10^{-2}$.
For rubidium ${}^{87}$Rb atoms used at JILA, with a scattering length of 
$110 \, a_{\mbox{\scriptsize Bohr}}$,
we obtain $\Delta k \, a =0.1$. 

In the case of an almost pure condensate in a trap, the typical $k$ is given by the inverse of the size
$R$ of the condensate, as the condensate wavefunction is not very far from a minimum uncertainty state.
Generally this results in a much smaller $\Delta k$ than Eq.(\ref{eq:dk}), as $R$ is much larger than 
the thermal de Broglie wavelength. One could however imagine a condensate in a very strongly confining
trap, such that $R$ would become close to $a$; in this case, not yet realized,
the mean field theory has to be revisited.

We turn to the second condition Eq.(\ref{eq:sgp}). The typical densities of condensates
are on the order of $2\times 10^{14}$ atoms per cm${}^3$. For the scattering
length of sodium this leads to $\rho^{1/3} a \simeq 0.015 \ll 1$. For the scattering
length of rubidium this leads to $\rho^{1/3} a \simeq 0.034 \ll 1$. 
Both conditions for the Born approximation applied
to the pseudo-potential are therefore satisfied.

\subsubsection{Relevance of the pseudo-potential beyond the Born approximation}
Let us try to determine necessary validity conditions for the substitution of the exact interaction
potential by the pseudo-potential.

First one should be in a regime dominated by $s$-wave scattering, as the pseudo-potential neglects
scattering in the other wave. This condition is easily satisfied in the $\mu$K temperature
range for Rb, Na.

Second the scattering amplitude of the exact potential in $s$-wave should be well approximated
by the pseudo-potential. For isotropic potentials vanishing for large $r$ as $1/r^n$, with $n>5$,
the $s$-wave scattering amplitude has the following low $k$ expansion:
\be
f_k^{s=0} = -\frac{1}{ a^{-1}+ik-\frac{1}{2} k^2r_e +\ldots}
\ee
where $r_e$ is the so-called effective range of the potential. To this order in $k$ the result of the pseudo-potential
corresponds to the approximation $r_e=0$. When $r_e$ is on the order of $a$
(which is the case for a hard sphere potential, but not necessarily true for a more general
potential) the term in $r_e$ can be neglected if $k^2 r_e \ll 1/a$, that is $(ka)^2\ll 1$;
there is therefore no meaning to use the pseudo-potential beyond the Born regime.

Consider now the case $r_e \ll |a|$. The term $r_e k^2$ remains small as compared to $1/a$ for $k|a|<1$.
For $k|a|\gg 1$ the term $ik$ dominates over $1/a$; $k^2 r_e$ remains small as compared 
to $i k$ as long as $k r_e \ll 1$. The use of the pseudo-potential may then extend beyond
the Born approximation. 

An example of a situation with $r_e \ll |a|$ is the so-called zero
energy resonance, where $a$ is diverging. When a bound state of the interaction
potential is arbitrarily
close to the dissociation limit, the scattering length diverges $a\rightarrow +\infty$, the bound
state has a large tail in $r$ scaling as $e^{-r/a}/r$ and the bound state energy scales
as $-\hbar^2/ma^2$ \cite{Landau2,Joachain}. These scaling laws hold for the pseudo-potential,
as we have seen. 
\section{Interacting Bose gas in the Hartree-Fock approximation}
\markright{Hartree-Fock approximation}
\label{Cap:HF}

Now that we have identified a simple model interaction potential treatable
in the Born approximation we use it in the simplest possible mean field approximation,
the so-called Hartree-Fock approximation. This approximation was applied
to trapped gases for the first time in 1981 (see \cite{Silvera})!

\subsection{BBGKY hierarchy}
The Hartree-Fock mean field approximation can be implemented in a variety of ways.
We have chosen here the approach in terms of
the BBGKY hierarchy, truncated to first order. 

\subsubsection{Few body-density matrices}
We have already introduced in \S\ref{Cap:ideal} the concept of the one-body density
matrix. We revisit here this notion and extend it to two-body density matrices.

\begin{itemize}
\item {\sl For a fixed total number of particles}
\end{itemize}
Let us first consider a system with a fixed total number of particles $N$ and
let $\sigma_{1,2...N}$ be the $N$-body density matrix. Starting from $\sigma_{1,2...N}$ we introduce simpler
objects as the one-body and two-body density matrices $\hat{\rho}_1$ and 
$\hat{\rho}_{12}$, by taking the trace
over the states of all the particles but one or two:
\bea
\hat{\rho}_1^{(N)} &=& N\, \mbox{Tr}_{2,3...N} (\sigma_{1,2,...N}) \\
\hat{\rho}_{12}^{(N)} &=& N (N-1)\, \mbox{Tr}_{3,4...N} (\sigma_{1,2,...N}) \, .
\eea
In practice the knowledge of $\hat{\rho}_1$ and $\hat{\rho}_{12}$ is sufficient to describe most
of the experimental results.
As you know, $\langle \vec{r}\, |\hat{\rho}_1 | \vec{r}\, \rangle$ is the density of particles and  
$\langle \vec{r_1},\vec{r_2} |\hat{\rho}_1 |  \vec{r_1},\vec{r_2} \rangle$ is the pair distribution
function.

\begin{itemize}
\item {\sl For a fluctuating total number of particles}
\end{itemize}
If $N$ fluctuates according to the probability distribution $P_N$, we define few-body density matrices
by the following averages over $N$:
\bea
\hat{\rho}_1 &=& \sum_N P_N \,\hat{\rho}_1^{(N)} \\
\hat{\rho}_{12} &=& \sum_N P_N \,\hat{\rho}_{12}^{(N)}
\eea

Alternatively on can define directly the one-body and two-body density matrices in second quantization:
\bea
\langle \vec{r_1} |\hat{\rho}_1 | \vec{r_2} \rangle &= &
 \langle \hat{\psi}^\dagger({\vec{r_2}})  \hat{\psi}({\vec{r_1}}) \rangle \\
\langle \vec{r_1},\vec{r_2} |\hat{\rho}_{12} |  \vec{r_3},\vec{r_4} \rangle &=  &
 \langle \hat{\psi}^\dagger({\vec{r_3}})  \hat{\psi}^\dagger({\vec{r_4}})
  \hat{\psi}({\vec{r_2}})  \hat{\psi}({\vec{r_1}}) \rangle.
\eea

Note that the few-body density matrices are normalized as
\bea
\mbox{Tr}[ \hat{\rho}_1] &=&\langle N \rangle \\
\mbox{Tr}[ \hat{\rho}_{12}] &=&\langle N(N-1) \rangle
\eea
so that one can obtain  the
variance of the
fluctuations in the number of atoms from the one-body and two-body density matrices.

\subsubsection{Equations of the hierarchy}
The idea of our derivation of the mean field approximation is to get 
an approximate closed equation for $\hat{\rho}_1$ 
by closing the hierarchy with some \lq\lq cooking recipe" giving $\hat{\rho}_{12}$ in terms of $\hat{\rho}_1$. 

To derive the first equation of the hierarchy we start from the exact master equation:
\be
i \hbar \frac{d}{dt} \sigma_{1,2..N} = [ H, \sigma_{1,2..N}] 
\ee
where the Hamiltonian is the sum of one-body and two-body terms:
\be
H = \sum_{i=1}^N h_i + \frac{1}{2} \sum_{i \neq j; i,j=1}^N V_{ij}.
\ee
The single particle Hamiltonian
$h_i$ contains the kinetic and trapping potential energy of the atom $i$ and 
$V_{ij}$ in the interaction potential between the atoms $i$ and $j$.
Now we take the trace of the master equation over the particles 2,3 ...$N$ and 
multiply it by $N$, obtaining
\be
i \hbar \frac{d}{dt} \hat{\rho}_1 =[h_1,\hat{\rho}_1] + N \mbox{Tr}_{2,...N} \left\{
\sum_{j=2}^N [V_{1j}, \sigma_{1,2....N}]\right\}.
\ee
We have kept here only the terms involving the atom 1, as the other terms are commutators
of vanishing trace. The sum over $j$ amounts to $N-1$ times the same contribution, e.g.\ the $j=2$
contribution, as the atoms are indiscernible. We finally obtain the first equation of the
hierarchy:
\be
i \hbar \frac{d}{dt} \hat{\rho}_1 =[h_1,\hat{\rho}_1] + \mbox{Tr}_{2} \{[ V_{12}, \hat{\rho}_{12}]\}.
\label{eq:rho1dot}
\ee

The equation Eq.(\ref{eq:rho1dot}) is not closed for $\hat{\rho}_1$, as it involves $\hat{\rho}_{12}$. 
The next equation of the hierarchy,
the equation for $\hat{\rho}_{12}$, involves
$\hat{\rho}_{123}$, etc, up to the $N$-body density matrix, where the hierarchy terminates. 
The mean field approximation consists in
replacing $\hat{\rho}_{12}$ by an ad hoc function of $\hat{\rho}_1$. 

\subsection{Hartree-Fock approximation for $T>T_c$}

\subsubsection{Mean field potential for the non-condensed particles}
We use the following simple approximation to break the hierarchy:
\be
\hat{\rho}_{12} \simeq \hat{\rho}_{12}^{HF} = 
\left(\frac{1+P_{12}}{\sqrt{2}} \right) \hat{\rho}_1 \otimes \hat{\rho}_1  \left(\frac{1+P_{12}}{\sqrt{2}} \right)
= (1+P_{12})  \hat{\rho}_1 \otimes \hat{\rho}_1 
\label{eq:sym_bad}
\ee
where $P_{12}$ is the permutation operator exchanging the states of the particles 1 and 2.
The last identity in (\ref{eq:sym_bad}) 
is obtained by using the commutation of $P_{12}$ and $\hat{\rho}_1 \otimes \hat{\rho}_1$, and
the fact that $P_{12}^2=1$.

The factorized prescription $\hat{\rho}_{12}=\hat{\rho}_1 \otimes \hat{\rho}_1$ is the Hartree approximation. It assumes weak
correlations between the particles. Indeed at short distances $r_{12}$, the real $\hat{\rho}_{12}$ is expected
to be a statistical mixture
of scattering states of the interaction potential. Neglecting the correlations in $\hat{\rho}_{12}$ between particles $1$ and
$2$ amounts to considering only separable, plane wave  scattering 
states, which corresponds to the zeroth order in the Born expansion of the scattering theory. 
Actually $\hat{\rho}_{12}$
appears in Eq.(\ref{eq:rho1dot}) inside a commutator with $V_{12}$, so that taking the zeroth order approximation
for the scattering states 
in $\hat{\rho}_{12}$ corresponds to the first order of the Born approximation in the equation for
$\hat{\rho}_1$. 

As we are dealing with bosons we have supplemented the Hartree approximation by a 
bosonic symmetrization procedure, involving
the permutation operator $P_{12}$.
Note that the symmetrization as it was written works only for particles 1 and 2 in orthogonal states:
\be
\frac{1+P_{12}}{\sqrt{2}}|\alpha\rangle|\beta\rangle = {|\alpha\rangle|\beta\rangle + |\beta\rangle|\alpha\rangle\over
\sqrt{2}}
\ee
as the factor $\sqrt{2}$ is the correct normalization factor only in this case.
This is almost true for a non-degenerate Bose gas.
This restriction forces us to treat separately the case in which a condensate is present ($T<T_c$).

We now insert the Hartree-Fock ansatz for $\hat{\rho}_{12}$ in the hierarchy
\footnote{Note that for the present
calculation the regularization of the pseudo-potential is not necessary. Indeed by considering plane
waves as scattering states in $\hat{\rho}_{12}$ we suppress any problem of divergences in the commutator
with $V_{12}$, and we can then take $V_{12}$ as a simple delta distribution.}
\be
i \hbar \frac{d}{dt} \hat{\rho}_1 =[h_1,\hat{\rho}_1] + \mbox{Tr}_{2} \{[ V_{12}, \hat{\rho}_{12}^{HF}]\}.
\ee
In the commutator with $V_{12}$ we will encounter
\be
\delta(\vec{r_1}-\vec{r_2}) (1+P_{12}) = (1+P_{12}) \delta(\vec{r_1}-\vec{r_2}) = 2  \delta(\vec{r_1}-\vec{r_2}).
\ee
The fact that $P_{12}$ commutes with $V_{12}$ is due to the parity of the delta distribution, and
$P_{12}$ acting on a state with two particles at the same position can be replaced by the identity. 
As a consequence, with our zero-range interaction potential, the Fock term simply doubles the Hartree term. 
We finally obtain 
\be i \hbar \frac{d}{dt} \hat{\rho}_1 = 
\left[\frac{\vec{p}\,^2}{2m} +U(\vec{r}\,) + {\cal V}(\vec{r}\,), \hat{\rho}_1\right]
\ee
where ${\cal V}(\vec{r}\,)$ is the mean field potential
\be
{\cal V}(\vec{r}\,) = 2 g \langle \vec{r}\, |\hat{\rho}_1| \vec{r}\, \rangle = 2 g \rho(\vec{r}\,).
\label{eq:mfpot}
\ee
The Hartree-Fock Hamiltonian is then
\be
h^{HF}(1) = \frac{\vec{p}\,^2}{2m} + U(\vec{r}\,) + 2 g \rho(\vec{r}\,).
\label{eq:hfh}
\ee

The problem is then formally reduced to the one of an ideal Bose gas moving in a self-consistent potential. For $g>0$ the
mean field corresponds to repulsive  interactions, as $ 2 g \rho(\vec{r}\,)$ expels the
atoms from the
region of high density, while for  $g<0$ the mean field corresponds to attractive interactions.

\subsubsection{Effect of interactions on $T_c$}
Let us now consider the Hartree-Fock one-body density matrix at thermal equilibrium; 
we use the same formula as the
ideal Bose gas Eq.(\ref{eq:rho1ideal}), replacing $h_1$ by the Hartree-Fock Hamiltonian:
\be
\hat{\rho}_1 = \{ \exp\left[\beta \left(h^{HF}(1)-\mu \right)\right] -1  \}^{-1}\,.
\ee

For $k_B T \gg \Delta E$ where where $\Delta E$ is the level spacing of $h^{HF}$ we can perform
the semiclassical approximation. We obtain for the spatial density as in Eq.(\ref{eq:secforrho}):
\be
\rho_{sc}(\vec{r}\,)=
\frac{1}{\ldB^3} g_{3/2} ({\mathrm z} \exp [ -\beta (U(\vec{r}\,) + 2 g \rho_{sc}(\vec{r}\,) )] )
\ee
At $T=T_c$ the argument of $ g_{3/2}$ goes to 1 in the point $\vec{r}_{\mbox{\scriptsize min}}$ 
where the potential is minimal,
so that Einstein's condition still holds in the Hartree-Fock approximation:
\be
\rho_{sc}(\vec{r}_{\mbox{\scriptsize min}}) \ldB^3 = \zeta(3/2).
\ee
For the harmonic trap $U(\vec{r}\,)=m\omega^2 r^2/2$
the minimum occurs at the center of the trap, $\vec{r}_{\mbox{\scriptsize min}}=\vec{0}$
so that the chemical potential at the phase transition  is given by
\be
\mu = 2 g \rho_{sc}(\vec{0}).
\ee
It is shifted by the mean field effect with respect to the ideal Bose gas.
Using as a small parameter $\rho_{sc}(\vec{0}) g/k_B T_c^0$, 
one can derive at constant $N$ \cite{StringTc}
the first order change in 
the critical temperature with respect to $T_c^0$, the transition temperature
of the ideal Bose gas:
\be
{\delta T_c\over T_c^0} = -2.5 \rho_{sc}^{1/3}(\vec{0}) a = 
-1.33 {a\over (\hbar/m\omega)^{1/2}}N^{1/6}.
\ee
For $N=10^7$ atoms of ${}^{23}$Na in a trap of harmonic frequency $\omega=2\pi\times 100$Hz, with a scattering
length $a=50\, a_{\mbox{\scriptsize Bohr}}$ we find
$T_c^0\simeq 1\mu$K, and $\delta T_c/ T_c^0\simeq -2.5\times 10^{-2}$, 
an effect for the moment smaller than the
experimental accuracy.
The fact that $\delta T_c$ is negative 
for effective repulsive interactions ($a>0$) is intuitive: for fixed
values of $N$ and $T$ the interacting gas has a lower density at the center of the trap than the ideal Bose gas,
so that one needs to further cool the gas to get Bose-Einstein condensation.

\begin{itemize}
\item {\sl A calculation of $\delta T_c$ beyond mean field}
\end{itemize}
The purest situation to study the effect of the interactions on the critical
temperature $T_c$ is the case of atoms trapped
in a flat bottom potential; in this case the density is uniform, the previously mentioned intuitive mean field
effect is suppressed, and our Hartree-Fock theory predicts the same critical temperature as the ideal
Bose gas. This prediction is actually not correct, and rigorous results for the first order correction
of $T_c$ in $a\rho^{1/3}$ have been obtained recently, by a combination of perturbative theory
and Quantum Monte Carlo calculations \cite{Werner2}:
\be
{\delta T_c^{\mbox{\scriptsize box}}\over T_c^0} = (2.2\pm 0.25) a\rho^{1/3} + o(a\rho^{1/3}).
\ee
Recent calculations in the many body Green's function formalism confirm this result
\cite{Laloe}.
This effect, if heuristically extended to the trap, is of opposite sign and of
the same order of magnitude as the mean-field prediction.

\subsection{Hartree-Fock approximation in presence of a condensate}
\subsubsection{Improved Hartree-Fock Ansatz}
As already emphasized in the previous subsection the symmetrization procedure of the Hartree-Fock
prescription Eq.(\ref{eq:sym_bad}) has to be modified in presence of a condensate.
To this end we split the one-body density matrix as 
\be
\hat{\rho}_1 = \langle N_0\rangle |\phi\rangle \langle\phi| + \hat{\rho}_1'
\ee
where $\phi$ is the condensate wavefunction, $\langle N_0\rangle$ is the mean number of particles
in the condensate and $\hat{\rho}_1'$ is the one-body density matrix of the non-condensed fraction.
The Hartree approximation for the two-body density matrix now reads:
\be
\hat{\rho}_1\otimes\hat{\rho}_1 = \langle N_0\rangle^2 |\phi,\phi\rangle \langle\phi,\phi| + 
\mbox{remaining Hartree part}.
\label{eq:remain}
\ee
The first term in the right hand size is already symmetrized; the second term can be symmetrized
as in  Eq.(\ref{eq:sym_bad}) as it does not involve coexistence of two atoms in  the (only)
macroscopically populated state $\phi$. We therefore put forward the following Hartree-Fock
ansatz:
\be
\hat{\rho}_{12}^{HF} = \langle N_0\rangle^2 |\phi,\phi\rangle \langle\phi,\phi| +
\left(\frac{1+P_{12}}{\sqrt{2}} \right) \mbox{remaining Hartree part}  
\left(\frac{1+P_{12}}{\sqrt{2}} \right).
\ee
Eliminating the remaining Hartree part with the help of Eq.(\ref{eq:remain}), we finally obtain
\be
\hat{\rho}_{12}^{HF} =
\left(\frac{1+P_{12}}{\sqrt{2}} \right) \hat{\rho}_1\otimes\hat{\rho}_1 \left(\frac{1+P_{12}}{\sqrt{2}} \right)
-\langle N_0\rangle^2 |\phi,\phi\rangle \langle\phi,\phi|.
\ee
In this way we have avoided the double counting of the condensate contribution that would
have resulted from the prescription Eq.(\ref{eq:sym_bad}).

\subsubsection{Mean field seen by the condensate}

We replace $\hat{\rho}_{12}$ in the first equation of the hierarchy by the improved
Hartree-Fock ansatz. The first bit of the ansatz gives the same result as in the case $T>T_c$,
the second bit involves the term:
\be
\mbox{Tr}_2\left([\delta(\vec{r_1}-\vec{r_2}),|\phi,\phi\rangle \langle\phi,\phi|]\right)= 
[|\phi(\vec{r_1})|^2,|\phi\rangle \langle\phi|].
\ee
Splitting $\hat{\rho}_1$ as condensate and non-condensed contribution we arrive at
\bea
i \hbar \frac{d}{dt} \hat{\rho}_1 &=& \left[
\frac{\vec{p}\,^2}{2m} +U(\vec{r}\,) + 2g\rho(\vec{r}\,)-\langle N_0\rangle g
|\phi(\vec{r}\,)|^2, \langle N_0\rangle |\phi\rangle \langle\phi| \right] \\
&+& \left[\frac{\vec{p}\,^2}{2m} +U(\vec{r}\,) + 2g\rho(\vec{r}\,), \hat{\rho}_1'\right].
\eea

The non-condensed particles still move in the mean field potential $2g\rho(\vec{r}\,)$. 
On the contrary
the atoms in the condensate see a different mean field potential:
\be
2g\rho(\vec{r}\,)-g\langle N_0\rangle |\phi(\vec{r}\,)|^2 = 2g \rho'(\vec{r}\,) +
g\langle N_0\rangle |\phi(\vec{r}\,)|^2
\ee
where $\rho'$ is the non-condensed density and $\langle N_0\rangle |\phi|^2$ is the condensate density.
\footnote{A careful reader may argue that we forget 
here the condition of orthogonality of the eigenstates of $\hat{\rho}_1'$ to $\phi$.
Inclusion of this condition 
is beyond accuracy of the Hartree-Fock approximation.
It will be carefully included in the more
precise number conserving Bogoliubov approach of
\S\ref{Cap:bogol}.}
This result can be interpreted as follows: An atom in the condensate interacts with non-condensed
particles with the effective coupling constant $2g$, and it 
interacts with another particle of the condensate with the effective coupling constant $g$.

For repulsive effective interactions ($g>0$) this is at the basis of Nozi\`eres'argument 
against fragmentation of the condensate in several orthogonal states: in a box of size $L$
in the thermodynamical limit, transferring a finite fraction of condensate particles from the
plane wave $\vec{p}=\vec{0}$ to an excited plane wave $p=O(\hbar/L)$ costs a negligible amount
of kinetic energy per particle but a finite amount of interaction energy. The transferred fraction
would indeed be repelled with a stronger amplitude ($2g$ rather than $g$) by the atoms remaining in the
condensate. 

\subsubsection{At thermal equilibrium}\label{subsubsec:hfte}
At thermal equilibrium the one-body density matrix of non-condensed atoms is given
by the usual Bose distribution for the ideal Bose gas, with the trapping potential
being supplemented by the mean-field potential:
\be
\hat{\rho}_1' = \frac{1}{\exp\left\{\beta\left[{\vec{p}\,^2\over 2m}
+U(\vec{r}\,)+2g\rho(\vec{r}\,)-\mu\right]\right\}-1}
\label{eq:self1}
\ee

The condensate wave function has to be a steady state of the total, mean field plus trapping 
potential seen by an atom in the condensate:
\be
\lambda\phi(\vec{r}\,) = -{\hbar^2\over 2m}\Delta \phi
+[U(\vec{r}\,)+g\langle N_0\rangle |\phi(\vec{r}\,)|^2+2g\rho'(\vec{r}\,)]\phi(\vec{r}\,).
\label{eq:self2}
\ee
The Hartree-Fock single particle energy $\lambda$ should not be confused with the energy per particle
in the condensate, as it will become clear in the next section. The occupation number of the
condensate is related to $\lambda$ by the Bose formula:
\be
\langle N_0\rangle = {1\over e^{\beta(\lambda-\mu)}-1}.
\label{eq:self3}
\ee
We now have to solve in a self consistent way the three equations Eq.(\ref{eq:self1},\ref{eq:self2},
\ref{eq:self3}).
In practice, when $\langle N_0\rangle$ is already large, one can assume $\lambda=\mu$, which eliminates
one unknown $\lambda$ and one equation Eq.(\ref{eq:self3}).

\subsection{Comparison of Hartree-Fock to exact results}
\subsubsection{Quantum Monte Carlo calculations}
The Quantum Monte Carlo method developed by David Ceperley and others allows to sample in an exact way
the $N$-body distribution function of a gas of $N$ interacting bosons at thermal equilibrium.
I.e.\ the algorithm generates random positions $\vec{r_1},...,\vec{r_N}$ for the $N$ 
particles
with a probability distribution given by the exact $N$-body distribution function of the atoms.

\begin{figure}[htb]
\centerline{
\epsfysize=6cm \epsfbox{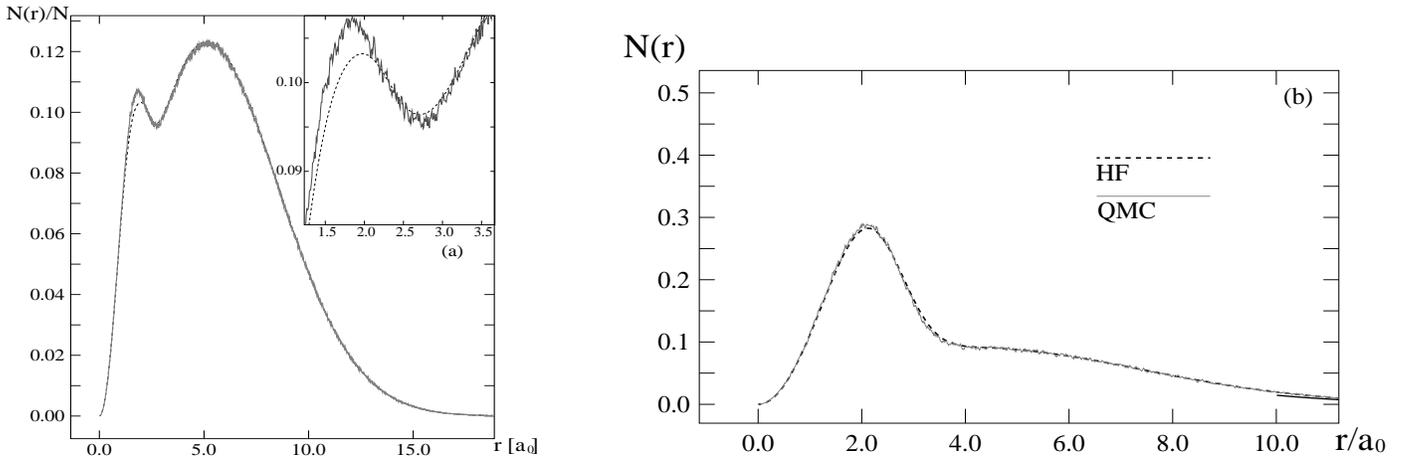} 
\epsfysize=6cm \epsfbox{fig08b.eps}}
\caption{\small Radial density of particles, $r^2\rho(r)$, for an interacting Bose gas 
at thermal equilibrium in an isotropic harmonic trap. Noisy lines: results of a Quantum
Monte Carlo simulation. Smooth solid lines: Hartree-Fock prediction. The curves corresponds
to the temperatures $T/T_c^0=0.88$ (a), $T/T_c^0=0.7$ (b).
The number of particles is $N=10^4$ and the parameters are the ones
of ${}^{87}$Rb. These figures are taken from \cite{Nara}.}\label{fig:QMC}
\end{figure}

On the figure
\ref{fig:QMC} 
the Hartree-Fock prediction for the radial density of particles in a spherical harmonic trap,
$r^2\rho(r)$, is compared to the Quantum Monte Carlo result for several temperatures below $T_c$.
The Hartree-Fock prediction is in good agreement with the exact result, except close to $T_c$
where it tends to underestimate the number of particles in the condensate \cite{Nara}.

\subsubsection{Experimental results for the energy of the gas}
\begin{figure}[htb]
\epsfysize=8cm \centerline{\epsfbox{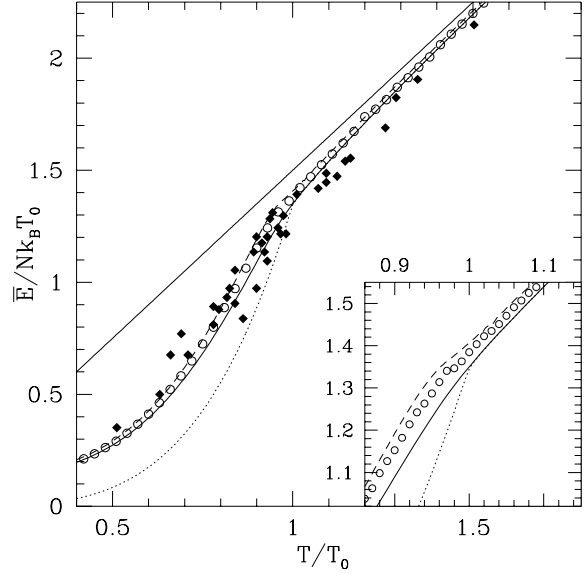}}
\caption{\small Expansion energy of the gas  per particle and
in units of $k_B T_c^0$ as function of the temperature in units of $T_c^0$.
The filled rhombi correspond to the experimental results of \cite{n0_sur_n}.
The straight solid line is the prediction of Boltzmann statistics. 
The dotted curve is the ideal Bose gas prediction. 
The circles are the numerical solution to the Hartree-Fock approach.
The curved solid line and the dashed line are approximate solutions to the
Hartree-Fock equations.
The inset is a magnification
showing the change of slope of the energy as function of $T$ close to $T=T_c^0$.
The figure is taken from \cite{Tosi}.}\label{fig:HF_JILA}
\end{figure}

At JILA the sum of kinetic and interaction energy of the atoms was measured as function of 
temperature, as we have already explained in \S\ref{subsec:verdict}.
Whereas the ideal Bose gas model was clearly getting wrong for $T<T_c$, 
the Hartree-Fock prediction \cite{Tosi}
is consistent with the experimental 
results over the whole considered temperature range (see figure \ref{fig:HF_JILA}).

At very low temperatures ($T<T_c/2$), 
measurements at MIT have shown that the same energy becomes mainly
a function of the number of particles $N_0$ in the condensate. 
By setting $\rho'=0$ in the Hartree-Fock
approximation, and using approximations presented in the coming section
\S\ref{subsec:TFA}, an analytical expression
can be obtained for the energy, in excellent agreement with the experimental results
(see figure \ref{fig:mit_ener}): 
the energy per particle has a power law dependence with $N_0$, with an exponent $2/5$,
to be contrasted
with the constant ideal Bose gas result, and has typical values an order of magnitude larger
than the zero-point energy of the harmonic oscillator.
\section{Properties of the condensate wavefunction}
\markright{The condensate wavefunction}
\label{Cap:GPE}

In this section we consider the regime of an almost pure condensate, where the non-condensed
cloud has a negligible effect on the condensate. At thermal equilibrium with temperature $T$
this regime corresponds to the limit $T\ll T_c$. 
As we shall see  most of the experimental results obtained with almost pure condensates
can be well reproduced by a single equation for the condensate wavefunction, the 
so-called Gross-Pitaevskii equation, derived independently by Gross 
\cite{Gross} and Pitaevskii \cite{Pita}.

\subsection{The Gross-Pitaevskii equation}
\subsubsection{From Hartree-Fock}
Let us assume that the density of non-condensed particles is much smaller than the
density of condensate particles over the spatial width of the condensate:
\be
\rho'(\vec{r}\,) \ll N_0 |\phi(\vec{r}\,)|^2
\ee
where $N_0$ is the mean number of particles in the condensate, $\phi$ is the
condensate wavefunction normalized to unity:
\be
\int d^3\vec{r}\; \phi(\vec{r},t)\phi^*(\vec{r},t)=1.
\ee
In the Hartree-Fock expression of the mean field potential
seen by the condensate, derived in the previous section \S\ref{Cap:HF}, 
we can drop the contribution
of the non-condensed particles, to get for the evolution of the condensate contribution
to the 1-body density matrix:
\be
i\hbar {d\over dt}(N_0|\phi\rangle\langle\phi|) = \left[{\vec{p}\,^2\over 2m}+U(\vec{r},t) + 
gN_0|\phi(\vec{r},t)|^2,
N_0|\phi\rangle\langle\phi|\right].
\label{eq:commu}
\ee
This equation leads to $N_0=$constant and to the evolution equation for the 
condensate wavefunction:
\be
i\hbar\partial_t\phi(\vec{r},t) = 
\left[-{\hbar^2\over 2m}\Delta + U(\vec{r},t)+N_0 g |\phi(\vec{r},t)|^2-\xi(t)\right]\phi(\vec{r},t).
\ee
This non-linear Schr\"odinger equation is the so-called time dependent Gross-Pitaevskii equation. This equation
is determined from our Hartree-Fock approach up to an
arbitrary real function of time, $\xi(t)$, as Eq.(\ref{eq:commu}) involves a commutator to which
$\xi(t)$ does not contribute. In general the precise value of $\xi(t)$ is considered
as a matter of convenience,
as it can be absorbed in a redefinition of the global phase of $\phi$. 
The knowledge of the value of $\xi(t)$ can become important when one is interested
in the evolution of the relative phase of {\sl two} Bose-Einstein condensates.
The value of $\xi(t)$ has been
derived in \cite{phase_mel} assuming a well defined number of particles in the condensate.
If the condensate is assumed to be in a Glauber coherent state that is
a quasi-classical state of the atomic field with a well defined relative phase
(see \S\ref{Cap:PC})
one obtains $\xi(t)=0$ as we will see in \S\ref{subsubsec:trick}.

When the gas is at thermal equilibrium, the only time dependence left for $\phi$ is a global
phase dependence. The most convenient choice is to assume $\partial_t\phi=0$
so that $\xi(t)$ is a constant. As shown in 
\S\ref{subsubsec:hfte} this constant is very close to the
chemical potential of the gas as $N_0$ is large 
so that we get the so-called time independent Gross-Pitaevskii
equation:
\be
\mu\phi(\vec{r}\,) = \left[-{\hbar^2\over 2m}\Delta + U(\vec{r}\,)+
N_0 g |\phi(\vec{r}\,)|^2\right]\phi(\vec{r}\,).
\ee

Both the time independent and the time dependent Gross-Pitaevskii equations can be solved
numerically. But, as explained in the next part of this section, the fact that the trap is harmonic
allows one to find very good approximate analytical solutions.

\subsubsection{Variational formulation}
Variational calculus turns out to be a very fruitful approximate technique 
in the solution of
the Gross-Pitaevskii equation. We therefore derive here a variational 
formulation of the Gross-Pitaevskii equation.
\begin{itemize}
\item {\sl Time independent case}
\end{itemize}
The time independent Gross-Pitaevskii equation can be obtained from extremalization over $\phi$
of the so-called Gross-Pitaevskii energy functional:
\be
E[\phi,\phi^*] = N_0 \int d^3\vec{r} \left[ {\hbar^2\over 2m}|\vec{\mbox{grad}}\,\phi|^2 +
U(\vec{r}\,)|\phi(\vec{r}\,)|^2 +{1\over 2}N_0g|\phi(\vec{r}\,)|^4\right]
\label{eq:GPef}
\ee
with the constraint that $\phi$ is normalized to unity.

{\bf Proof:} We take into account the normalization constraint with the method of Lagrange multiplier,
so that we simply have to express the fact that $\phi$ extremalizes without constraint
the functional:
\be
X[\phi,\phi^*] = E[\phi,\phi^*] - \lambda N_0 \int d^3\vec{r}\;\phi(\vec{r}\,)\phi^*(\vec{r}\,).
\ee
The parameter $\lambda$ is the Lagrange multiplier. We calculate the first order variation of $X$
due to an infinitesimal arbitrary variation of the condensate wavefunction:
\be
\phi(\vec{r}\,) \rightarrow \phi(\vec{r}\,) + \delta\phi(\vec{r}\,).
\ee
We obtain:
\be
\delta X =N_0 \int d^3\vec{r} \;
\left[{\hbar^2\over 2m}\vec{\mbox{grad}}\,\delta\phi^*\cdot
\vec{\mbox{grad}}\,\phi+
U(\vec{r}\,)\delta\phi^*\phi+
N_0g\delta\phi^*\phi^*\phi^2-\lambda\delta\phi^*\phi\right]+\mbox{c.c.}
\ee
We modify the variation of the kinetic energy term by integrating by part, assuming that
$\phi$ vanishes at infinity:
\be
 \int d^3\vec{r} \; (\vec{\mbox{grad}}\,\delta\phi^*\cdot\vec{\mbox{grad}}\,\phi
+\mbox{c.c.}) =  -\int d^3\vec{r}\; (\delta\phi^*\Delta\phi+\mbox{c.c.}).
\ee
The variation $\delta X$ has to vanish for any $\delta\phi$. We can take as independent variables
the real part and the imaginary part of $\delta\phi$, or equivalently $\delta\phi$ and $\delta\phi^*$
as it amounts to considering independent linear superpositions of the real and imaginary part.
We therefore obtain:
\be
N_0 \left[-{\hbar^2\over 2m}\Delta + U(\vec{r}\,)+N_0 g |\phi(\vec{r}\,)|^2-
\lambda\right]\phi(\vec{r}\,)=0.
\ee
We recover the time independent Gross-Pitaevskii equation, 
with $\lambda=\mu$, which gives  a physical interpretation
to the Lagrange multiplier $\lambda$.

\begin{itemize}
\item {\sl Time dependent case}
\end{itemize}
The time dependent Gross-Pitaevskii equation 
with the choice $\xi(t)\equiv 0$ is obtained over a time interval $[t_1,t_2]$
from extremalization of the action:
\be
A =\int_{t_1}^{t_2}dt\left[{i\hbar\over 2}\left(\langle\phi|{d\over dt}|\phi\rangle-\mbox{c.c.}\right)N_0
-E[\phi(t),\phi^*(t)]\right]
\label{eq:action}
\ee
with fixed values of $|\phi(t=t_1)\rangle$ and $|\phi(t=t_2)\rangle$.

\begin{itemize}
\item {\sl Physical interpretation of the Gross-Pitaevskii energy functional}
\end{itemize}
We now show that $E[\phi,\phi^*]$ is simply the mean energy of the gas in the Hartree-Fock
approximation in the limit of a pure condensate. 
As the $N$-body Hamiltonian is a sum of one-body and two-body (binary interaction) terms,
\be
H =\sum_{i=1}^{N} h_i +{1\over 2}\sum_{i\ne j} V_{ij}
\ee
the mean energy of the gas involves the one-body and two-body density matrices:
\be
\langle H\rangle = \mbox{Tr} [h_1\hat{\rho}_1] +{1\over 2} \mbox{Tr}[V_{12}
\hat{\rho}_{12}].
\ee
In the limit of a pure condensate we keep only the condensate contribution to $\hat{\rho}_1$:
\be
\hat{\rho}_1 \simeq N_0 |\phi\rangle\langle\phi|
\ee
and we approximate $\hat{\rho}_{12}$ by the Hartree ansatz
\be
\hat{\rho}_{12} \simeq \hat{\rho}_1\otimes\hat{\rho}_1.
\ee
We then obtain $E[\phi,\phi^*]=\langle H\rangle$. 
It was actually clear from the start that $E[\phi,\phi^*]$
was the sum of kinetic energy, trapping potential energy and mean field interaction energy
of the condensate.

A different and interesting point of view at zero temperature
is to use directly a Hartree-Fock ansatz for
the ground state wavefunction $|\Psi\rangle$ of the gas, assuming that all the particles are
in the condensate:
\be
|\Psi\rangle = |N:\phi\rangle= |\phi\rangle\otimes \ldots \otimes |\phi\rangle.
\ee
The mean energy of $|\Psi\rangle$ for the interaction potential $g\delta(\vec{r}_1-
\vec{r}_2)$ is then
\be
E[\phi,\phi^*] = N \int d^3\vec{r} \left[ {\hbar^2\over 2m}|\vec{\mbox{grad}}\,\phi|^2 +
U(\vec{r}\,)|\phi(\vec{r}\,)|^2 +{1\over 2}(N-1)g|\phi(\vec{r}\,)|^4\right],
\label{eq:GPef_bis}
\ee
which differs from Eq.(\ref{eq:GPef}) in the limit $N_0=N$ only by the occurrence
of a factor $(N-1)$ rather than $N$ in front of the coupling constant $g$, ensuring
that the interaction term disappears for $N=1$! 

\begin{itemize}
\item {\sl What is the chemical potential ?}
\end{itemize}
At zero temperature, assuming a pure condensate $N_0\simeq N$, the usual thermodynamical definition of
the chemical potential $\mu$ reduces to:
\be
\mu = {d\langle H\rangle\over dN} \simeq {d\over dN_0} E[\phi,\phi^*,N_0]
\ee
where we have made appear the explicit dependence of $E$ on $N_0$. When one takes the total
derivative of $E$ with respect to $N_0$, one gets in principle a contribution from the
implicit dependence of $E$ on $N_0$ through the $N_0$ dependence of $\phi,\phi^*$;
actually this contribution vanishes as the variation of $E$ due to a change in $\phi,\phi^*$
vanishes to first order in this change. We therefore get
\bea
{d\over dN_0} E[\phi,\phi^*,N_0] &=& {\partial\over\partial N_0} E[\phi,\phi^*,N_0] 
\nonumber \\
&=& \int d^3\vec{r} \left[ {\hbar^2\over 2m}|\vec{\mbox{grad}}\,\phi|^2 +
U(\vec{r}\,)|\phi(\vec{r}\,)|^2 +N_0g|\phi(\vec{r}\,)|^4\right].
\label{eq:quevautmu}
\eea
This quantity coincides with the chemical potential indeed, as can be checked
by multiplying the time independent Gross-Pitaevskii equation 
by $\phi^*$ and integrating over the whole space.
As $g$ does not have the factor $1/2$ in 
Eq.(\ref{eq:quevautmu}), whereas it is multiplied
by $1/2$ in the expression for $E[\phi,\phi^*]$, we see that in the interacting
case $g\ne 0$:
\be
\mu \ne {E\over N_0}
\ee
that is the chemical potential $\mu$ differs from the mean energy per particle.

\subsubsection{The fastest trick to recover the Gross-Pitaevskii equation}\label{subsubsec:trick}
Starting from the second quantized form of the Hamiltonian,
\be
H = \int\,d1\,\hat{\psi}^{\dagger}(1)h_1\hat{\psi}(1) +
\frac{1}{2}\int\,d1\!\int\,d2\,\hat{\psi}^{\dagger}(1)\hat{\psi}^{\dagger}(2)V_{12}
\hat{\psi}(2)\hat{\psi}(1)
\ee
where $1$ and $2$ stand for three-dimensional coordinates in real space,
one first derives the Heisenberg equation of motion for the field operator:
\bea
i\hbar {d\over dt} \hat{\psi}(1) &=&
[\hat{\psi}(1),H] = \partial_{\hat{\psi}^{\dagger}(1)}H \\
&=& h_1\hat{\psi}(1)+\int\,d2\,\hat{\psi}^{\dagger}(2)V_{12}\hat{\psi}(2)\hat{\psi}(1)
\eea
and then replaces the quantum field operator by a classical field:
\be
\hat{\psi} \rightarrow \psi=\sqrt{N_0}\phi.
\ee
As $V_{12}$ is the pseudo-potential, the equation that we get for $\phi$ is the 
time dependent Gross-Pitaevskii equation
with $\xi(t)\equiv 0$.

This sheds a new light on the Gross-Pitaevskii equation: 
the Gross-Pitaevskii equation is the equation of motion of the atomic field in the
classical approximation, neglecting quantum fluctuations of the field.  A Bose-Einstein condensate
is a classical state of the atomic field, in a way similar to the laser being a classical
state of the electromagnetic field.

\subsection{Gaussian Ansatz}
In this subsection we look for a variational solution to the Gross-Pitaevskii equation 
in a harmonic trap,
using a Gaussian ansatz for $\phi$ \cite{Pethick}. The choice of a Gaussian is 
natural in the non-interacting limit
$g\rightarrow 0$, where it becomes exact. 
It turns out to give also interesting results in presence of strong interactions.

\subsubsection{Time independent case} \label{subsubsec:tic}
Consider for simplicity an isotropic harmonic trap, where the atoms have the oscillation frequency $\omega$.
We assume the following Gaussian for the condensate wavefunction:
\be
\phi(\vec{r}\,) = {1\over \pi^{3/4}\sigma^{3/2}}e^{-r^2/2\sigma^2}
\ee
the spatial width $\sigma$ being the only variational parameter.
The mean energy per particle can be calculated exactly for this ansatz:
\be
\varepsilon\equiv {E[\phi,\phi^*]\over N_0} =
{3\hbar^2\over 4m\sigma^2} +{3\over 4} m\omega^2\sigma^2 +{\hbar^2\over m} {N_0a\over\sigma^3}{1\over
\sqrt{2\pi}}.
\ee
The form of the result is intuitive: the kinetic energy term 
scales as $\Delta p_x^2$, where $\Delta p_x=\hbar/(2\Delta x)=\hbar/(\sqrt{2}\sigma)$;
the trapping potential energy scales as $\sigma^2$ and the interaction energy per particle is
proportional to the coupling constant $g=4\pi\hbar^2a/m$ and to the typical density of atoms
in the gas, $N_0/\sigma^3$. Taking the harmonic oscillator length $(\hbar/m\omega)^{1/2}$
as a unit of length and the harmonic quantum of vibration $\hbar\omega$ as a unit of energy
we get the simple form:
\be
\varepsilon = {3\over 4}\left[{1\over\sigma^2}+\sigma^2\right] +{\chi\over 2\sigma^3}
\ee
where the only physical parameter left is 
\be
\chi= \sqrt{2\over\pi}{N_0a\over \sqrt{\hbar/m\omega}}.
\label{eq:def_chi}
\ee
This parameter $\chi$ measures the effect of the interactions on the condensate density:
The case $\chi\ll 1$ corresponds to the weakly interacting regime, close to the ideal Bose gas limit
$\chi=0$; the case $\chi\gg 1$ corresponds
to the strongly interacting regime.

\begin{figure}[htb]
\epsfysize=8cm \centerline{\epsfbox{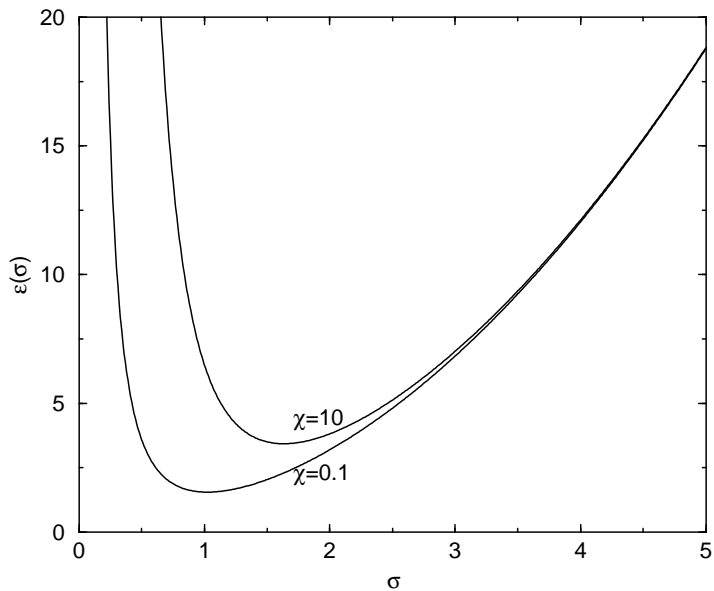}}
\caption{\small Energy per particle in the condensate in units of $\hbar\omega$
as function of the variational width $\sigma$ in units of $(\hbar/m\omega)^{1/2}$.
Case of effective repulsive interactions $a>0$.}\label{fig:e_v_s}
\end{figure}

\begin{itemize}
\item {\sl case $a>0$}
\end{itemize}
In the case of effective repulsive interactions between the particles, the dependence of $\varepsilon$
with $\sigma$ is plotted in figure
\ref{fig:e_v_s}. In the limit $\sigma\rightarrow 0$, the energy $\epsilon$ is
dominated by the positively diverging repulsive interaction ($\sim 1/\sigma^3$). For large $\sigma$
the trapping potential term $\sim \sigma^2$ dominates. The function $\varepsilon$ has a single
minimum, in $\sigma=\sigma_0$, solving
\be
{d\varepsilon\over d\sigma}(\sigma_0)=0 \rightarrow \sigma_0^5 = \sigma_0 + \chi.
\ee
For $\chi\ll 1$ one recovers the ground state of the harmonic trap, with $\sigma_0=1$. For $\chi\gg 1$
the condensate cloud becomes much broader than the ground state of the harmonic trap,
\be
\sigma_0 \simeq \chi^{1/5} \propto N_0^{1/5}.
\ee
In this regime one can check that the kinetic energy term becomes negligible as compared
to the trapping energy:
\be
{E_{\mbox{\scriptsize kin}}\over E_{\mbox{\scriptsize trap}}} = 
{1\over \sigma^4} \simeq {1\over \chi^{4/5}}
\ee
so that the steady state of the condensate is an equilibrium between the trapping potential
and the repulsive interactions between particles. This regime will be studied in detail in
the next subsection.

\begin{figure}[htb]
\centerline{
\epsfysize=7cm \epsfbox{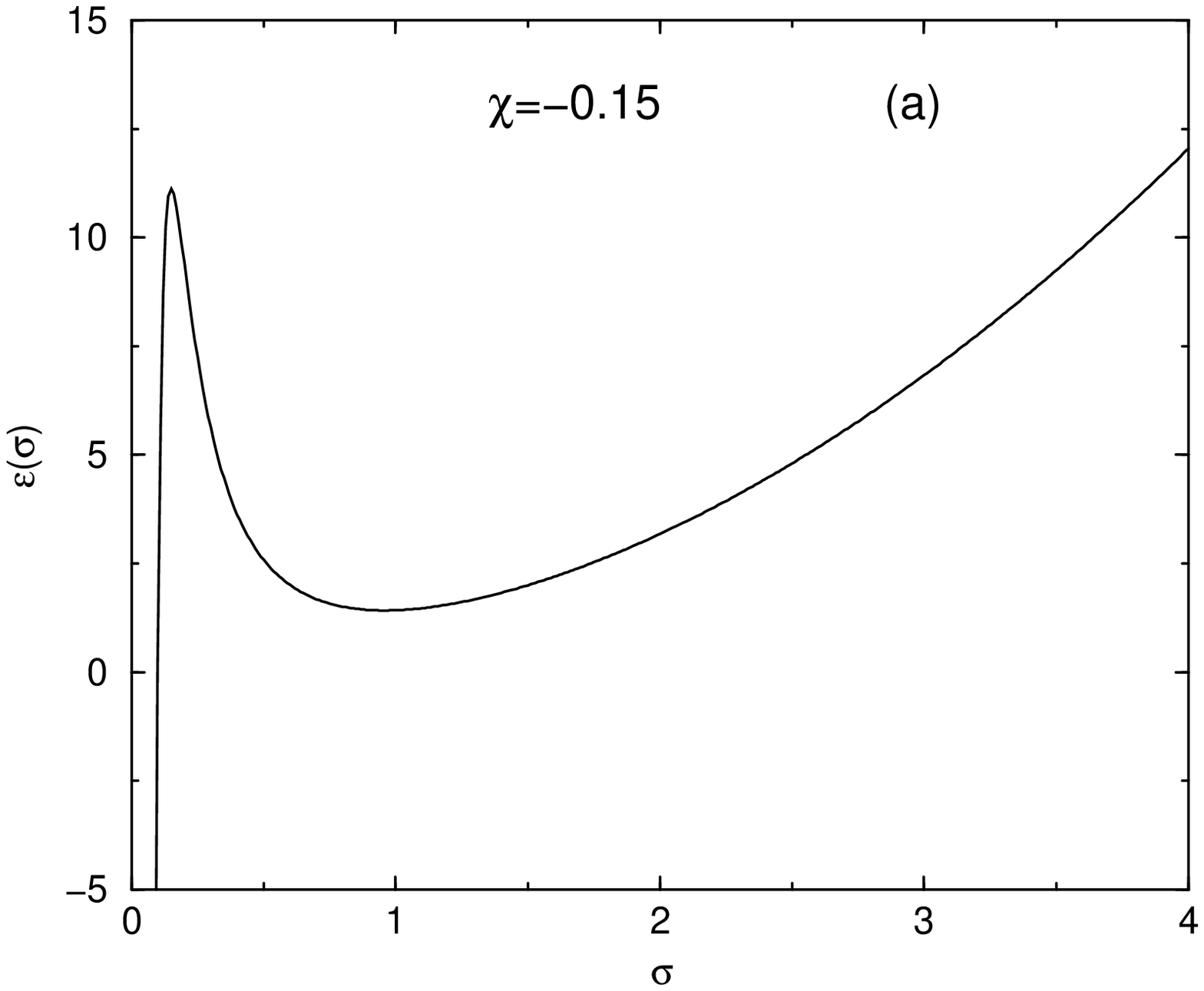} \ \ \ \
\epsfysize=7cm \epsfbox{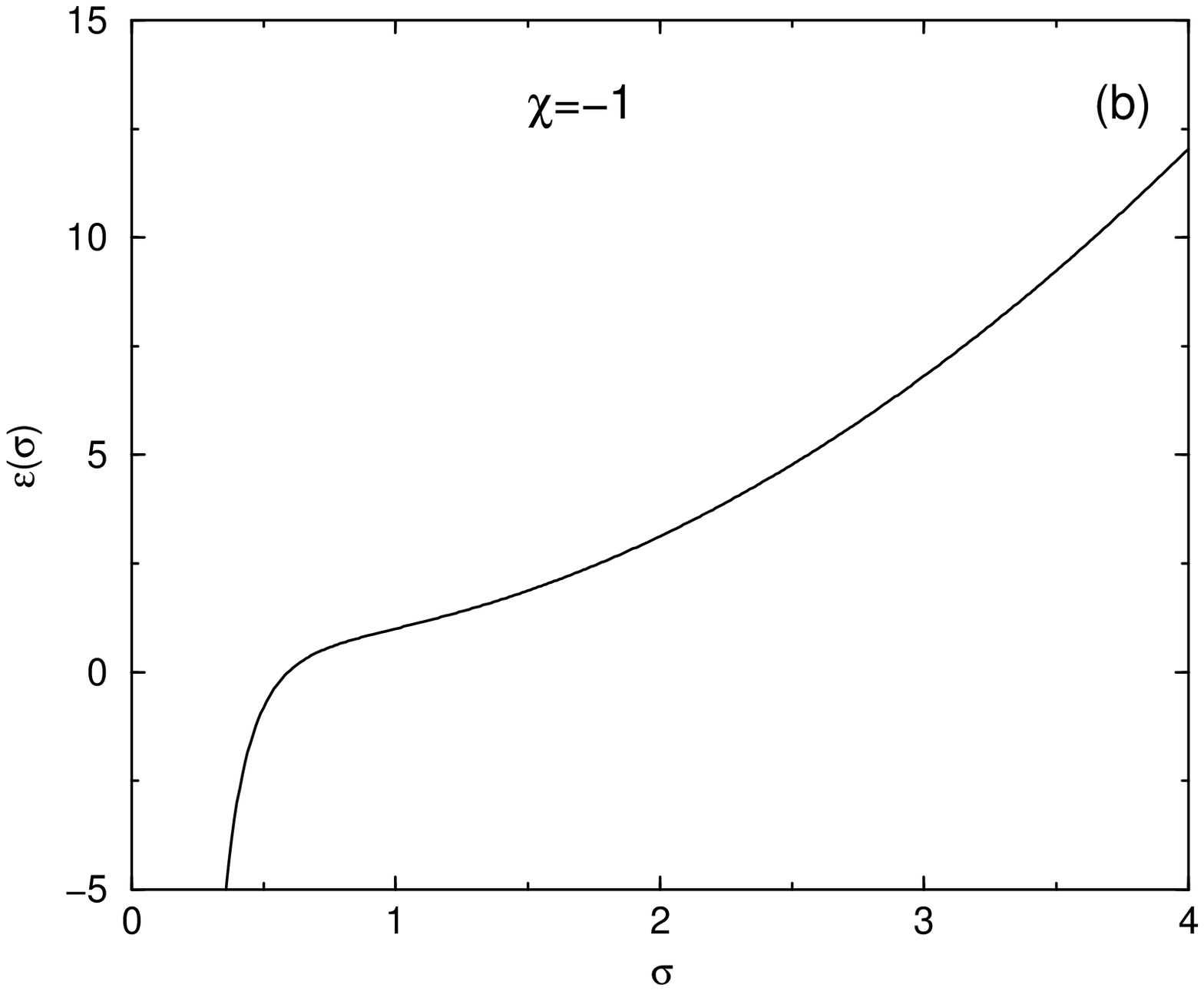}}
\caption{\small 
For an effective attractive interaction $a<0$ between the particles
energy per particle in the condensate in units of $\hbar\omega$
as function of the variational width $\sigma$ in units of $(\hbar/m\omega)^{1/2}$.
The curve has two possible shapes, (a)  with two minima
when $|\chi|$ is smaller than a critical value $|\chi_c|$, and (b) with a single
minimum for $|\chi|>|\chi_c|$.  }\label{fig:coll}
\end{figure}

\begin{itemize}
\item {\sl case $a<0$}
\end{itemize}
For effective attractive interactions between the particles the shape of $\varepsilon$ as function
of $\sigma$ depends on the balance between kinetic and interaction energy (see figure
\ref{fig:coll}).
The interaction energy is negatively diverging as $\sigma\rightarrow 0$ always faster than 
the positively diverging kinetic energy so that $\sigma=0$ is always a minimum of
$\varepsilon$, with $\varepsilon=-\infty$: the condensate is in a spatially collapsed state !
Of course the Gross-Pitaevskii equation not longer applies for a 
too small $\sigma$, as the validity of the Born approximation
requires $k|a|\simeq |a|/\sigma\ll 1$.
For $|\chi|$ larger than some critical value $|\chi_c|$, this collapsed minimum is the only one
of $\varepsilon(\sigma)$ so that we do not find any stable solution for the condensate
wavefunction.
When $|\chi|$ is smaller than $|\chi_c|$  the kinetic energy term, which is opposed to spatial
compression of the gas, is able to beat the attractive energy over some range of $\sigma$,
so that a local minimum of $\varepsilon(\sigma)$ appears, in $\sigma=\sigma_0$, separated
from the collapsed minimum by a barrier.

To calculate $|\chi_c|$ we express the fact that the stationary point of $\varepsilon$ in $\sigma=\sigma_0$
has now a vanishing curvature (inflexion point of $\varepsilon$):
\bea
\left({d\varepsilon\over d\sigma}\right)_{\sigma=\sigma_0}^{\chi=\chi_c} &=& 0 \\
\left({d^2\varepsilon\over d\sigma^2}\right)_{\sigma=\sigma_0}^{\chi=\chi_c} &=& 0 
\eea
By eliminating $\sigma_0$ between these two equations we obtain
\be
\chi_c = - {4\over 5^{5/4}}=-0.5350...
\ee
This result can be rephrased in terms of a maximal number of atoms $N_0^c$ that can be put in the
condensate without inducing a collapse, according to a Gaussian ansatz:
\be
{N_0^c |a|\over \sqrt{\hbar/ m\omega}} = \left({\pi\over 2}\right)^{1/2}|\chi_c|\simeq -0.67.
\ee

A more precise result has been obtained by a numerical solution of 
Gross-Pitaevskii equation, not restricting to
the subspace of Gaussian wavefunctions
\cite{Burnett}:
no solution of the time independent Gross-Pitaevskii equation 
is obtained for $N_0>N_0^c$, where
\be
{N_0^c |a|\over \sqrt{\hbar/ m\omega}} \simeq -0.57.
\ee

By a generalization of the Gaussian ansatz to the case of a non-isotropic harmonic trap
one can also get a prediction of $N_0^c$ for the parameters of the lithium experiment
of Hulet's group \cite{Hulet2}. In the experiment the traps frequencies are 
$\omega_z = 2\pi\times 117$Hz and $\omega_{x,y}= 2\pi\times 163$Hz, and the scattering length
is $a=-27$ Bohr radii. The Gaussian prediction is then $N_0^c \simeq 1500$,
consistent with the experimental results.

\begin{itemize}
\item {\sl Physical origin of the stabilization for $a<0$}
\end{itemize}
In a harmonic trap, the energy of the ground state level is separated from the energy of excited
states by $\hbar\omega$. At low values of $\chi$ the mean interaction energy per
particle, $\sim \rho |g|$, where $\rho$ is the density, is much smaller than $\hbar\omega$
so that it cannot efficiently induce a transition from the ground harmonic level
to excited harmonic levels. Initiation of collapse on the contrary requires
that the wavefunction $\phi$ expands on many excited levels in the trap, so that the density $|\phi|^2$
can exhibit a high density peak narrower than $\sqrt{\hbar/m\omega}$. We therefore intuitively
reformulate the non-collapse condition as
\be
\rho |g| < \hbar\omega.
\ee
Estimating $\rho$ as $N_0/(\hbar/m\omega)^{3/2}$ we recover a $N_0^c$ 
scaling as $\sqrt{\hbar/m\omega}/|a|$.
This reasoning also applies to the gas confined in a cubic box with periodic boundary conditions,
as we shall see in section \S\ref{Cap:linear} of the lecture.

\subsubsection{Time dependent case}
As done in \cite{Zoller_PRA,Zoller_PRL}
the Gaussian ansatz can be
generalized to the time dependent case. We assume here for simplicity that the condensate,
initially in steady state, is excited only by a temporal variation of the trap frequencies
$\omega_\alpha(t)$; then no oscillation of the center of mass motion of the condensate
takes place, $\phi$ remaining of vanishing mean position and momentum. The Gaussian ansatz
then contains only exponential of terms quadratic with position, its does not
involve exponential of terms linear with position:
\be
\phi(\vec{r},t) = {e^{i\delta(t)}\over \pi^{3/4}\left[\prod_{\alpha}\sigma_\alpha(t)\right]^{1/2}}
\exp\left[-\sum_\alpha {r_\alpha^2\over 2\sigma_\alpha^2(t)}
+i\sum_\alpha r_\alpha^2\gamma_\alpha(t)\right].
\ee
We do not assume that the trap is isotropic, so we have as variational parameters
3 spatial widths $\sigma_\alpha$ ($\alpha=x,y,z$), 3 factors $\gamma_\alpha$ governing the
spatially quadratic phase and a global phase $\delta$.

One gets time evolution equations for the variational parameters by inserting the ansatz for
$\phi$ in the action $A$ of Eq.(\ref{eq:action}) and by writing the Lagrange equations
expressing the stationarity condition. It turns out that $\gamma_\alpha$ can be expressed
in terms of the widths and their time derivatives:
\be
\gamma_\alpha = -{m\dot\sigma_\alpha\over 2\hbar\sigma_\alpha}
\ee
so that one is left with equations for the $\sigma_\alpha$'s. Taking $\omega^{-1}$ as a unit of
time, $\sqrt{\hbar/m\omega}$ as a unit of length, where $\omega$ is an arbitrary reference frequency,
we get:
\be
\ddot\sigma_\alpha + \nu_\alpha^2\sigma_\alpha = {1\over\sigma_\alpha^3} +{\chi\over\sigma_\alpha
\sigma_x\sigma_y\sigma_z}
\label{eq:ev_sig}
\ee
where the trap frequencies are $\omega_\alpha = \nu_a \omega$ and
$\chi$ is defined in Eq.(\ref{eq:def_chi}).
In the absence of interaction ($\chi=0$) these evolution equations become
exact, and give a remarkable (and known !) result for the time dependent harmonic oscillator.
In the interacting case ($\chi\neq 0$) these equations can be cast in Hamiltonian
form as the \lq\lq force" seen by the variable $\sigma_\alpha$ derives from
a potential. The corresponding dynamics is non linear and non trivial;
chaotic behavior has been obtained in \cite{Kagan_chaos} in the limiting regime of 
$\chi\gg 1$ where the $1/\sigma_\alpha^3$ can be neglected.

\begin{figure}[htb]
\epsfysize=8cm \centerline{\epsfbox{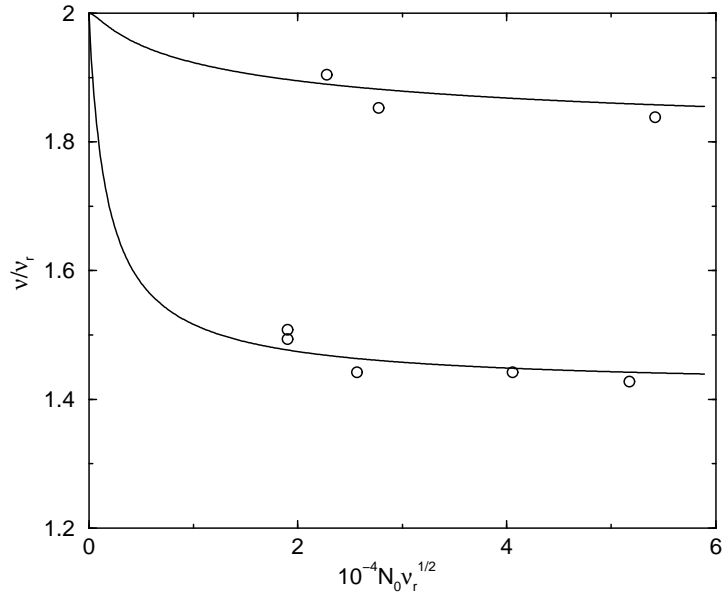}}
\caption{\small Frequencies of two eigenmodes of a condensate in a cylindrically symmetric
harmonic trap,
in units of the radial trap frequency $\nu_r$, as a function of a parameter proportional
to $\chi$ measuring the strength of the interactions.
Plotting symbols: measurements at JILA \cite{JILA_exc}. Solid lines :
predictions of the Gaussian Ansatz \cite{Zoller_PRL}.
\label{fig:Z_v_J} }
\end{figure}

One can use Eq.(\ref{eq:ev_sig}) to study the response of the condensate to a weak
excitation, the trap frequency $\omega_\alpha$ in the experiments
being typically slightly perturbed from its steady state value $\omega_\alpha(0)$ 
for a finite excitation time. Linearizing the evolution equations in terms
of the deviations of the $\sigma$'s from their steady state value:
\be
\sigma_\alpha(t) = \sigma_\alpha^{\mbox{\scriptsize st}} +\delta \sigma_\alpha(t)
\ee
one gets a three by three system of second order differential equations
for the $\delta\sigma$'s. Looking for eigenmodes of this system, one finds three
eigenfrequencies \cite{Zoller_PRL}. Their values have been compared to experimental
results at JILA \cite{JILA_exc}, see Fig.\ref{fig:Z_v_J}: the agreement is
very good, not only in the weakly interacting regime $\chi\ll 1$
but also in the regime $\chi \gg 1$, where the Gaussian ansatz for
the condensate wavefunction has no reason to be a good one! The explanation
of this mystery is given in \S\ref{subsubsec:sca}.

\subsection{Strongly interacting regime: Thomas-Fermi approximation}\label{subsec:TFA}
In this subsection we focus on the strongly interacting regime: the scattering length
is positive, with the dimensionless parameter $\chi$ of Eq.(\ref{eq:def_chi})
much larger than one. This regime is the so-called Thomas-Fermi regime.
As we now see analytical results can be obtained in this limit.

\subsubsection{Time independent case}
If we put a large enough number of particles into the condensate the atoms
will experience repulsive interactions that will increase the spatial
radius of the condensate to a value $R$ much larger than the one of
the ground state of the harmonic trap:
\be
R \gg \left(\frac{\hbar}{m\omega}\right)^{1/2}.
\ee
For increasing value of $N_0$, $R$ increases so that the momentum width
of the condensate, scaling as $\hbar/R$ as $\phi_0$ is a non-oscillating
function of the position, is getting smaller and smaller. More precisely
we find that the typical kinetic energy of the condensate becomes much
smaller than the typical harmonic potential energy of the condensate:
\be
\frac{E_{\mbox{\scriptsize kin}}}{E_{\mbox{\scriptsize harm}}} \simeq
\frac{\frac{\hbar^2}{mR^2}}{m\omega^2 R^2} \simeq \left(
\frac{\hbar}{m\omega R^2}\right)^2 \ll 1.
\ee
The mechanical equilibrium of the condensate in the trap then comes mainly
from the balance between the expelling effect of the repulsive interactions and 
the confining effect of the trap.

In this large $R$ regime we neglect the kinetic energy term in the Gross-Pitaevskii
energy functional, which leads to a functional of the condensate density only
(similarly to the Thomas-Fermi approximation for electrons).
This approximation amounts to neglecting the $\Delta\phi$ term in the
Gross-Pitaevskii equation:
\be
\mu\phi(\vec{r}\,) \simeq U(\vec{r}\,)\phi(\vec{r}\,) + N_0 g |\phi(\vec{r}\,)|^2\phi(\vec{r}\,).
\ee
Taking $\phi$ to be real we find that 
\be
\phi(\vec{r}\,) = \left(\frac{\mu-U(\vec{r}\,)}{N_0 g}\right)^{1/2}
\label{eq:tf}
\ee
in the points of space where $\mu > U(\vec{r}\,)$, otherwise we have
$\phi(\vec{r}\,)=0$.

This very important, Thomas-Fermi result Eq.(\ref{eq:tf}) can also be obtained
in a local density approximation point of view. A spatially uniform condensate
with a chemical potential $\mu$ and in presence of a uniform external potential
$U$ has a density $N_0|\phi|^2=(\mu-U)/g$. Applying this formula with a 
$\vec{r}$ dependent potential $U$ gives again Eq.(\ref{eq:tf}). A local density
approximation can be used only if the density of the condensate varies
slowly at the scale of the so-called \lq\lq healing length" $\xi$, introduced
in \S\ref{subsubsec:cha}; one can check that the condition $\xi\ll R$ is indeed satisfied
in the Thomas-Fermi regime.

We specialize Eq.(\ref{eq:tf}) to the case of a harmonic but not necessarily
isotropic trap:
\be
U(\vec{r}\,) = \frac{1}{2} m \sum_{\alpha} \omega_\alpha^2 r_\alpha^2
\ee
where $\alpha= x,y,z$ label the eigenaxis of the trap. The boundary of the condensate
$\mu=U(\vec{r}\,)$ is then an ellipsoid with a radius $R_\alpha$
along axis $\alpha$ given by:
\be
\mu = \frac{1}{2} m\omega_\alpha^2 R_\alpha^2.
\label{eq:radii}
\ee
The condensate wavefunction can be rewritten in terms of these radii:
\be
\phi(\vec{r}\,) = \left(\frac{\mu}{N_0g}\right)^{1/2}
\left(1-\sum_\alpha \frac{r_\alpha^2}{R_\alpha^2} \right)^{1/2}.
\ee
Using the normalization condition of $\phi$ to unity we can also express
the \lq\lq normalization" factor $\sqrt{\mu/N_0 g}$ in terms of the radii.
The integral of $|\phi|^2$  can be calculated in spherical coordinates
after having made the change of variable $u_\alpha = r_\alpha / R_\alpha$.
This leads to
\be
\left(\frac{\mu}{N_0g}\right)^{1/2} = \left(\frac{15}{8\pi \displaystyle\prod_{\alpha}R_\alpha}
\right)^{1/2}.
\ee
Eliminating $R_\alpha$ in terms of $\mu$ thanks to Eq.(\ref{eq:radii}) we can
calculate the chemical potential:
\be
\mu = \frac{1}{2} \hbar\bar{\omega}
\left[15\frac{N_0a}{\left(\hbar/m\bar{\omega}\right)^{1/2}}\right]^{2/5}
\ee
where $\bar{\omega}$ is the geometrical mean of the trap frequencies:
\be
\bar{\omega} = (\omega_x \omega_y \omega_z)^{1/3}.
\ee
We can now see that in the limit $\chi\gg1$ the chemical
potential $\mu$ satisfies
\be
\mu \gg \hbar \bar{\omega},
\ee
which is a convenient way of defining the Thomas-Fermi regime.

We can now compare these Thomas-Fermi predictions
to the MIT experimental results on the
energy of the condensate \cite{MIT_expans}. In the experiment the trapping
potential is switched off abruptly, so that the energy of the gas abruptly reduces to
$E_{\mbox{\scriptsize red}}= E_{\mbox{\scriptsize kin}} + E_{\mbox{\scriptsize int}} \simeq E_{\mbox{\scriptsize int}}$;
afterwards the cloud ballistically expands, $E_{\mbox{\scriptsize int}}$ is
converted in kinetic expansion energy that can be measured.
In the Thomas-Fermi approximation the integral of $N_0^2 g |\phi|^4/2$
can be done, which leads to
\be
E_{\mbox{\scriptsize int}} \simeq \frac{2}{7} N_0 \mu \propto N_0^{7/5}.
\ee
The resulting dependence in $N_0$ is in good agreement with the MIT
results, see Fig.\ref{fig:mit_ener}.

From the expression of the chemical potential we can also calculate
the total energy of the condensate in the trap , as $\mu =\partial_{N_0} E$:
integrating over $N_0$ gives
\be
E \simeq \frac{5}{7}\mu N_0.
\ee
One can then check explicitly that $\mu \neq E/N_0$!

\subsubsection{How to extend the Thomas-Fermi approximation to the time dependent case ?}
We would like to analyze time dependent situations encountered in the experiments,
{\sl e.g.} 
\begin{itemize}
\item ballistic expansion of the gas: this is a crucial example, as it is a standard
experimental imaging technique of the condensate
\item collective excitations: response of the condensate to a modulation of the trap
frequencies
\end{itemize}
in the strongly interacting regime.
An immediate generalization of the Thomas-Fermi approximation consisting in neglecting
the kinetic energy of the condensate is now too naive! In the case of ballistic expansion
for example the interaction energy is gradually transformed into kinetic energy when
the cloud expands so kinetic energy becomes important!

The trick is actually to split the kinetic energy in two contributions, 
one of them remaining small and negligible in the time dependent case.
This is performed using the so-called  hydrodynamic representation
of the condensate classical field, split in a modulus and a phase:
\be
N_0^{1/2}\phi(\vec{r}\,) = \rho^{1/2}(\vec{r}\,) e^{iS(\vec{r}\,)/\hbar}
\label{eq:hydro_forme}
\ee
where $S$ has the dimension of an action and $\rho$ is simply the condensate density.
The mean kinetic energy of the condensate then writes
\bea
E_{\mbox{\scriptsize kin}}[\phi,\phi^*]  &=& \int d^3\vec{r}\; \frac{\hbar^2}{2m}
|\vec{\mbox{grad}}\,\phi|^2 \nonumber \\
&=& \int d^3\vec{r}\; \left[\frac{\hbar^2}{2m}\left(\vec{\mbox{grad}}\sqrt{\rho}\right)^2
+\rho\frac{\left(\vec{\mbox{grad}}\,S\right)^2}{2m}\right].
\eea
As we shall see
during ballistic expansion of the condensate the density $\rho$ remains a smooth,
slowly varying function of the position so that it has a very small contribution to the
kinetic energy; most of the kinetic energy induced from interaction energy
is stored in the spatial variation of the phase of the condensate wavefunction.

\subsubsection{Hydrodynamic equations}\label{subsubsec:hye}
In this subsection we rewrite the time dependent Gross-Pitaevskii equation
in terms of the density $\rho$ and the phase $S$. This can be done of course
by a direct insertion of Eq.(\ref{eq:hydro_forme}) in the Gross-Pitaevskii
equation.

A more elegant way is to use the covariant nature of the Lagrangian
formulation of the Gross-Pitaevskii equation, Eq.(\ref{eq:action}). We rewrite
the density of Lagrangian in terms of $\rho$ and $S$:
\be
{\cal L} = -\left[\rho\partial_t S +\frac{\hbar^2}{2m} \left(\vec{\mbox{grad}}\sqrt{\rho}\right)^2
+\rho\frac{\left(\vec{\mbox{grad}}\,S\right)^2}{2m}+ U(\vec{r},t)\rho +\frac{g}{2}\rho^2\right].
\ee
An evolution equation for an arbitrary coordinate $Q(\vec{r},t)$ of the field is
obtained from the Lagrange equation:
\be
\partial_t\left(\frac{\partial{\cal L}}{\partial(\partial_t Q)}\right)+
\sum_\alpha \partial_{r_\alpha}\left(\frac{\partial{\cal L}}{\partial(\partial_{r_\alpha} Q)}\right)=
\frac{\partial{\cal L}}{\partial Q}.
\ee

We first specialize the Lagrange equations to the choice $Q=\sqrt{\rho}$; dividing
the resulting equation by $2\sqrt{\rho}$ we obtain
\be
\partial_t S +\frac{1}{2m}\left(\vec{\mbox{grad}}\,S\right)^2+U+\rho g =
\frac{\hbar^2}{2m} \frac{\Delta \sqrt{\rho}}{\sqrt{\rho}}.
\label{eq:q=rac_rho}
\ee
Then we set $Q=S$ in the Lagrange equations which leads to
\be
\partial_t\rho + \mbox{div}\left[\frac{\rho}{m}\vec{\mbox{grad}}\,S\right]=0.
\label{eq:q=s}
\ee

This last equation looks like a continuity equation. This is confirmed by the following
physical interpretation of $\vec{\mbox{grad}}\,S$. It is known in basic quantum mechanics
that the probability current density associated to a single particle wavefunction
$\phi$ is
\be
\vec{j}_{\mbox{\scriptsize proba}} = \frac{\hbar}{2im}\left[\phi^*\vec{\mbox{grad}}\,\phi
-\mbox{c.c.}\right].
\ee
Multiplying this expression by $N_0$, as there are $N_0$ particles in the condensate,
and introducing the $(\rho,S)$ representation of $\phi$ we get the following
expression for the current density of condensate particles:
\be
\vec{j} = \rho \frac{\vec{\mbox{grad}}\,S}{m} \equiv \rho\,\vec{v}
\ee
where $\vec{v}$ is the so-called local velocity field in the gas.

Equation (\ref{eq:q=s}) is therefore the usual continuity equation:
\be
\partial_t\rho + \mbox{div}\left[\rho\,\vec{v}\,\right]=0.
\label{eq:conti}
\ee
The other equation (\ref{eq:q=rac_rho}) can be turned into an evolution equation for
the velocity field by taking its spatial gradient:
\be
m\partial_t\vec{v}+\vec{\mbox{grad}}\,\left[\frac{1}{2}mv^2 + U(\vec{r}\,) + g\rho(\vec{r}\,)
-\frac{\hbar^2}{2m}\frac{\Delta\sqrt{\rho}}{\sqrt{\rho}}\right] =0.
\label{eq:ns}
\ee

This looks like the Navier-Stockes equation used in classical hydrodynamics, in the limiting
case of a fluid with no viscosity.  The term $\vec{\mbox{grad}}(\frac{1}{2}mv^2)$ looks
unusual but using the fact that $\vec{v}$ is the gradient of a function $S/m$ 
one can put it in the usual form of a convective term:
\be
\vec{\mbox{grad}}\left(\frac{1}{2}mv^2\right) = m(\vec{v}\cdot\vec{\mbox{grad}})\vec{v}.
\ee
A difference with classical hydrodynamics is the so-called quantum pressure term,
involving
\be
-\frac{\hbar^2}{2m}\frac{\Delta\sqrt{\rho}}{\sqrt{\rho}},
\label{eq:qp}
\ee
the only term in the equations (\ref{eq:conti},\ref{eq:ns}) where $\hbar$ appears.

\subsubsection{Classical hydrodynamic approximation} \label{subsubsec:cha}
The classical hydrodynamic approximation consists precisely in neglecting
the quantum pressure term Eq.(\ref{eq:qp}) in the equation (\ref{eq:ns})
for the velocity field of the condensate.

We can estimate simply the validity condition of this approximation.
Denoting $d$ a typical length scale for the variation of the condensate
density $\rho(\vec{r}\,)$ we obtain the estimate
\be
\frac{\Delta\sqrt{\rho}}{\sqrt{\rho}} \sim \frac{1}{d^2}.
\ee
Comparing the quantum pressure term Eq.(\ref{eq:qp}) to the classical mean
field term $\rho g$ yields the condition
\be
\frac{\hbar^2}{m d^2} \ll g\rho(\vec{r}\,) \sim \rho_{\mbox{\scriptsize max}} g
\ee
where $\rho_{\mbox{\scriptsize max}}$ is the maximal density (usually at the
center of the trap). This validity condition can be reformulated
in terms of the healing length,
\be
d \gg \xi \equiv \left(\frac{\hbar^2}{2 m \rho_{\mbox{\scriptsize max}} g}\right)^{1/2}.
\ee
Note that $\xi$ is sometimes also called coherence length, which can be confusing.

Why this name of healing length for $\xi$?  Imagine that you cut with an infinite
wall a condensate in an otherwise uniform potential. Right at the wall the
condensate density vanishes; far away from the wall the density of the condensate
is uniform. The condensate density adapts from zero to its constant bulk
value over a length typically on the order of $\xi$. This can be checked by 
an explicit solution of the Gross-Pitaevskii equation: 
\be
N_0^{1/2}\phi(x,y,z) = \rho_{\mbox{\scriptsize max}}^{1/2}\tanh\left(
\frac{z}{\sqrt{2}\xi}\right)
\ee
where $z=0$ is the plane of the infinite wall. This explicit solution shows
that at a distance $z\gg \xi$ from the infinite wall there is no more any effect 
of the boundary condition $\phi(x,y,z=0)=0$. This is to be contrasted with the
case of the ideal Bose gas: the ground state between infinite walls
separated by the length $L$ then scales as $\sin(\pi z/L)$ and depends dramatically
on $L$.

For a moderate excitation of the condensate by a modulation of the trap
frequencies, or in the course of ballistic expansion of the condensate,
we shall see that the only typical length scale for the variation of the
condensate density is the radius $R$ of the condensate itself. One can
then check that in the Thomas-Fermi regime the classical hydrodynamic
approximation indeed applies:
\be
\frac{R}{\xi} \simeq \left(\frac{2\mu}{m \omega^2}\right)^{1/2} 
\times \left(\frac{2m\mu}{\hbar^2}\right)^{1/2}  = \frac{2\mu}{\hbar\omega} \gg 1.
\ee
In the Thomas-Fermi regime we therefore neglect the quantum pressure
term to obtain
\be
m\left[\partial_t+\left(\vec{v}\cdot\vec{\mbox{grad}}\right)\right]\vec{v}(\vec{r},t)
=  -\vec{\mbox{grad}}\left( U(\vec{r},t) + g\rho(\vec{r},t)\right) \equiv \vec{F}(\vec{r},t).
\label{eq:euler}
\ee
This equation is then a purely classical equation, Newton's equation in presence
of the force field $\vec{F}$ written in Euler's point of view. The operator
between square brackets is simply the so-called convective derivative.

It is instructive to rewrite Eq.(\ref{eq:euler}) in Lagrange's point of view.
One then follows a small piece of the fluid in course of its motion. Denoting
$\vec{r}(t)$ the trajectory of the small piece of fluid we directly write
Newton's equation:
\be
m\frac{d^2}{dt^2}\vec{r}(t) = \vec{F}(\vec{r}(t),t) = 
-\left[\vec{\mbox{grad}}\left( U + g\rho\right)\right](\vec{r}(t),t).
\label{eq:lagrange}
\ee
This equation automatically implies the continuity equation (\ref{eq:conti}) and
the Euler equation (\ref{eq:euler}). 
The unusual feature
is that the force field depends itself on the density of the gas, so
that we are facing here a self-consistent classical problem, corresponding formally
to the mean field approximation for a collisionless classical gas! A surprising
conclusion, knowing that we are actually studying the motion of a Bose-Einstein
condensate!

\subsection{Recovering time dependent experimental results}

\subsubsection{The scaling solution} \label{subsubsec:sca}
It turns out that the self-consistent classical problem Eq.(\ref{eq:lagrange})
can be solved exactly for the particular conditions of a gas initially at rest
and in a harmonic trap.

At time $t=0$ we assume a steady state Bose-Einstein condensate in the trap, of course
in the Thomas-Fermi regime so that the classical hydrodynamic approximation is
reasonable. The steady state of Eq.(\ref{eq:lagrange}) corresponds to a
force field $\vec{F}$ vanishing everywhere, so that
\be
U(\vec{r}\,) + g\rho(\vec{r}\,) = \mbox{constant}.
\ee
One recovers the stationary Thomas-Fermi density profile, the constant being
determined from the normalization condition of $\rho$ and therefore coinciding
with the Thomas-Fermi approximation for $\mu$.

At time $t>0$ the trapping potential remains harmonic with the same eigenaxis
\cite{generalisation}
but the eigenfrequencies of the trap can have any time dependence:
\be
U(\vec{r},t) = \frac{1}{2}\sum_{\alpha=x,y,z} m\omega_\alpha^2(t)r_\alpha^2.
\ee
Then any small piece of the fluid with initial positions $r_\alpha(0)$ along
axis $\alpha$ will move according to the trajectory
\be
r_\alpha(t) = \lambda_\alpha (t) r_\alpha(0)
\label{eq:sca}
\ee
where the scaling factors $\lambda_\alpha(t)$ depend only on time,
not on the initial position of the small piece of fluid. In other words
the density of the gas will experience a mere (possibly anisotropic) dilatation
\be
\rho(\vec{r},t) = \frac{1}{\lambda_x(t)\lambda_y(t)\lambda_z(t)}
\rho\left(\{\frac{r_\alpha}{\lambda_\alpha(t)}\},t=0\right).
\ee

We can see simply why the ansatz Eq.(\ref{eq:sca}) solves indeed Eq.(\ref{eq:lagrange})
for a harmonic trap.
As the initial density in the trap has a quadratic dependence on position, so will
have the density at time $t$. The gradient $-\vec{\mbox{grad}}(\rho g)$  appearing in
the expression of the force field will then be a linear function of the coordinates;
so is the harmonic force $-\vec{\mbox{grad}}\,U(\vec{r},t)$. Newton's equation
is therefore linear in the coordinates; dividing it by  $r_\alpha(0)$  one then gets
equations for $\lambda_\alpha(t)$ irrespective of the initial coordinates
$r_\alpha(0)$! 

More details are given in \cite{Dum_sca,Kagan_sca}, we give here the equations for the
scaling parameters:
\be
\frac{d^2}{dt^2}\lambda_\alpha(t) = \frac{\omega_\alpha^2(0)}{\lambda_\alpha \lambda_x
\lambda_y \lambda_z} -\omega_\alpha^2(t) \lambda_\alpha(t), \ \ \ \alpha=x,y,z
\label{eq:scal_par}
\ee
with the initial conditions
\bea
\lambda_\alpha(0) &=& 0 \\
\frac{d}{dt}\lambda_\alpha(0) &=& 0
\eea
since the condensate is initially at rest.

Finally we make the connection between these scaling solutions and the equations
for the spatial widths $\sigma_\alpha$ obtained in Eq.(\ref{eq:ev_sig}) from a time
dependent variational Gaussian ansatz for the condensate wavefunction. We
are here in the Thomas-Fermi regime $\chi\gg 1$ so that the $1/\sigma_\alpha^3$ terms
can be neglected in Eq.(\ref{eq:ev_sig}). The steady state solutions for the $\sigma_\alpha$'s
are then $\sigma_\alpha{(0)}\simeq \chi^{1/5}\bar{\nu}^{3/5}/\nu_\alpha(0)$ where
$\bar{\nu}$ is the geometrical mean of the initial frequencies $\nu_\alpha(0)$,
and the quantities $\sigma_\alpha(t)/\sigma_\alpha(0)$ then obey the
same equations as the $\lambda_\alpha$'s! The Gaussian ansatz,
which has the wrong shape in the Thomas-Fermi regime, is however able to capture
the right scaling nature of the solution! This explains why the collective mode
frequencies obtained from Eq.(\ref{eq:ev_sig}) are a good approximation,
not only when $\chi\ll 1$, where the Gaussian ansatz was expected to hold,
but also in the strongly interacting regime $\chi\gg 1$.

\subsubsection{Ballistic expansion of the condensate}
At time $t=0^+$ the trapping potential is turned off suddenly. The scaling parameters
then satisfy the simpler equations
\be
\frac{d^2}{dt^2}\lambda_\alpha(t) = \frac{\omega_\alpha^2(0)}{\lambda_\alpha \lambda_x
\lambda_y \lambda_z}.
\ee
These equations are still difficult to solve analytically. In the experimentally
relevant regime of cigar-shaped traps, with $\omega_z(0)\ll\omega_x(0)=\omega_y(0)$,
one can find an approximate solution \cite{Dum_sca}.

Experimentally the scaling predictions have been tested carefully. First
one can see if the ballistically expanded condensate density has indeed the
shape of an inverted parabola \cite{Dum_sca}. Second one can measure the radii of the condensate
as function of time to see if they fit the scaling predictions \cite{Rempe}. Both tests confirm
the scaling predictions in the Thomas-Fermi regime.

\subsubsection{Breathing frequencies of the condensate}\label{subsubsec:breath}
A typical excitation sequence of breathing modes of the condensate proceeds as
follows. One starts with a steady state condensate in the trapping potential.
Then one modulates one of the trap frequencies for some finite time
$t_{\mbox{\scriptsize exc}}$, at a frequency close to an expected resonance
of the condensate. Then one lets the excited condensate
evolve in the unperturbed trap for some adjustable time $t_{\mbox{\scriptsize osc}}$.
Finally one can perform imaging of the cloud, {\sl e.g.} by performing a ballistic
expansion of the condensate and measuring the aspect ratio of the expanded 
cloud. By repeating the whole sequence for different values of $t_{\mbox{\scriptsize osc}}$
one can reconstruct the aspect ratio as function of $t_{\mbox{\scriptsize osc}}$.

\begin{figure}[htb]
\epsfysize=8cm \centerline{\epsfbox{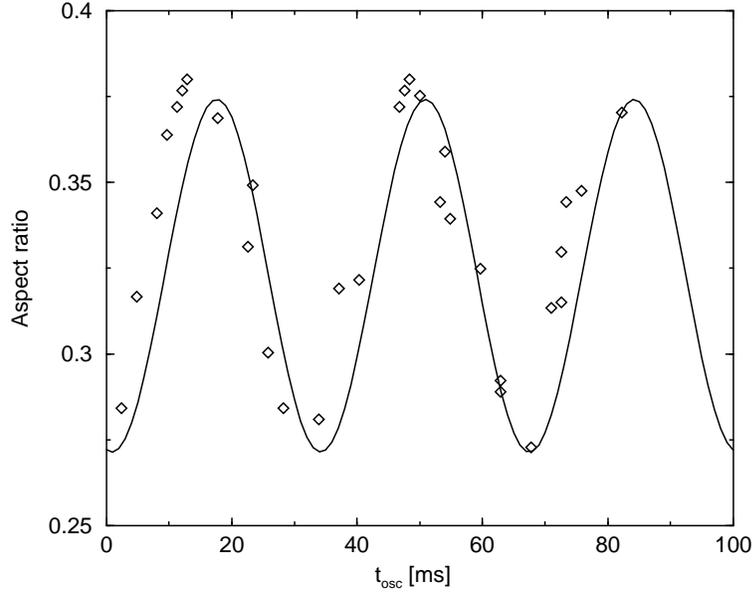}}
\caption{\small Aspect ratio of the excited and ballistically expanded condensate
as a function of free oscillation time $t_{\mbox{\scriptsize osc}}$.
The expansion time is 40 ms, the unperturbed  trap frequencies are
$\omega_\perp(0)=2\pi\times 250$ Hz, $\omega_z(0)=2\pi
\times 19$ Hz. Solid line: theory.
Diamonds: experimental
data obtained at MIT.}\label{fig:MIT_exc}
\end{figure}

Such a procedure has been used at JILA and at MIT. In figure \ref{fig:MIT_exc} are shown
results obtained at MIT in a cigar-shaped trap, for a modulation of the trap
frequency along the slow (that is weakly confining) axis $z$. The solid line corresponds
to the prediction of scaling theory, the input parameters being (i) the oscillation frequencies
$\omega_\alpha$ of the atoms in the trap, (ii) the precise way the excitation is performed,
and (iii) the duration of ballistic expansion; 
the agreement between theory and experiment is good, considering
the fact that there is no fitting parameter \cite{Dum_sca}.

If one is interested only in the frequencies of the breathing eigenmodes of the condensate
it is sufficient to linearize the equations of the scaling parameters around their
steady state value:
\be
\frac{d^2}{dt^2} \delta\lambda_\alpha = -2\omega_\alpha^2(0)\delta\lambda_\alpha
-\omega_\alpha^2(0)\sum_\beta \delta\lambda_\beta
\label{eq:lin_sca}
\ee
and find the eigenvalues of the corresponding three by three linear
system (see also \S\ref{subsubsec:asht}).
For a trap with cylindrical symmetry one gets the eigenfrequencies
$\Omega = \sqrt{2}\omega_\perp(0) $ and
\be
\Omega^2 = \frac{1}{2}\left[3\omega_z^2(0) +4\omega_\perp^2(0)
\pm \left(9\omega_z^4(0)+16\omega_\perp^4(0)-16\omega_z^2(0)\omega_\perp^2(0)\right)^{1/2}\right].
\ee
The mode observed at MIT corresponds to the $-$ sign in the above expression;
as the trap was cigar-shaped in the experiment, $\omega_\perp(0)\gg\omega_z(0)$
so that one has the approximate formula
\be
\Omega \simeq \left(\frac{5}{2}\right)^{1/2}\omega_z(0).
\label{eq:omz}
\ee

\section{What we learn from a linearization of the Gross-Pitaevskii equation}
\markright{Linearization of Gross-Pitaevskii equation}
\label{Cap:linear}

There are several important motivations to perform a linearization
of the Gross-Pitaevskii equation around a steady state solution $\phi_0$:
\begin{itemize}
\item as the Gross-Pitaevskii equation is a non-linear equation it
is crucial to check the so-called \lq\lq dynamical"
stability of the steady state solution. More precisely one has to check
with a linear stability analysis that any small deviation $\delta\phi$ 
of the condensate wavefunction from $\phi_0$ does not diverge exponentially
with time. Otherwise $\phi_0$ may not be physically considered as a steady state
as even very small perturbations will eventually induce an evolution
of the condensate wavefunction far from $\phi_0$.
\item as a byproduct of linear stability analysis we obtain a linear
response theory for the condensate very useful to interpret experiments
which apply a weak perturbation to the condensate.
\item another important byproduct is the Bogoliubov approach which gives
a description of the state of the non-condensed particles that is still approximate
but more accurate at low temperature  (typically $k_B T <\mu$) that the
Hartree-Fock approach. This allows to check the so-called \lq\lq thermodynamical
stability" of the condensate and will be the subject of \S\ref{Cap:bogol}.
\end{itemize}

\subsection{Linear response theory for the condensate wavefunction}
\label{subsec:linear_response}
\subsubsection{Linearize the Gross-Pitaevskii solution around a steady state
solution}
Let $\phi_0(\vec{r}\,)$ be a steady state solution of the
Gross-Pitaevskii solution 
in the time independent trapping potential $U_0(\vec{r}\,)$:
\be
0 = \left[-\frac{\hbar^2}{2m}\Delta + U_0 + gN_0 |\phi_0|^2-\mu\right]
\phi_0.
\ee

The trapping potential is then slightly modified by a time
dependent perturbation $\delta U(\vec{r},t)$, resulting in
a total trapping potential
\be
U(\vec{r},t) = U_0(\vec{r}\,) + \delta U(\vec{r},t).
\ee
The condensate wavefunction, initially equal to $\phi_0$,
evolve according to
\be
i\hbar\partial_t \phi = \left[-\frac{\hbar^2}{2m}\Delta + U + gN_0 |\phi|^2
-\mu\right] \phi.
\label{eq:tblin}
\ee
As $\delta U$ is so small 
we assume that $\phi$ experiences only a small deviation
from $\phi_0$:
\be
\phi(\vec{r},t) = \phi_0(\vec{r}\,) + \delta\phi(\vec{r},t)
\ee
so that we can linearize Eq.(\ref{eq:tblin}) in terms of $\delta \phi$.
Neglecting the second order product of $\delta\phi$ and $\delta U$ we obtain:
\be
i\hbar\partial_t\delta\phi =  \left[-\frac{\hbar^2}{2m}\Delta + U_0-\mu\right]
\delta\phi + 2 g N_0 \phi_0^*\phi_0\delta\phi + gN_0\phi_0^2\delta\phi^*
+\delta U\phi_0.
\label{eq:gpelin}
\ee
Note the presence of the factor 2 in front of the term proportional
to $g\delta\phi$; it turns out (and this should become clear in
the Bogoliubov approach)
that this factor 2 has the same origin as the one in the Hartree-Fock
potential Eq.(\ref{eq:mfpot}) for the non-condensed particles.
As $\phi$ remains normalized to unity, as $\phi_0$ was,
we note that to first order in $\delta \phi$,
\be
\int d^3\vec{r}\; \left[\delta\phi(\vec{r},t)\phi_0^*(\vec{r}\,) +
\phi_0(\vec{r}\,)\delta\phi^*(\vec{r},t)\right] = 0.
\label{eq:norma}
\ee

A peculiar feature of Eq.(\ref{eq:gpelin}) is that, though it is obtained
from a linearization procedure, it is not a linear equation for $\delta\phi$
in the strict mathematical sense: if $\delta\phi$ is a particular
solution of the homogeneous part of this equation (set $\delta U=0$), 
the function $\alpha \delta\phi$ (where
$\alpha$ is a constant complex number) is generally not a solution of
the homogeneous part anymore because of the coupling of $\delta\phi$ to 
$\delta\phi^*$. There are several possibilities to restore this
linearity. A first one is to consider as unknown functions the
real part and the imaginary part of $\delta\phi$. A second, more elegant method, 
more common in the literature, is to introduce formally as unknown
the two-component column vector:
\be
\left(
\begin{tabular}{c}
$\delta\phi(\vec{r},t)$\\ $\delta\phi^*(\vec{r},t)$
\end{tabular}
\right)
\ee
the functions $\delta\phi$ and $\delta\phi^*$ being now considered
as independent. We then rewrite Eq.(\ref{eq:gpelin}) as the linear
system:
\be
\label{eq:gpelinsys}
i\hbar\partial_t \left(
\begin{tabular}{c}
$\delta\phi(\vec{r},t)$\\ $\delta\phi^*(\vec{r},t)$
\end{tabular}
\right) ={\cal L}_{GP}
\left(
\begin{tabular}{c}
$\delta\phi(\vec{r},t)$\\ $\delta\phi^*(\vec{r},t)$
\end{tabular}
\right)
+\left(
\begin{tabular}{c}
$S(\vec{r},t)$ \\ $-S^*(\vec{r},t)$
\end{tabular}
\right)
\ee
with a source term $S(\vec{r},t)=\delta U(\vec{r},t)\phi_0(\vec{r}\,)$
and a linear operator
\be
{\cal L}_{GP} = \left(
\begin{tabular}{cc}
$\displaystyle H_{GP}+gN_0|\phi_0|^2 $
& $g N_0 \phi_0^2$ \\
$-g N_0 \phi_0^{*2}$ &
$\displaystyle -\left[H_{GP}+gN_0|\phi_0|^2\right]^*$
\end{tabular}
\right)
\ee
where we have introduced the Gross-Pitaevskii Hamiltonian:
\be
\label{eq:GPH}
H_{GP} \equiv -\frac{\hbar^2}{2m}\Delta + U_0 + gN_0|\phi_0|^2
-\mu.
\ee
Note the presence of complex conjugation in the second line
of ${\cal L}_{GP}$; it also applies to the potential $U_0$, without
effect here as $U_0$, hermitian function of $\vec{r}$,
is real; it should not be forgotten if situations where the potential
contains a complex term such as $-\Omega L_z$ where $L_z$
is the angular momentum operator (inertial term in a frame
rotating at angular velocity $\Omega$).

As the operator ${\cal L}_{GP}$ is time independent the general
method to determine the time evolution of $\delta\phi$ is
to diagonalize ${\cal L}_{GP}$ and expand $\delta\phi$
on the corresponding eigenmodes. At this stage one faces a slight difficulty:
it turns out that ${\cal L}_{GP}$ is not diagonalizable, that is the set
of all eigenvectors of ${\cal L}_{GP}$ does not form a basis (in general
one vector is missing to span the whole functional space). Mathematically
this can be solved by putting ${\cal L}_{GP}$ into the so-called
Jordan normal form. Here we use the more physical following procedure.

\subsubsection{Extracting the \lq\lq relevant part" from $\delta\phi$}
We split $\delta\phi$ on a component along $\phi_0$ and a part
orthogonal to $\phi_0$:
\be
\delta\phi(\vec{r},t) = \eta(t) \phi_0(\vec{r}\,) +\delta\phi_\perp(\vec{r},t).
\ee
From Eq.(\ref{eq:norma}) valid to first order in $\delta\phi$
we realize that $\eta=\langle\phi_0|\delta\phi\rangle$ is
such that $\eta(t)+\eta^*(t)=0$, so that $\eta(t)$ is purely imaginary
and can be reinterpreted as a change of phase of $\phi_0$:
\be
\phi(\vec{r},t) \simeq e^{\eta(t)}\phi_0(\vec{r}\,) + \delta\phi_\perp
(\vec{r},t).
\ee
One then sees that this change of phase has no consequence on the one-body
density matrix of the condensate, up to first order in $\delta \phi$:
\be
|\phi\rangle\langle\phi| \simeq |\phi_0\rangle\langle\phi_0|
+|\phi_0\rangle\langle\delta\phi_\perp|+
|\delta\phi_\perp\rangle\langle\phi_0|.
\ee

After a little algebra we turn Eq.(\ref{eq:gpelinsys}) into a closed equation
for $\delta\phi_\perp$:
\be
\label{eq:gpelinperp}
i\hbar\partial_t \left(
\begin{tabular}{c}
$\delta\phi_\perp(\vec{r},t)$\\ $\delta\phi^*_\perp(\vec{r},t)$
\end{tabular}
\right) ={\cal L}
\left(
\begin{tabular}{c}
$\delta\phi_\perp(\vec{r},t)$\\ $\delta\phi^*_\perp(\vec{r},t)$
\end{tabular}
\right)
+\left(
\begin{tabular}{c}
$S_\perp(\vec{r},t)$ \\ $-S_\perp^*(\vec{r},t)$
\end{tabular}\right).
\ee
Introducing the projection operators orthogonally to $\phi_0$
and $\phi_0^*$:
\bea
Q &=& 1 -|\phi_0\rangle\langle\phi_0| \\
Q^* &=& 1 -|\phi_0^*\rangle\langle\phi^*_0|
\eea
we have $S_\perp= Q S$ and
\be
{\cal L} = \left(
\begin{tabular}{cc}
$\displaystyle H_{GP}
+gN_0Q|\phi_0|^2Q$
& $g N_0 Q\phi_0^2Q^*$ \\
$-g N_0 Q^*\phi_0^{*2}Q$ &
$\displaystyle-\left[H_{GP}
+gN_0Q|\phi_0|^2Q\right]^*$
\end{tabular}
\right)
\ee
where the Gross-Pitaevskii Hamiltonian $H_{GP}$
is defined in Eq.(\ref{eq:GPH}).
In general the operator ${\cal L}$ is diagonalizable.

\subsubsection{Spectral properties of $\cal L$ and dynamical stability}
Consider an eigenvector of $\cal L$ with eigenvalue
$\epsilon_k$:
\be
{\cal L}\left(
\begin{tabular}{c}
$u_k$ \\ $v_k$
\end{tabular}\right) =
\epsilon_k
\left(
\begin{tabular}{c}
$u_k$ \\ $v_k$
\end{tabular}\right)
\ee
The free evolution of this mode, that is for $\delta U=0$, is given by the phase
factor $\exp(-i\epsilon_k t/\hbar)$. This factor remains bounded
in time provided that the imaginary part of $\epsilon_k$ is
negative, which leads to the dynamical stability condition
\be
\label{eq:stab1}
\mbox{Im}(\epsilon_k) \leq 0 \ \ \mbox{for all}\ \ k.
\ee

One has the following three interesting spectral properties.
\begin{enumerate}
\item $\epsilon_k^*$ is also an eigenvalue of $\cal L$.
\item $\left( \begin{tabular}{c} $v_k^*$ \\ $u_k^*$ \end{tabular}\right)$
is also an eigenvector of $\cal L$, with eigenvalue $-\epsilon_k^*$.
\item $\left( \begin{tabular}{c} $u_k$ \\ $-v_k$ \end{tabular}\right)$
is an eigenvector of ${\cal L}^\dagger$ with eigenvalue $\epsilon_k$.
\end{enumerate}

The last two of these three properties can be checked by direct substitution.
They can be viewed more elegantly as a consequence of the symmetry
properties:
\bea
\left(\begin{array}{rr}0 & 1 \\ 1 & 0
\end{array}\right)
{\cal L} \left(\begin{array}{rr}0 & 1 \\ 1 & 0\end{array}\right)^{-1}
= -{\cal L}^* \\
\left(\begin{array}{rr}1 & 0 \\ 0 & -1
\end{array}\right){\cal L} \left(\begin{array}{rr}1 & 0 \\ 0 & -1
\end{array}\right)^{-1}
= {\cal L}^\dagger.
\label{eq:sigma_3}
\eea
As we shall see this last property involving ${\cal L}^\dagger$
is useful to write $\cal L$ in diagonal form.

The first of these properties is easy to prove when $\phi_0$ is real.
In this case the operator $\cal L$ is a real operator (as $Q$,
$U_0$, $\phi_0^2$, $\Delta$ are real) so that complex eigenvalues 
come by pairs of complex conjugates. 
When $\phi_0$ is complex one can use the following mathematical fact:
if $\epsilon_k$ is an eigenvalue of an arbitrary operator $\cal L$,
$\epsilon_k^*$ is an eigenvalue of the operator ${\cal L}^\dagger$.
Here ${\cal L}^\dagger$ differs from $\cal L$ by a unitary transform
(\ref{eq:sigma_3}) so that it has the same spectrum as $\cal L$.
This property of the spectrum is actually known 
in classical mechanics for a linearized Hamiltonian system, and the
Gross-Pitaevskii equation can be viewed as a classical Hamiltonian
equation for a continuous set of conjugate coordinates $q=\mbox{Re}(\phi),
p=\mbox{Im}(\phi)$. 

As a consequence of this first spectral property
of $\cal L$ the dynamical stability condition Eq.(\ref{eq:stab1})
can be reformulated as 
\be
\mbox{Im}(\epsilon_k)= 0  \ \ \mbox{for all}\ \ k
\ee
that is all the eigenvalues of $\cal L$ have to be real
to have a dynamically stable condensate wavefunction.
We assume that this property is satisfied in the remaining part
of this subsection \ref{subsec:linear_response}.

\subsubsection{Diagonalization of $\cal L$}
As $\cal L$ is not a Hermitian operator the eigenbasis of $\cal L$
is not orthogonal (see the minus sign in the
second line of $\cal L$, due to Bose statistics; one does not have
this sign in the BCS theory for interacting fermions).

To write $\cal L$ in diagonal form the knowledge of the eigenvectors
is then {\sl a priori} not sufficient. One generally proceeds
as follows:

\noindent {\bf Reminder: } {\sl Let $M$ be a diagonalizable but
not necessarily Hermitian operator. Then the diagonal form of 
$M$ can be written as}
\be
M = \sum_k m_k|\psi_k^R\rangle\langle\psi_k^L|
\ee
{\sl where $|\psi_k^R\rangle$ is a right eigenvector of $M$ with
eigenvalue $m_k$:}
\be
M|\psi_k^R\rangle = m_k |\psi_k^R\rangle
\ee
{\sl and $\langle\psi_k^L|$ is a left eigenvector of $M$ with eigenvalue
$m_k$:}
\be
\langle\psi_k^L|M = m_k \langle\psi_k^L|
\ee
{\sl or equivalently}
\be
M^\dagger |\psi_k^L\rangle = m_k^* |\psi_k^L\rangle.
\ee
{\sl The normalization of the left and right eigenvectors is
such that }
\be
\langle\psi_k^L|\psi_{k'}^R\rangle = \delta_{k,k'}.
\label{eq:cond_norm}
\ee
{\sl $|\psi_k^L\rangle$ is then called the adjoint vector
of $|\psi_k^R\rangle$.}

We apply this reminder to $\cal L$. We have already defined
the right eigenvector:
\be
|\psi_k^R\rangle = \left(\begin{tabular}{c}
$|u_k\rangle$ \\ $|v_k\rangle$\end{tabular}\right),
\ \ \mbox{eigenvalue of $\cal L$}\ : \ \epsilon_k.
\ee
From the third of the above spectral properties of $\cal L$
we can easily obtain the corresponding left eigenvector
up to a normalization factor:
\be
|\psi_k^L\rangle = {\cal N}_k\left(\begin{tabular}{c}
$|u_k\rangle$ \\ $-|v_k\rangle$\end{tabular}\right),
\ \ \mbox{eigenvalue of ${\cal L}^\dagger$}\ : \ \epsilon_k^*=\epsilon_k.
\ee
The normalization condition Eq.(\ref{eq:cond_norm}) imposes
\be
\langle\psi_k^L|\psi_{k}^R\rangle=1 = {\cal N}_k^*
[\langle u_k|u_k\rangle -\langle v_k|v_k\rangle].
\ee
It is then natural to normalize the right eigenvectors
in such a way that the quantity between square brackets is
$\pm 1$, leading to ${\cal N}_k =\pm 1$. 

We therefore group the eigenvectors of $\cal L$ in three
families:
\begin{itemize}
\item the $+$ family, such that 
$\langle u_k|u_k\rangle -\langle v_k|v_k\rangle=+1$
\item the $-$ family, such that 
$\langle u_k|u_k\rangle -\langle v_k|v_k\rangle=-1$
\item the $0$ family, such that 
$\langle u_k|u_k\rangle -\langle v_k|v_k\rangle=0$.
\end{itemize}

From the spectral property number 2 we see that there
is a duality between the $+$ family and the $-$ family. Conventionally
we will refer to eigenvectors of the $+$ family as
$(u_k,v_k)$ (eigenvalue $\epsilon_k$)
and the eigenvectors of the $-$ family will be expressed
as $(v_k^*,u_k^*)$ (eigenvalue $-\epsilon_k$).

Generally there are only the following two members in the $0$ family: 
\be
\label{eq:fzero}
\left(\begin{tabular}{c}$\phi_0$ \\ 0\end{tabular}\right)
\ \ \ \mbox{and} \ \ \
\left(\begin{tabular}{c}0 \\ $\phi_0^*$\end{tabular}\right).
\ee
One can check that these two vectors are eigenvectors of $\cal L$
with the eigenvalue zero. [In the case of ${\cal L}_{GP}$ one
finds in general only one zero-energy eigenmode, the missing one
leads to the non-diagonalizability]. 
In general these two vectors span the whole $0$ family. As they are also
eigenmodes of the operator ${\cal L}^\dagger$ with the eigenvalue
zero they are actually their own conjugate vectors!
From Eq.(\ref{eq:cond_norm}) we then get the important property:
\be
\langle\phi_0|u_k\rangle = \langle \phi_0^*|v_k\rangle =0
\ \ \mbox{for all}\ \ k \ \ \mbox{in $+$ family}.
\ee

It is important to note that the $+$ in the denomination \lq\lq $+$ family"
refers {\sl a priori}
to the sign of $\langle u_k|u_k\rangle -\langle v_k|v_k\rangle$
and {\bf not} to the sign of $\epsilon_k$ !

\subsubsection{General solution of the linearized problem}
We expand the unknown column vector of Eq.(\ref{eq:gpelinperp})
onto eigenmodes of $\cal L$. We assume that the only modes
of the $0$ family are the ones of Eq.(\ref{eq:fzero}); these modes
do not contribute to the expansion as $\langle\phi_0|\delta\phi_\perp\rangle
=\langle\phi_0^*|\delta\phi_\perp^*\rangle =0$.
We then get the expansion
\be
\left(\begin{tabular}{c}
$\delta\phi_\perp(\vec{r},t)$ \\ $\delta\phi_\perp^*(\vec{r},t)$ 
\end{tabular}\right)
= \sum_{k \in + \mbox{\scriptsize family}}
b_k(t) \left(\begin{tabular}{c} $u_k(\vec{r}\,)$ \\ $v_k(\vec{r}\,)$
\end{tabular}\right)
+b_k^*(t) \left(\begin{tabular}{c} $v_k^*(\vec{r}\,)$ \\ $u_k^*(\vec{r}\,)$
\end{tabular}\right)
\label{eq:expang}
\ee
with the coefficients
\be
b_k(t) = (\langle u_k|,-\langle v_k|) 
\left(\begin{tabular}{c}
$|\delta\phi_\perp(t)\rangle$ \\
$|\delta\phi_\perp^*(t)\rangle$
\end{tabular}\right)
=\int d^3\vec{r}\; \left[u_k^*(\vec{r}\,)\delta\phi_\perp(\vec{r},t)
-v_k^*(\vec{r}\,)\delta\phi_\perp^*(\vec{r},t)\right].
\ee
As the second component of the expanded column vector is the complex
conjugate of the first component the amplitudes on the $-$ family
modes are the complex conjugates of the amplitudes $b_k$ on the
$+$ family modes, that is $b_k^*$.

Similarly one expands the source term of Eq.(\ref{eq:gpelinperp})
on the eigenmodes of $\cal L$. The components on the $+$ family
modes are given by
\be
s_k(t) = \int d^3\vec{r}\; \left[u_k^*(\vec{r}\,)S_\perp(\vec{r},t)+
v_k^*(\vec{r}\,)S_\perp^*(\vec{r},t)\right].
\ee
Note the absence of $-$ sign here, due to the fact that the second
component of the source column vector is the opposite of the complex
conjugate of the first component.
Finally the projection of Eq.(\ref{eq:gpelinperp}) 
on the eigenmodes of the $+$ family gives the set of equations:
\be
i\hbar \frac{d}{dt}b_k(t) = \epsilon_k b_k(t) + s_k(t)
\ee
which can be integrated including the initial condition
$\delta\phi_\perp(t_0)=0$:
\be
b_k(t) = \int_0^{t-t_0} \frac{d\tau}{i\hbar}s_k(t-\tau)
e^{-i\epsilon_k\tau/\hbar}.
\ee

\subsubsection{Link between eigenmodes of ${\cal L}_{GP}$
and eigenmodes of $\cal L$}

The linear operators $\cal L$ and ${\cal L}_{GP}$ describe
the same physical problem, so that one expects  in
particular that their spectra, which correspond to the
linear response frequencies of the condensate, are the same.

This expectation is confirmed by the simple result, that one can
check by direct substitution: if $(|u_k^{GP}\rangle,
|v_k^{GP}\rangle)$ is an eigenvector
of ${\cal L}_{GP}$ with the eigenvalue $\epsilon_k$
then $(Q|u_k^{GP}\rangle,Q^*|v_k^{GP}\rangle)$ is an eigenvector
of $\cal L$ with the {\sl same} eigenvalue $\epsilon_k$.

\subsection{Examples of dynamical instabilities}
We consider simple stationary solutions of the Gross-Pitaevskii
equations that are dynamically unstable, that is with
non-real eigenfrequencies $\epsilon_k/\hbar$.
The situations considered correspond to a gas trapped in a cubic box
with periodic boundary conditions; analytical calculations can then
be performed. The conclusions remain qualitatively correct 
for harmonic traps.

\subsubsection{Condensate in a box}\label{subsubsec:bogs}
The atoms are trapped in a cubic box of size $L$, and we assume
periodic boundary conditions. An obvious solution of the time
independent Gross-Pitaevskii equation is then the plane wave
with vanishing momentum,
\be
\phi_0(\vec{r}\,) = \frac{1}{L^{3/2}}.
\ee
It has a chemical potential
\be
\mu = gN_0 |\phi_0|^2 =\rho_0 g \ee
where $\rho_0= N_0/L^3$ is the density of condensate atoms.

To obtain the linear response frequencies of the condensates
we calculate the spectrum of ${\cal L}_{GP}$, this operator
takes here the very simple form:
\be
{\cal L}_{GP} = \left(
\begin{tabular}{cc}
$\displaystyle-\frac{\hbar^2}{2m}\Delta + \rho_0 g $
& $\rho_0 g$ \\
$-\rho_0 g$ &
$\displaystyle-\left[-\frac{\hbar^2}{2m}\Delta + \rho_0 g\right]$
\end{tabular}
\right).
\ee
Using the translational invariance of this operator we seek
its eigenvectors in the form of plane waves:
\be
\left(\begin{tabular}{c}
$u_{\vec k}^{GP}(\vec{r}\,)$ \\ \\ $v_{\vec k}^{GP}(\vec{r}\,)$
\end{tabular} \right)=
\frac{e^{i\vec{k}\cdot\vec{r}}}{L^{3/2}}
\left(\begin{tabular}{c}
$U_{\vec k}$ \\ \\ $V_{\vec k}$
\end{tabular} \right).
\label{eq:planew}
\ee
Within the subspace of plane waves with wave vector $\vec{k}$,
${\cal L}_{GP}$ is represented by the 2$\times$2 non-Hermitian
matrix:
\be
{\cal L}_{GP}[\vec{k}]=
\left(\begin{array}{cc}\displaystyle\frac{\hbar^2k^2}{2m}+\rho_0 g & \rho_0 g \\
-\rho_0 g & -\left[\displaystyle\frac{\hbar^2k^2}{2m} +\rho_0 g\right]\end{array}\right).
\label{eq:2by2}
\ee
For $\vec{k}\neq \vec{0}$ this matrix can be diagonalized, giving
one eigenvector of $+$ family, with the eigenvalue
\be
\epsilon_{\vec k}=\left[\frac{\hbar^2k^2}{2m}\left(
\frac{\hbar^2 k^2}{2m}+2\rho_0 g\right) \right]^{1/2}
\label{eq:bogspec}
\ee
and one eigenvector of $-$ family with the eigenvalue $-\epsilon_{\vec k}$.
The eigenvector of the $+$ family can be chosen as
\bea
U_{\vec k} + V_{\vec k} &=& \left(
\frac{\displaystyle\frac{\hbar^2 k^2}{2m}}{\displaystyle\frac{\hbar^2 k^2}{2m}+2\rho_0 g}
\right)^{1/4} \nonumber\\
U_{\vec k} - V_{\vec k} &=& \left(
\frac{\displaystyle\frac{\hbar^2 k^2}{2m}}{\displaystyle\frac{\hbar^2 k^2}{2m}+2\rho_0 g}
\right)^{-1/4}
\label{eq:u_et_v}
\eea
with the correct normalization $U_{\vec{k}}^2 - V_{\vec{k}}^2 =1$.

The spectrum Eq.(\ref{eq:bogspec}) is the so-called Bogoliubov
spectrum, as it was first derived by Bogoliubov. Physically it
is a very important result.
In the limiting case of the ideal Bose gas ($g=0$) the spectrum is
the usual parabola; for $g\neq 0$ the spectrum is very different,
and this deserves a more detailed discussion

\begin{itemize}
\item {\sl Bogoliubov spectrum for $g>0$}
\end{itemize}
In the case of effective repulsive interactions
the Bogoliubov spectrum strongly differs from the one of a free
particle in the low momenta domain $\hbar^2 k^2/2m\ll 2\rho_0 g$
as it scales linearly with $k$:
\be
\epsilon_{\vec k} \simeq \hbar k \sqrt{\frac{\rho_0g}{m}}.
\ee
This linear dispersion relation leads to a propagation of low
energy excitations in the  condensate in the form of sound waves
with a sound velocity $c_s$ given by
\be
c_s = \frac{d\omega_{\vec k}}{dk}=\frac{1}{\hbar}
\frac{d\epsilon_{\vec k}}{dk} = \sqrt{\frac{\rho_0g}{m}}
\ee
or equivalently by the relativistic type formula
\be
mc^2_s = \rho_0 g.
\label{eq:vs}
\ee

Superfluidity is an important consequence of this linear behavior
of the spectrum at low $k$, as shown by an argument due to Landau and that
we explain briefly. Consider
a particle (of mass $M$) sent in the atomic gas
with an initial velocity $\vec{u}$. The motion of this particle
can be damped by interaction with the condensate only if the particle
can create some excitation of the condensate. Imagine that such
an excitation is produced, with momentum $\vec{k}$; the 
particle experiences a momentum recoil of $-\hbar\vec{k}$ and
conservation of energy imposes
\be
\epsilon_{\vec{k}} = \frac{1}{2}M \vec{u}\,^2-
\frac{1}{2M} \left[M\vec{u}-\hbar\vec{k}\,\right]^2= \hbar\vec{k}\cdot\vec{u}-
\frac{\hbar^2k^2}{2M}.
\ee
The velocity $u$ has then to satisfy the inequality:
\be
u\geq \left|\frac{\vec{k}\cdot\vec{u}}{k}\right| \geq \frac{\epsilon_{\vec k}}{k}
\geq c_s.
\ee
So a particle with an incoming velocity smaller than the sound velocity
can move through the condensate without damping, only scattering on
thermal excitations of the gas can damp its motion. This prediction
has received an experimental confirmation at MIT \cite{MIT_Landau}.

At high momenta ($\hbar^2 k^2/2m \gg \rho_0 g$) 
corresponding to a velocity $\hbar k/m$ much larger
than the sound velocity  the Bogoliubov spectrum reduces to a shifted
parabola
\be
\epsilon_{\vec k} \simeq \frac{\hbar^2k^2}{2m}+\rho_0 g
= \frac{\hbar^2k^2}{2m}+2\rho_0 g-\mu.
\label{eq:highk}
\ee

This approximate form can be obtained by a series expansion
of the general formula Eq.(\ref{eq:bogspec}). It can also
be derived more instructively from the observation that
the off-diagonal coupling $\rho_0 g$
between the $U_{\vec k}$ component
and the $V_{\vec k}$  component in the 2$\times$2 matrix
Eq.(\ref{eq:2by2}) becomes very non-resonant at high $k$
(because the diagonal terms for $U_{\vec k}$ and $V_{\vec k}$
have opposite signs); neglecting this coupling
one recovers Eq.(\ref{eq:highk}) with $U_{\vec k}\simeq 1$,
$V_{\vec k}\simeq 0$.

This last high energy property applies also in a non-uniform
trapping potential:
neglecting the off-diagonal coupling between
$u_k$ and $v_k$  one approximates the high energy part 
of the Bogoliubov spectrum by the eigenvalues of
\be
-\frac{\hbar^2}{2m}\Delta + U_0 + 2gN_0|\phi_0|^2-\mu
\ee
which (up to the shift $-\mu$) is the Hartree-Fock Hamiltonian for non-condensed
particles Eq.(\ref{eq:hfh}) in a regime of an almost pure condensate where
the density of non-condensed particles is negligible
as compared to the condensate density $N_0|\phi_0|^2$.

From this we expect that the Hartree Fock approach is invalid
for the low energy fraction of the non-condensed gas (energy
typically less than $\mu$); this may become a problem
at temperatures $k_B T < \mu$, where one has to use the
more precise Bogoliubov approach of \S\ref{Cap:bogol}.

\begin{itemize}
\item{\sl Case of a negative $g$}
\end{itemize}

In the case of effective attractive interactions between particles
the dynamical
stability condition $\mbox{Im}(\epsilon_{\vec{k}})=0$ is satisfied
if and only if
\be
\frac{\hbar^2k^2}{2m}+2g\rho_0 \geq 0 \ \ \mbox{for all}\ \
k>0.
\label{eq:stco}
\ee
If one considers the thermodynamical limit of an infinite
number of condensate atoms with a fixed mean density
$\rho_0=N_0/L^3$ the stability condition cannot be satisfied
as $k$ can be arbitrarily close to $0$ in an infinite box. One
may then be tempted to conclude that condensates with
effective attractive interactions cannot be obtained
experimentally, attractive interactions leading to a spatial
collapse of the gas.

Experiments with atomic gases can deal however with small
number of atoms and the simplifying assumption of a
thermodynamical limit is not necessarily a good approximation.
In the cubix box of size $L$ with periodic boundary conditions
the components of the wavevector $\vec{k}$ of an atom are integer
multiples of $2\pi/L$. The smallest but non zero modulus of wavevector
that can be achieved is therefore $2\pi/L$ (by taking e.g.\
$k_x=2\pi/L,k_y=k_z=0$).  Dynamical stability condition Eq.(\ref{eq:stco})
can then be rewritten as
\be
\frac{\hbar^2}{2m} \left(\frac{2\pi}{L}\right)^2 \geq 2 |g|
\frac{N_0}{L^3}
\label{eq:stcoex}
\ee
or equivalently in terms of the scattering length as
\be
\frac{N_0|a|}{L}\leq \frac{\pi}{4}.
\ee
Condensates for $a<0$ can contain a limited number of atoms
proportional to the size of the condensate.

Condition Eq.(\ref{eq:stcoex})
has a clear physical interpretation: the energy
gap in the spectrum of a particle in the box between the ground
state and the first excited states should be larger than the mean
field energy per particle: stabilization against collapse is
thus provided by the discrete spectrum of the atoms in the trapping
potential. This condition can thus be qualitatively extended
to the case of an isotropic harmonic trap, $\hbar\omega > |g|
N_0/a_{\mbox{\scriptsize ho}}^3$ where $\omega$ is the oscillation frequency
of the atoms in the trap and $a_{\mbox{\scriptsize ho}}=(\hbar/m\omega)^{1/2}$ is the
typical spatial extension of the ground state of the trap.
One then recovers up to a numerical factor the results of
\S\ref{subsubsec:tic}.

\subsubsection{Demixing instability}
We consider here atoms with two internal states $a,b$; this model
is relevant for experiments performed at JILA on binary mixtures of
$^{87}$Rb condensates, and also (if one includes a third atomic
internal level) experiments at MIT in Ketterle's group on spinor
$^{23}$Na condensates.

To describe the elastic interactions between the atoms with two internal
states one needs three coupling constants,
all positive in the case of $^{87}$Rb:
$g_{aa}$ and $g_{bb}$
for interactions between atoms in the same internal state,
$g_{ab}$ for interactions between atoms in different internal states:
\bea
g_{aa}&:&\ \ a+a\rightarrow a+a \nonumber \\
g_{bb}&:&\ \ b+b\rightarrow b+b \label{eq:inter_elas}\\
g_{ab}&:&\ \ a+b\rightarrow a+b \nonumber.
\eea
In the JILA experiment internal states $a$ and $b$ correspond
to different hyperfine levels of the atoms so that inelastic
collisions such as $a+a\rightarrow a+b$ are either strongly
endothermic (and do not take place) or strongly exothermic
(and result in losses of atoms from the trap); we neglect these
inelastic processes.

Omitting for simplicity the regularizing operator in the
pseudo-potential we write the interaction Hamiltonian between
the atoms in second quantized form as
\bea
H_{\mbox{\scriptsize int}}=\int d^3\vec{r}\; &&
\left[ \frac{g_{aa}}{2}\hat{\psi}^\dagger_a(\vec{r}\,) \hat{\psi}^\dagger_a(\vec{r}\,)
\hat{\psi}_a(\vec{r}\,) \hat{\psi}_a(\vec{r}\,)+
\frac{g_{bb}}{2}\hat{\psi}^\dagger_b(\vec{r}\,) \hat{\psi}^\dagger_b(\vec{r}\,)
\hat{\psi}_b(\vec{r}\,) \hat{\psi}_b(\vec{r}\,)\right.\nonumber \\
&+& \left. g_{ab}\hat{\psi}^\dagger_b(\vec{r}\,) \hat{\psi}^\dagger_a(\vec{r}\,)
\hat{\psi}_a(\vec{r}\,) \hat{\psi}_b(\vec{r}\,)\right]
\eea
where $\hat{\psi}_a$ and $\hat{\psi}_b$ are the atomic field operators
for atoms in state $a$ and $b$ respectively. Note the absence of
factor $1/2$ in the $a-b$ interaction term, which is best understood
in first quantization point of view: all the pairs of atoms of the
form $a:i,b:j$, where $i$ running from 1 to $N_a$ labels the atoms
in $a$ and $j$ running from 1 to $N_b$ labels the atoms in $b$, are
different so that there is no double counting
of these interaction terms and no factor $1/2$ is required.

Using the trick of \S\ref{subsubsec:trick} we can rapidly derive the Gross-Pitaevskii
equations for the condensate wavefunctions $\phi_a$ in state $a$
and $\phi_b$ in state $b$, both wavefunctions being normalized
to unity. One simply has to write the Heisenberg equations of
motion for the field operators and perform the substitution
\bea
\hat{\psi}_a &\rightarrow& N_a^{1/2}\phi_a \\
\hat{\psi}_b &\rightarrow& N_b^{1/2}\phi_b
\eea
where $N_{a,b}$ are the number of particles in condensates $a,b$.
As in \S\ref{subsubsec:bogs} we restrict
to the case of atoms trapped in a cubix box of size $L$ with
periodic boundary conditions. We obtain the coupled time dependent
Gross-Pitaevskii equations
\bea
i\hbar\partial_t\phi_a &=& \left[-\frac{\hbar^2}{2m}\Delta
+N_a g_{aa} |\phi_a|^2+ N_b g_{ab} |\phi_b|^2\right]\phi_a \nonumber\\
i\hbar\partial_t\phi_b &=& \left[-\frac{\hbar^2}{2m}\Delta
+N_b g_{bb} |\phi_b|^2+ N_a g_{ab} |\phi_a|^2\right]\phi_b.
\label{eq:cgpe}
\eea

Consider now a steady state solution of these coupled Gross-Pitaevskii
equations. As we have not introduced any coherent coupling between
the internal states $a$ and $b$ (no $\hat{\psi}^\dagger_b\hat{\psi}_a$
term in the Hamiltonian) $\phi_a$ and $\phi_b$ can have in steady state
time dependent phase factors evolving with {\sl different} frequencies:
\bea
\phi_a(\vec{r},t)&=&\phi_{a,0}(\vec{r}\,)e^{-i\mu_a t/\hbar} \\
\phi_b(\vec{r},t)&=&\phi_{b,0}(\vec{r}\,)e^{-i\mu_b t/\hbar}
\eea
From a more thermodynamical perspective we can also observe
that the number of particles $N_a$ and $N_b$ are separately
conserved by the three elastic interactions of Eq.(\ref{eq:inter_elas})
so that {\sl two} distinct chemical potentials $\mu_a$ and $\mu_b$
are required to describe the equilibrium state.

The time independent Gross-Pitaevskii equations for
$\phi_{a,0}$ and $\phi_{b,0}$ in the box have the natural
solutions
\be
\phi_{a,0}(\vec{r}\,)=\phi_{b,0}(\vec{r}\,)=\frac{1}{L^{3/2}}
\label{eq:nsss}
\ee
leading to the following expressions for the chemical potentials:
\bea
\mu_a &=& \rho_{a,0} g_{aa} + \rho_{b,0} g_{bb} \\
\mu_b &=& \rho_{b,0} g_{bb} + \rho_{a,0} g_{ab}
\eea
where $\rho_{a,b}$ are the condensate densities in $a,b$.
Although we assume here that all the coupling constants
are positive it is physically intuitive
that these spatially uniform solutions should become instable
when the interactions between $a$ and $b$ are very repulsive;
one then feels that the two condensates $a$ and $b$ have a tendency
to spatially separate.

To test this expectation
we linearize the time dependent coupled Gross-Pitaevskii
equations around the steady state, setting
\bea
\phi_a(\vec{r},t) &=& e^{-i\mu_a t/\hbar}\left[\phi_{a,0}(\vec{r}\,)
+\delta\phi_a(\vec{r},t)\right] \\
\phi_b(\vec{r},t) &=& e^{-i\mu_b t/\hbar}\left[\phi_{b,0}(\vec{r}\,)
+\delta\phi_b(\vec{r},t)\right] .
\eea
We obtain
\be
i\hbar\partial_t\delta\phi_a = -\frac{\hbar^2}{2m}\Delta\delta\phi_a
+\rho_{a,0} g_{aa}[\delta\phi_a+\delta\phi_a^*]
+\rho_{b,0} g_{ab}[\delta\phi_b+\delta\phi_b^*]
\ee
and a similar equation for $\delta\phi_b$ exchanging the role
of $a$ and $b$ indices. We look for eigenmodes of these linear
equations, with eigenfrequency $\epsilon/\hbar$ and
a well defined wavevector $\vec{k}$. This amounts to performing the
substitutions
\bea
\delta\phi_a(\vec{r},t) &\rightarrow & u_a e^{i(\vec{k}\cdot\vec{r}
-\epsilon t/\hbar)} \nonumber \\
\delta\phi_a^*(\vec{r},t) &\rightarrow & v_a e^{i(\vec{k}\cdot\vec{r}
-\epsilon t/\hbar)} \nonumber 
\eea
and equivalent changes for the $b$ components. This leads to the eigensystem
\be
\left[\epsilon-\frac{\hbar^2k^2}{2m}\right] u_a =
\left[-\epsilon-\frac{\hbar^2k^2}{2m}\right] v_a =
\rho_{a,0}g_{aa}\left[u_a+v_a\right]+\rho_{b,0}g_{ab}\left[u_b+v_b\right]
\label{eq:sysc}
\ee
and similar equations obtained by exchanging the indices $a$
and $b$. Taking as new variables the sums and the differences
between $u$ and $v$
and using the first identity in Eq.(\ref{eq:sysc}) we eliminate
the differences as functions of the sums:
\be
u_a-v_a = \frac{2m\epsilon}{\hbar^2k^2}\left[u_a+v_a\right]\,.
\ee
We get from the second equality in Eq.(\ref{eq:sysc})
(and $a\leftrightarrow b$) a two by two system for the sums
$u_a+v_a$, $u_b+v_b$:
\be
\left\{\frac{\hbar^2 k^2}{2m} \left[ 1 -\left(\frac{2m\epsilon}{\hbar^2 k^2}
\right)^2\right] \; \mbox{Id} + M
\right\} 
\left(\begin{array}{c} u_a + v_a \\ u_b + v_b \end{array}\right)
=0
\ee
where we have introduced the two by two matrix
\be
M = \left(\begin{array}{cc} \rho_{a,0}g_{aa} & \rho_{b,0}g_{ab} \\
\rho_{a,0}g_{ab} & \rho_{b,0}g_{bb}\end{array}\right).
\ee
This leads to the following condition for the spectrum:
\be
\epsilon^2 = \left[\frac{\hbar^2k^2}{2m}
\left(\frac{\hbar^2k^2}{2m}+2\eta_{1,2}\right)\right]
\label{eq:la_condi}
\ee
where $\eta_{1,2}$ are the eigenvalues of $M$.

In the thermodynamical limit the mixture of condensates with
uniform densities is dynamically stable provided that
both eigenvalues $\eta_{1,2}$ are positive. This is equivalent
to the requirement that the symmetric matrix $M$ is positive.
As $g_{aa}>0$ here this is ensured provided that the determinant
of $M$ is positive:
\be
\det M=\rho_{a,0}\rho_{b,0}\left[g_{aa}g_{bb}-g_{ab}^2\right] \geq 0.
\ee
The mixture of spatially uniform condensates is therefore
stable if
\be
g_{ab}\leq (g_{aa} g_{bb})^{1/2}.
\label{eq:stabc}\ee
In this case one can check that the spectrum of the $+$ family
is given by the positive solutions $\epsilon$ to Eq.(\ref{eq:la_condi}).
The spectrum of the binary mixtures of condensates is
then made of two branches, which are both linear at low
momenta with sound velocities $(\eta_{1,2}/m)^{1/2}$.

What happens when this stability condition is not satisfied ?
The condensates $a$
and $b$ have a tendency to separate spatially.
This happens in the JILA experiment \cite{JILA_demix}.
Actually our model in
a box is too crude to be applied to the experimental case
of particles in a harmonic trap, the stability
condition Eq.(\ref{eq:stabc}) being marginally satisfied
for $^{87}$Rb; a numerical solution of the time dependent
Gross-Pitaevskii equations is required in this case
\cite{fed:demix}.

The occurrence of demixing when Eq.(\ref{eq:stabc}) is violated
can be connected with the following simple energy argument.
Consider a demixed configuration with all the $N_a$ atoms
in the left part of the box in a volume $\nu L^3$ and all the
$N_b$ atoms in the right part of the box in the complementary
volume $(1-\nu)L^3$. The condensate
densities vanish on a scale on the order of
the healing length $\xi$ of the gas, this leads to \lq\lq surface" kinetic
and interaction energies negligible in the thermodynamical limit
as compared to the volume interaction energy
\be
 \frac{N_a^2g_{aa}}{2\nu L^3}
+\frac{N_b^2g_{bb}}{2(1-\nu) L^3}.
\ee
We minimize this energy over $\nu$ to obtain
\be
\label{eq:demix}
E_{\mbox{\scriptsize demix}} = \frac{1}{2L^3}
\left[N_a^2g_{aa}+N_b^2 g_{bb}+2N_a N_b (g_{aa}g_{bb})^{1/2}\right].\ee
We find that the demixed configuration has an energy lower than
the one of the spatially uniform configuration
\be
\label{eq:unif}
E_{\mbox{\scriptsize unif}} = \frac{1}{2L^3}
\left[N_a^2g_{aa}+N_b^2 g_{bb}+2N_a N_b g_{ab}\right]\ee
precisely when the stability condition Eq.(\ref{eq:stabc})
of the uniform configuration is violated.    

\subsection{Linear response in the classical hydrodynamic approximation}
We consider in this subsection the case of a condensate in a
harmonic trap. The eigenmodes of the linearized Gross-Pitaevskii
equation can then in general be determined only numerically.
In the Thomas-Fermi regime however approximate results can be obtained
for the low energy eigenmodes of the system from the classical
hydrodynamic approach, as we shall see now.

\subsubsection{Linearized classical hydrodynamic equations}
The classical hydrodynamic equations for the position dependent
condensate
density $\rho(\vec{r}\,)$ and velocity field $\vec{v}(\vec{r}\,)$
have been derived in \S\ref{subsubsec:cha}:
\bea
\partial_t\rho+\mbox{div}\left[\rho\,\vec{v}\,\right] &=& 0 \\
m\left(\partial_t+\vec{v}\cdot\vec{\mbox{grad}}\right)\vec{v} &=&
-\vec{\mbox{grad}}[U+\rho g].
\eea
We linearize these equations around their steady state solution
with density $\rho_0$ and vanishing velocity field $\vec{v_0}=
\vec{0}$ in the unperturbed trapping potential $U_0$. Writing
the trap potential as a perturbation of $U_0$:
\be
U(\vec{r},t) = U_0(\vec{r}\,) + \delta U(\vec{r},t)
\ee
and splitting $\rho$ and $\vec{v}$ as
\bea
\rho(\vec{r},t) &=& \rho_0(\vec{r}\,) + \delta\rho(\vec{r},t) \\
\vec{v}(\vec{r},t) &=& \vec{0} + \delta\vec{v}(\vec{r},t)
\eea
we obtain the linearized equations:
\bea
\partial_t\delta\rho+\mbox{div}\left[\rho_0\delta\vec{v}\,\right] &=& 0 \nonumber \\
m\partial_t\delta\vec{v} +\vec{\mbox{grad}}[\delta\rho g] &=& -\vec{\mbox{
grad}}\,\delta U.
\label{eq:lche}
\eea
Taking the time derivative of the first equation we obtain a term
$\partial_t\delta\vec{v}$ that we can eliminate with the second equation.
This results in a closed equation for the perturbation of density:
\be
\partial_t^2\delta\rho - \mbox{div}\left[\frac{\rho_0 g}{m}\vec{\mbox
{grad}}\,\delta\rho\right] = \mbox{div}\left[\frac{\rho_0}{m}\vec{\mbox{grad}}\,\delta U
\right].
\label{eq:prop}
\ee
The homogeneous part of this equation can be rewritten in a more suggestive
way by introducing the position dependent velocity  $c_s$ given by
\be
mc_s^2(\vec{r}\,)= \rho_0(\vec{r}\,)g.
\label{eq:sv}
\ee
The homogeneous part of Eq.(\ref{eq:prop})
then reads
\be
\partial_t^2\delta\rho - \mbox{div}\left[c_s^2(\vec{r}\,)\vec{\mbox
{grad}}\,\delta\rho\right] 
\ee
which corresponds to the propagation of sound waves with a position
dependent sound velocity $c_s(\vec{r}\,)$. Note that the expression
(\ref{eq:sv}) could be expected from the result
Eq.(\ref{eq:vs}) obtained in the spatially homogeneous
case.

The propagation of sound waves in a cigar shaped trapped condensate has been
observed in Ketterle's group at MIT; the condensate was excited
mechanically by the dipole force induced by a far detuned
laser beam focused at the center of the trap. It is instructive
to note that to predict theoretically the velocity of sound obtained in
the experiment one has 
to carefully solve Eq.(\ref{eq:prop}), rather than to take
naively the sound velocity on the axis of the trap; the naive
expectation would be wrong by a factor of $\sqrt{2}$ 
\cite{Stringari_son1,Stringari_son2}.

\subsubsection{Validity condition of the linearized classical
hydrodynamic equations}

The classical hydrodynamic equations have been obtained from
the time dependent Gross-Pitaevskii equation in hydrodynamic
point of view by neglecting the quantum pressure term
(see \S\ref{subsubsec:cha}). If we keep this term and linearize
the resulting equation we get extra terms with respect to 
Eq.(\ref{eq:lche}). One of the extra terms is
\be
\frac{\hbar^2}{m\rho_0}\Delta\delta\rho
\ee
which should be small as compared to the \lq\lq classical" pressure
term $g\delta\rho$:
\be
\frac{\hbar^2}{m\rho_0}|\Delta\delta\rho| \ll g|\delta\rho|.
\ee
If we denote by $k$ a typical wavevector for the spatial variation
of $\delta\rho$, $\Delta\delta\rho \sim -k^2 \delta\rho$
and the condition for neglecting the quantum pressure
term reads
\be
\frac{\hbar^2k^2}{m} \ll g\rho_0.
\ee

This condition can be rewritten in a variety of ways.
It claims that the wavevector $k$ should satisfy
\be
k \xi \ll 1
\ee
where $\xi=\hbar/(2m\rho_0 g)^{1/2}$ is the healing length
of the condensate. Eq.(\ref{eq:prop}) cannot be used to describe
perturbations of the condensate at a length scale on the order of
$\xi$ or smaller.

The validity condition can also be written as 
\be
\hbar k\ll m c_s\ \ \mbox{or}\ \ \hbar k c_s \ll mc_s^2 = \rho_0 g\sim\mu
\ee
In terms of the Bogoliubov spectrum for the homogeneous condensate
this means that the wavevector $k$ has to be in the {\sl linear}
part of the excitation spectrum. The energy
of the corresponding eigenmode $\sim\hbar k c_s$ has to be
much smaller than $\mu$. Therefore only the eigenmodes of
Eq.(\ref{eq:prop}) with eigenenergy much less than $\mu$
are relevant:
\be
\epsilon \ll \mu
\label{eq:vc}
\ee
the higher energy modes are not an acceptable
approximation of the exact eigenmodes of the condensate.
Note that as shown in \cite{Stringari_not_suff1,Stringari_not_suff2}, Eq.(\ref{eq:vc})
is a necessary but not sufficient condition in a trap.

\subsubsection{Approximate spectrum in a harmonic trap}\label{subsubsec:asht}
We look for eigenmodes of the homogeneous part of Eq.(\ref{eq:prop})
with eigenenergy $\epsilon$. They solve the eigenvalue equation
\be
-\epsilon^2\delta\rho = \mbox{div}\left[c_s^2(\vec{r}\,)
\vec{\mbox{grad}}\,\delta\rho\right].
\label{eq:eig}
\ee
In the present Thomas-Fermi regime $c_s^2$ is a quadratic
function of the coordinates as it is proportional to the condensate
density $\rho_0$. We can therefore solve the eigenvalue equation
using an ansatz for $\delta\rho$ polynomial in the spatial coordinates
$x,y,z$. If $\delta\rho$ is a polynomial of total degree $n$,
$\vec{\mbox{grad}}\,\delta\rho$ is a polynomial of total degree
$n-1$; after multiplication by $c_s^2$ and action of
the div operator we get a polynomial of total degree $(n-1)+2-1=n$.
The subspace of polynomials of degree $\leq n$ is therefore stable.

For low values of the total degree $n$ an analytical calculation is possible.
For example in a general harmonic trap with atomic oscillation
frequencies $\omega_x,\omega_y,\omega_z$ one can check that the
polynomials $\delta\rho=x,y,z$ (respectively) are eigenvectors
with eigenvalues $\epsilon=\hbar\omega_x,\hbar\omega_y,\hbar\omega_z$ (respectively).
These three modes correspond to the oscillation of the center of
mass of the gas, which is exactly decoupled
from all the relative coordinates of the particles in a harmonic trap;
for this specific example the frequencies (but not the modes!) predicted
by classical hydrodynamics are exact. These \lq\lq sloshing" modes
are used experimentally to determine accurately the trap frequencies
$\omega_{x,y,z}$.

Another important example is the case of a total degree $n=2$. One can check that
the subclass of polynomials involving the monomials $x^2$, $y^2$, $z^2$ and $1$
is stable, which corresponds to the ansatz
\be
\delta\rho(\vec{r}\,) = B + \sum_{\alpha=x,y,z} A_\alpha r_\alpha^2.
\label{eq:ans_q}
\ee
By inserting this ansatz into Eq.(\ref{eq:eig}) we arrive at the eigenvalue
system
\be
(\epsilon/\hbar)^2 A_\alpha = 2\omega_\alpha^2 A_\alpha + \omega_\alpha^2
\sum_\beta A_\beta.
\ee
This eigenvalue system can be obtained more directly 
as in Eq.(\ref{eq:lin_sca})
by a linearization 
of the equations Eq.(\ref{eq:scal_par}) for the scaling parameters  $\lambda_\alpha$
around their steady state value $\lambda_\alpha =1$. Physically the modes
identified by the ansatz (\ref{eq:ans_q}) are therefore breathing modes
of the condensate.
The frequency of one of these modes (the one of \S\ref{subsubsec:breath})
has been measured with high precision at MIT in a cigar-shaped trap, 
it differs from the Thomas-Fermi prediction 
$\simeq (5/2)^{1/2}\omega_z$ (see Eq.(\ref{eq:omz}))
by less than one percent \cite{1569}.

In the case of an isotropic harmonic trap all the eigenenergies
$\epsilon$ can be calculated analytically
(keeping in mind that modes with $\epsilon>\mu$
are not properly described by classical hydrodynamics !).
As shown in \cite{Stringari_modes} one uses the rotational symmetry
of the problem, as in the case of Schr\"odinger's equation for
the hydrogen atom, with the ansatz
\be
\delta\rho(\vec{r}\,) = Y_l^m(\theta,\phi)r^l P_{l,n}(r)
\ee
where $\theta,\phi$ are the polar and azimuthal angles of spherical
coordinates, $Y_l^m$ is the spherical harmonic of angular momentum
$l$. The last factor $P_{l,n}(r)$ is a polynomial of degree $n$ in
$r$. As $P_{l,n}$ is a polynomial the recurrence relation
obtained from Eq.(\ref{eq:eig}) for the coefficients of the monomials
$r^j$ should terminate at $j=n$. This leads to the eigenfrequencies
$\Omega=\epsilon/\hbar$ such that
\be
\Omega^2 = (2n^2+2nl+3n+l)\omega^2 \ \ \mbox{for any}\ \ n,l\geq 0
\ee
where $\omega$ is the oscillation frequency of the atoms in the trap.
For $l=1,n=0$ we recover the sloshing modes with frequency
$\Omega=\omega$.
\section{Bogoliubov approach and thermodynamical stability}
\markright{Bogoliubov approach}
\label{Cap:bogol}

Imagine that we have already checked the dynamical stability
of a steady state solution ${\phi_0}$ of the Gross-Pitaevskii
equation for the condensate wavefunction. We cannot relax yet and
be sure that the condensate will remain in state ${\phi_0}$ in the
long run.

What is missing is a check that interaction of the condensate
with the non-condensed cloud does not induce an evolution of the
condensate far from the predicted ${\phi_0}$. We have to check what
is called thermodynamical stability of the condensate as it
involves the \lq\lq thermal", non-condensed component of the gas.

This check will be performed in the low temperature domain
($T\ll T_c$) using the Bogoliubov approach. We will then present 
examples of thermodynamical instabilities.

We give here a summarized account of the $U(1)$-symmetry preserving
Bogoliubov approach developed in \cite{Gardiner,Dum_U1}.
In contrast to the almost general attitude in the literature
this approach does not assume a symmetry breaking state
with $\langle\hat{\psi}\rangle\neq 0$ but considers instead a state
with a fixed total number of particles. A different $U(1)$-symmetry
preserving Bogoliubov approach was developed long ago in
\cite{Girardeau}.

This allows to eliminate
a technical problem of the symmetry breaking approach: if
$\langle\hat{\psi}\rangle(t=0)\neq 0$ the state of the system necessarily
involves a coherent superposition of states with different total number
of particles; such a state {\sl cannot be stationary}
(as states with different
number of particles have also different energies) and it experiences
a phase collapse $\langle\hat{\psi}\rangle(t)\rightarrow 0$ making
the description of the evolution of the system more involved.

\subsection{Small parameter of the theory}
We restrict in this section to a steady state
regime where most of the atoms of the
gas are in the condensate. We split the atomic field operator
as
\be
\hat{\psi}(\vec{r}\,) = {\phi_0}(\vec{r}\,) \hat{a}_{\phi_0}+\delta\hat{\psi}(\vec{r}\,)
\label{eq:expan}
\ee
where ${\phi_0}$ is the condensate wavefunction and $\hat{a}_{\phi_0}$ annihilates
a particle in the mode ${\phi_0}$.
The idea is to treat  $\delta\hat{\psi}$ as a perturbation with
respect to ${\phi_0}\hat{a}_{\phi_0}$: Let us compare indeed the typical matrix
elements of these two operators:
\be
\delta\hat{\psi} \sim \langle \delta\hat{\psi}^\dagger\delta\hat{\psi}
\rangle ^{1/2}
\sim (\rho')^{1/2}
\ee
where $\rho'$ is the density of non-condensed particles
whereas
\be
{\phi_0}\hat{a}_{\phi_0}\sim N_0^{1/2}{\phi_0}  \sim \rho_0^{1/2}
\ee
where $\rho_0$ is the condensate density.
We will therefore assume
\be
\rho' \ll \rho_0
\ee
and even more
\be
N'=\int d^3\vec{r}\, \rho'(\vec{r}\,) \ll N_0 \simeq N
\ee
where $N$ is the total number of particles in the gas. Using these
two assumptions a systematic expansion of the field equations in
powers of the small parameter
\be
\varepsilon=(N'/N_0)^{1/2} \ll 1
\label{eq:varepsilon}
\ee
can
be performed.  We give here a somewhat simplified presentation.

The following identities are important properties of $\delta\hat{\psi}$.
First, $\delta\hat{\psi}$ is the part of the atomic field operator
transverse to the condensate wavefunction:
\be
\int d^3\vec{r}\,{\phi_0}^*(\vec{r}\,)\delta\hat{\psi}(\vec{r}\,)=0.
\label{eq:prop1}
\ee
As a consequence the bosonic commutation relation obeyed by $\delta\hat{\psi}$
involves matrix elements of the projector $Q$ orthogonal to $\phi_0$ rather
than the identity operator:
\begin{equation}
[\delta\hat{\psi}(\vec{r}_1\,),\delta\hat{\psi}^\dagger(\vec{r}_2\,)]
= \langle\vec{r}_1\,|Q|\vec{r}_2\rangle =\delta(\vec{r}_1-\vec{r}_2\,)
-\phi_0(\vec{r}_1\,)\phi_0^*(\vec{r}_2).
\label{eq:commut}
\end{equation}
Second, there should be no coherence in the one-body density matrix
between the condensate and the non-condensed modes, or equivalently
${\phi_0}$ should be an eigenstate of the one-body density matrix
(with eigenvalue $N_0$):
\be
\langle\hat{a}_{\phi_0}^\dagger \delta\hat{\psi}\rangle =0.
\label{eq:prop2}
\ee
This last identity is used to calculate the condensate wavefunction
order by order in the small parameter
\cite{Gardiner,Dum_U1}.
\subsection{Zeroth order in $\varepsilon$: Gross-Pitaevskii equation}
To zeroth order in $\varepsilon$ all the particles are supposed to
be in the condensate. In steady state one then recovers the time
independent Gross-Pitaevskii equation with the further approximation
$N_0\simeq N$:
\be
\mu{\phi_0} = \left[-\frac{\hbar^2}{2m} \Delta + U + gN |{\phi_0}|^2\right]
{\phi_0}.
\label{eq:zero}
\ee

\subsection{Next order in $\varepsilon$: linear dynamics of non-condensed
particles}
Calculation to first order in $\varepsilon$ corresponds to a linearization
of the Heisenberg field equations around ${\phi_0}\hat{a}_{\phi_0}$ keeping
terms up to first order in $\delta\hat{\psi}$. Equivalently it corresponds
to a quadratization of the Hamiltonian around ${\phi_0}\hat{a}_{\phi_0}$
keeping terms up to second order in $\delta\hat{\psi}$.

We use this quadratization approach here. At this order of the calculation
the regularizing operator of the pseudo-potential can be neglected,
divergences due the use of the non-regularized $g\delta$ potential
coming at the next order $\varepsilon^2$. We therefore take the
Hamiltonian
\be
\hat{H} = \int d^3\vec{r}\, \left[\hat{\psi}^\dagger h_1\hat{\psi}
+\frac{g}{2} \hat{\psi}^\dagger\hat{\psi}^\dagger\hat{\psi}\hat{\psi}\right].
\label{eq:ham}
\ee
The one-body Hamiltonian $h_1$ contains the kinetic energy and
the trapping potential energy:
\be
h_1 = -\frac{\hbar^2}{2m} \Delta + U.
\ee
It does not contain any $-\mu$ term as we use here in the canonical
rather than grand canonical point of view, the total number
of particles being fixed to $N$.

We substitute expansion (\ref{eq:expan}) for Eq.(\ref{eq:ham}) 
and we keep terms up to quadratic in $\delta\hat{\psi}$.

\noindent $\bullet$ The contribution of $h_1$ is quadratic in $\hat{\psi}$ so that
all the terms should be kept. One of the contributions is
\be
(\int d^3\vec{r}\,{\phi_0}^* h_1{\phi_0})\hat{a}_{\phi_0}^\dagger\hat{a}_{\phi_0}.
\ee
One can use the following trick to express this quantity in terms
of $\delta\hat{{\psi}}$: from Eq.(\ref{eq:prop1}) one sees that
the total number of particles operator can be written as
\be
\hat{N}\equiv \int d^3\vec{r}\,\hat{\psi}^\dagger\hat{\psi}
= \hat{n}_{\phi_0}+\delta\hat{N}
\ee
that is the sum of the number operator of condensate particles:
\be
\hat{n}_{\phi_0} = \hat{a}_{\phi_0}^\dagger\hat{a}_{\phi_0}
\ee
and the number operator of non-condensed particles:
\be
\delta\hat{N} = \int d^3\vec{r}\,\delta\hat{\psi}^\dagger\delta\hat{\psi}.
\ee
We therefore obtain
\be
\hat{n}_{\phi_0}=\hat{N}-\delta\hat{N}.
\label{eq:elim}
\ee

\noindent $\bullet$ The expansion of the interaction term gives
the following up to quadratic contributions:
\bea
\frac{g}{2} \hat{\psi}^\dagger\hat{\psi}^\dagger\hat{\psi}\hat{\psi}
\rightarrow &&
\frac{g}{2}|{\phi_0}|^4 \hat{a}_{\phi_0}^\dagger\hat{a}_{\phi_0}^\dagger
\hat{a}_{\phi_0}\hat{a}_{\phi_0}         \nonumber \\
&+&g[{\phi_0}^*{\phi_0}^*{\phi_0}\hat{a}_{\phi_0}^\dagger\hat{a}_{\phi_0}^\dagger\hat{a}_{\phi_0}
\delta\hat{\psi} +\mbox{h.c.}] \nonumber \\
&+&\frac{g}{2}[{\phi_0}^*{\phi_0}^*\hat{a}_{\phi_0}^\dagger\hat{a}_{\phi_0}^\dagger
\delta\hat{\psi}\delta\hat{\psi}+\mbox{h.c.}]\nonumber \\
&+&2g {\phi_0}^*{\phi_0}\hat{a}_{\phi_0}^\dagger\delta\hat{\psi}^\dagger
\delta\hat{\psi}\hat{a}_{\phi_0}.
\label{eq:uni}
\eea

The first line of this expression is transformed using Eq.(\ref{eq:elim}):
\be
\hat{a}_{\phi_0}^\dagger\hat{a}_{\phi_0}^\dagger
\hat{a}_{\phi_0}\hat{a}_{\phi_0} =  \hat{n}_{\phi_0} (\hat{n}_{\phi_0}-1)=
\hat{N}(\hat{N}-1) -2\hat{N}\delta\hat{N}+\ldots
\ee
with $N\simeq N-1$. When we group the above term in $\delta\hat{N}$
with the one coming from $h_1$ and we replace 
$\hat{N}$ by $N$ we obtain
\be
-\delta\hat{N}\left[\int d^3\vec{r}\, {\phi_0}^* h_1 {\phi_0} + 
{N} g|{\phi_0}|^4
\right]
=-\mu\delta\hat{N}
\ee
as ${\phi_0}$ solves the Gross-Pitaevskii equation (\ref{eq:zero}).
This is amusing: we obtain formally a grand canonical Hamiltonian
for the non-condensed particles, the reservoir being formed
by the condensate particles! In the second line of Eq.(\ref{eq:uni})
we replace $\hat{a}_{\phi_0}^\dagger\hat{a}_{\phi_0}$ by ${N}$ as
the corrective term $\delta\hat{N}$ would lead to a cubic contribution
in $\delta\hat{\psi}$.

Another important transformation is performed by
collecting the terms linear in $\delta\hat{\psi}$ from Eq.(\ref{eq:uni})
and from the contribution of $h_1$, leading to
\be
\int d^3\vec{r}{\phi_0}^*\hat{a}_{\phi_0}^\dagger [h_1 + g |{\phi_0}|^2{N}]
\delta\hat{\psi} + \mbox{h.c.}
\ee
As ${\phi_0}$ solves the Gross-Pitaevskii equation,
the operator between brackets, when acting on the left on ${\phi_0}^*$,
gives a contribution $\mu {\phi_0}^*$ orthogonal to $\delta\hat{\psi}$
(see Eq.(\ref{eq:prop1})). The sum of all the terms linear in $\hat{
\delta\psi}$ therefore vanishes! We shall see later that this could be
expected from Eq.(\ref{eq:prop2}).

\subsection{Bogoliubov Hamiltonian}

We now collect all the terms of $\hat{H}$ up to quadratic in $\delta\hat{\psi}$
and express them in terms of the field operator
\be
\hat{\Lambda}(\vec{r}\,) = \frac{1}{\hat{N}^{1/2}}
\hat{a}_{\phi_0}^\dagger\delta\hat{\psi}(\vec{r}\,).
\ee
This operator commutes with the operator $\hat{N}$ giving the total number of
particles: it transfers one non-condensed particle into
the condensate. As the operator $\delta\hat{\psi}$,
it is transverse to $\phi_0$ (see Eq.(\ref{eq:prop1})).
By definition of the condensate wavefunction $\hat{\Lambda}$ has
a zero mean (see Eq.(\ref{eq:prop2})).

In general it is difficult to exactly eliminate $\delta\hat{\psi}$
in terms of $\hat{\Lambda}$. Fortunately
at the present order of the calculation 
we can use the assumption of a very small non-condensed fraction
so that one has for example
\be
\delta\hat{\psi}^\dagger\delta\hat{\psi} \simeq
\delta\hat{\psi}^\dagger\hat{a}_{\phi_0}\hat{N}^{-1}\hat{a}_{\phi_0}^\dagger
\delta\hat{\psi}=
\hat{\Lambda}^\dagger\hat{\Lambda}\, .
\ee
To the same order of approximation
$\hat{\Lambda}$ obeys the same commutation relation Eq.(\ref{eq:commut})
as $\delta\hat{\psi}$.

The final result can be written in term of the operator $\cal L$
introduced in \S\ref{Cap:linear}, with the approximation $N_0\simeq N$:
\be
\hat{H}_{\mbox{\scriptsize quad}} = f(\hat{N}) +
\frac{1}{2} \int d^3\vec{r}\, (\hat{\Lambda}^\dagger,-\hat{\Lambda})
{\cal L}
\left( \begin{tabular}{c}
$\hat{\Lambda}$ \\ $\hat{\Lambda}^\dagger$
\end{tabular}
\right)
\label{eq:hquad}
\ee
where the function $f$ is specified in \cite{Dum_U1}.

From this quadratic Hamiltonian the Heisenberg equations of motion
for the field $\hat{\Lambda}$ have the suggestive form
\be
i\hbar \frac{d}{dt}
\left( \begin{tabular}{c}
$\hat{\Lambda}$ \\ $\hat{\Lambda}^\dagger$
\end{tabular}
\right)=
{\cal L}
\left( \begin{tabular}{c}
$\hat{\Lambda}$ \\ $\hat{\Lambda}^\dagger$
\end{tabular}
\right).
\label{eq:jolie}
\ee
We note  that an hypothetic term of $\hat{H}_{\mbox{\scriptsize quad}}$
linear in $\hat{\Lambda}$ would give rise to a source term in Eq.(\ref{eq:jolie})
preventing one from satisfying Eq.(\ref{eq:prop2}) at all times!

The result Eq.(\ref{eq:jolie}) is really a crucial one.
It shows that the linearized evolution of the non-condensed part
$\delta\hat{\psi}$ of the atomic field is formally equivalent to
the linearized response of the condensate to a classical perturbation
(e.g.\ of the trapping potential) derived from the Gross-Pitaevskii equation;
both treatments indeed exhibit the same operator $\cal L$
(see Eq.(\ref{eq:gpelinperp})).

All the machinery of \S\ref{Cap:linear} can therefore be used. We expand
the field operator $\hat{\Lambda}$ on the eigenmodes of
$\cal L$. We assume here dynamical stability 
and that the only eigenmodes in the $0$ family
are the zero-energy modes $({\phi_0},0)$ and $(0,{\phi_0}^*)$ to which
the field $\hat{\Lambda}$ is orthogonal according to Eq.(\ref{eq:prop1}).
We therefore get an expansion similar to Eq.(\ref{eq:expang}):
\be
\left(\begin{tabular}{c}
$\hat{\Lambda}(\vec{r}\,)$ \\ $\hat{\Lambda}^\dagger(\vec{r}\,)$ 
\end{tabular}\right)
= \sum_{k \in + \mbox{\scriptsize family}}
\hat{b}_k\left(\begin{tabular}{c} $u_k(\vec{r}\,)$ \\ $v_k(\vec{r}\,)$
\end{tabular}\right)
+\hat{b}_k^\dagger \left(\begin{tabular}{c} $v_k^*(\vec{r}\,)$ \\ $u_k^*(\vec{r}\,)$
\end{tabular}\right)
\ee
with the important difference that the coefficients in the expansion
are now operators:
\be
\hat{b}_k=
\int d^3\vec{r}\, \left[u_k^*(\vec{r}\,)\hat{\Lambda}(\vec{r}\,)
-v_k^*(\vec{r}\,)\hat{\Lambda}^\dagger(\vec{r}\,)\right].
\ee
From the normalization condition between the eigenvectors of
$\cal L$ and their adjoint vectors (see \S\ref{Cap:linear})
one shows that the $\hat{b}_k$ obey bosonic commutation relations:
\be
[\hat{b}_k,\hat{b}_{k'}^\dagger]=\delta_{k,k'} \ \ \mbox{and}\ \  [\hat{b}_k,\hat{b}_{k'}] = 0.
\ee
$\hat{b}_k$ corresponds formally to an annihilation operator;
as $|v_k\rangle  \neq 0$ in general
$\hat{b}_k$ does not simply annihilate a particle as it is a coherent
superposition of $\hat{\Lambda}$ (which transfers one non-condensed
particle to the condensate) and of $\hat{\Lambda}^\dagger$ (which
transfers one condensate particle to the non-condensed
fraction). One then says that $\hat{b}_k$ annihilates a
{\sl quasi-particle} in mode $k$.

Finally we rewrite the Hamiltonian Eq.(\ref{eq:hquad}) in terms of
the $\hat{b}_k$'s:
\be
\hat{H}_{\mbox{\scriptsize quad}} = E_0(\hat{N}) +
\sum_{k \in + \mbox{\scriptsize family}} \epsilon_k
\hat{b}_k^\dagger \hat{b}_k.
\label{eq:hbd}
\ee
We recall that $\epsilon_k$ is the eigenenergy of the mode
$(u_k,v_k)$ of the $+$ family. Our quadratic Hamiltonian describes
a gas of non-interacting quasi-particles: this is the so-called
Bogoliubov Hamiltonian.

The ground state of
$\hat{H}_{\mbox{\scriptsize quad}}$ is obtained when no quasi-particle
is present, it corresponds to all the modes $k$ being in vacuum state:
\be
\hat{H}_{\mbox{\scriptsize quad}} |0\rangle = E_0(N) |0\rangle
\ee
where
\be
\hat{b}_k|0\rangle = 0 \ \ \forall k\, .
\ee
$E_0$ is therefore the Bogoliubov approximation for the
ground state energy of the gas. To get a finite expression for
$E_0$ one has to include the regularizing operator in the
pseudo-potential.

The excited states of the system are obtained in the Bogoliubov approximation
by successive actions of the $\hat{b}_k^\dagger$'s. For this reason
$\hat{b}_k^\dagger$ is said to create an {\sl elementary} excitation
$k$ in the system, to distinguish with {\sl collective} excitations
involving all the particles of the condensate (induced e.g.\ by
a perturbation $\delta U$ of the trapping potential).
We emphasize again that the elementary excitations of the gas have the same
frequency $\epsilon_k/\hbar$ as the collective excitations in the
linear response domain ($\delta U$ small enough). This intriguing property
is valid only at the presently considered regime of an almost pure
condensate ($T\ll T_c$).

\subsection{Order $\varepsilon^2$: corrections to the Gross-Pitaevskii
equation}
Expanding the Heisenberg field equations for
$\hat{\psi}$ keeping terms up to $N^{1/2}\varepsilon^2$
one can calculate the first correction to the prediction
Eq.(\ref{eq:zero}) for the condensate wavefunction. This correction
includes (i) the fact that the number of condensate particles
$N_0$ rather than the total number of particles $N$ should appear
in the Gross-Pitaevskii equation, and (ii) the mechanical back-action
of the non-condensed particles
on the condensate in the form of mean field potentials.

The calculations
are a bit involved \cite{Dum_U1} and require the use
of the regularizing operator in the pseudo-potential (a fact realized
in \cite{BCS} but not yet in \cite{Dum_U1}).
We give here only the result. The condensate wavefunction is
given by an expansion
\be
{\phi_0}={\phi_0}^{(0)} + {\phi_0}^{(2)} + o(\varepsilon^2)
\ee
where ${\phi_0}^{(0)}$, zeroth order approximation in $\varepsilon$,
is the solution of Eq.(\ref{eq:zero}). The correction ${\phi_0}^{(2)}$ is
of order $\varepsilon^2$; its component on ${\phi_0}^{(0)}$ is purely imaginary
(as both ${\phi_0}$ and ${\phi_0}^{(0)}$ are normalized to unity) and can be
considered as a (not physically relevant)
change of global phase of ${\phi_0}^{(0)}$; the part of ${\phi_0}^{(2)}$ orthogonal to
${\phi_0}^{(0)}$ is given by:
\be
-{\cal L}
\left(\begin{tabular}{c}
$Q{\phi_0}^{(2)}$ \\ $Q^*{\phi_0}^{(2)*}$
\end{tabular}\right)=
\left(\begin{tabular}{c}
$Q S$ \\ $-Q^*S^*$
\end{tabular}\right)
\ee
where $Q$ projects orthogonally to ${\phi_0}$ and where the source
term $S$ is equal to
\bea
S(\vec{r}\,)=&-&(1+\langle\delta\hat{N}\rangle)g|{\phi_0}^{(0)}(\vec{r}\,)|^2
{\phi_0}^{(0)}(\vec{r}\,) \nonumber \\
&+& 2g\langle\hat{\Lambda}^\dagger(\vec{r}\,)\hat{\Lambda}(\vec{r}\,)
\rangle{\phi_0}^{(0)}(\vec{r}\,)+
g\left[\partial_s\left(s
\langle\hat{\Lambda}(\vec{r}-\vec{s}/2)\hat{\Lambda}(\vec{r}+\vec{s}/2)
\rangle\right)\right]_{s\rightarrow 0}{\phi_0}^{(0)*}(\vec{r}\,)\nonumber \\
&-&g \int d^3\vec{s}\,|{\phi_0}^{(0)}(\vec{s}\,)|^2\langle\left[
\hat{\Lambda}^\dagger(\vec{s}\,){\phi_0}^{(0)}(\vec{s}\,)+
\hat{\Lambda}(\vec{s}\,){\phi_0}^{(0)*}(\vec{s}\,)\right]
\hat{\Lambda}(\vec{r}\,)\rangle.
\eea
The first line of this expression contains the effect of the depletion 
of the condensate, the number of non-condensed particles
$\langle\delta\hat{N}\rangle$ being
calculated in the Bogoliubov approximation, see discussion
in \S\ref{subsec:deplet}. The other terms are mean field terms, among which
one recognizes the Hartree-Fock contribution
$2g\langle\hat{\Lambda}^\dagger(\vec{r}\,)\hat{\Lambda}(\vec{r}\,)
\rangle$ already obtained in \S\ref{Cap:HF}.

\subsection{Thermal equilibrium of the gas of quasi-particles}
In the Bogoliubov approximation the quasi-particles behave as an
ideal Bose gas; such a gas can reach thermal equilibrium
only by contact with a thermostat. There is no such thermostat
in the experiments on trapped Bose gas, relaxation to thermal
equilibrium has to be provided instead by interactions
between the atoms.

Fortunately the full Hamiltonian Eq.(\ref{eq:ham}) contains terms
cubic and quartic in $\delta\hat{\psi}$: when expressed in terms
of the $\hat{b}$'s and $\hat{b}^\dagger$'s they correspond to interactions
between the quasi-particles which will provide thermalization.
Two situations can then be considered, depending on the sign
of $\epsilon_k$.

$\bullet$ \lq\lq Good" case: $\epsilon_k>0$ for all $k$ in $+$
family. We assume for simplicity thermal equilibrium in the canonical
point of view (which should be equivalent to the micro-canonical
point of view in the limit of large number of particles) with a $N$-body
density matrix
\be
\hat{\rho}_{1,\ldots,N} = \frac{1}{Z}e^{-\beta \hat{H}} \simeq
\frac{1}{Z_{\mbox{\scriptsize quad}}}
e^{-\beta \hat{H}_{\mbox{\scriptsize quad}}}.
\label{eq:te}
\ee
We suppose therefore that the interactions between quasi-particles,
essential to ensure thermalization, have a weak effect on the thermal
equilibrium state.
From Eq.(\ref{eq:te}) we finally obtain the mean number of
quasi-particles in mode $k$:
\be
\langle\hat{b}_k^\dagger\hat{b}_k\rangle =
\frac{1}{e^{\beta\epsilon_k}-1}.
\label{eq:occup}
\ee
This good case corresponds to a thermodynamically stable condensate
in state ${\phi_0}$.

$\bullet$ \lq\lq Bad" case: there is a mode $k_0$ in the $+$ family
such that $\epsilon_{k_0}<0$. In this case the quadratic Hamiltonian
$\hat{H}_{\mbox{\scriptsize quad}}$ contains a harmonic oscillator
of frequency $\omega_{k_0}=|\epsilon_{k_0}|/\hbar$
having formally a negative mass $M$:
\be
-|\epsilon_{k_0}|\hat{b}_{k_0}^\dagger\hat{b}_{k_0} 
= \frac{1}{2M}[\hat{P}_{k_0}^2+\omega_{k_0}^2\hat{Q}_{k_0}^2]
\ee
where $\hat{P}_{k_0}$ and $\hat{Q}_{k_0}$ correspond formally to a momentum
and position operator.
By collisions with the quasi-particles of positive energy 
the mode $k_0$ can loose energy which increases its own excitation;
if the number of quanta in the mode can become comparable to $N$
the process of thermalization of the gas may lead the condensate
to a state different from the predicted ${\phi_0}$.

This phenomenon of thermodynamical instability should not be confused
with dynamical instability of the condensate, where $\epsilon_{k_0}$
is complex; e.g.\ the case of a purely imaginary $\epsilon_{k_0}$
corresponds formally to an oscillator in an expelling potential,
$[\hat{P}_{k_0}^2-|\omega_{k_0}|^2\hat{Q}_{k_0}^2]/(2M)$.

\subsection{Condensate depletion and the small parameter
$(\rho a^3)^{1/2}$} \label{subsec:deplet}
We assume the situation of thermodynamical stability. We calculate
the mean number $\langle\delta \hat{N}\rangle$ of particles
out of the condensate, that is in all the modes orthogonal
to the condensate wavefunction ${\phi_0}$:
\be
\langle\delta \hat{N}\rangle \equiv  \int d^3\vec{r}\, \langle
\delta\hat{\psi}^\dagger(\vec{r}\,)\delta\hat{\psi}(\vec{r}\,)\rangle.
\ee
In this way we can calculate explicitly the small parameter
of the present theory given in Eq.(\ref{eq:varepsilon}).

To lowest order in the Bogoliubov approximation we can replace
$\delta\hat{\psi}$ by $\hat{\Lambda}$ in the above expression:
\be
\langle\delta \hat{N}\rangle \simeq \int d^3\vec{r}\, \langle 
\hat{\Lambda}^\dagger(\vec{r}\,) \hat{\Lambda}(\vec{r}\,)\rangle.
\ee
We replace $\hat{\Lambda}$ by its modal expansion; using the approximation
in Eq.(\ref{eq:te}) the only terms with non-zero mean are
$\langle \hat{b}_k^\dagger\hat{b}_k\rangle$ given by Eq.(\ref{eq:occup})
and 
\be
\langle \hat{b}_k\hat{b}_k^\dagger\rangle=
\langle \hat{b}_k^\dagger\hat{b}_k\rangle+1. 
\label{eq:plus_un}
\ee
We therefore obtain:
\bea
\langle\delta \hat{N}\rangle &\simeq& \sum_{k\in + \mbox{\scriptsize family}}
\langle \hat{b}_k^\dagger\hat{b}_k\rangle [\langle u_k|u_k\rangle +
\langle v_k|v_k\rangle] \nonumber \\
&&+\sum_{k\in + \mbox{\scriptsize family}}
\langle v_k|v_k\rangle.
\label{eq:dnex}
\eea
The contribution of the occupation numbers
$\langle \hat{b}_k^\dagger\hat{b}_k\rangle$ corresponds to the so-called
thermal depletion of the condensate, as it is non-zero only at finite temperature.
The contribution of the $+1$ coming from the identity (\ref{eq:plus_un})
is the so-called quantum depletion of the condensate: it expresses the fact
that, even at zero temperature, there is a finite number of particles
out of the condensate due to atomic interactions, contrarily to the ideal
Bose gas case where all the $v_k$'s vanish.

It is instructive to calculate the  depletion of the condensate
in the homogeneous case in the thermodynamical limit, replacing the
sum over the wavevector $\vec{k}$ characterizing each mode of the $+$ family
by an integral.
From Eq.(\ref{eq:u_et_v}) we calculate
\be
\langle u_k|u_k\rangle+\langle v_k|v_k\rangle = 1 + 2 \langle v_k|v_k\rangle=
\frac{\frac{\hbar^2 k^2}{2m}+\rho g}{\left[ 
\frac{\hbar^2 k^2}{2m}\left(\frac{\hbar^2 k^2}{2m}+2\rho g\right)
\right]^{1/2}}.
\ee
At zero temperature we obtained for the non-condensed fraction of particles
the following integral:
\begin{equation}
\frac{\langle\delta \hat{N}\rangle}{N}(T=0)
= \frac{16}{\pi^{1/2}}\left(\rho a^3\right)^{1/2}
\int_0^{+\infty} dq\; q^2\left[\frac{q^2+1/2}{q\left(1+q^2\right)^{1/2}}-1
\right]
\end{equation}
where we have integrated over the solid angle in spherical coordinates
and we have made the change of variable 
\begin{equation}
\frac{\hbar^2 k^2}{2m} = 2\rho g q^2.
\label{eq:cov}
\end{equation}
This integral can be calculated exactly:
\be
\frac{\langle\delta \hat{N}\rangle}{N}(T=0) = \frac{8}{3\pi^{1/2}}
\left(\rho a^3\right)^{1/2}.
\label{eq:deplq}
\ee
In this way we obtain a very important small parameter of the
theory, $\left(\rho a^3\right)^{1/2}$, which characterizes the
domain of a weakly interacting Bose gas. This small parameter
is similar to the condition
obtained in Eq.(\ref{eq:sgp}) of \S\ref{Cap:model} already
with a totally different point of view, a condition of Born approximation
on the pseudo-potential!
In the typical experimental conditions of MIT for sodium atoms
we have $\rho \sim 10^{14}$ cm$^{-3}$ and $a \sim 25$ \AA, which
leads to a small parameter $\left(\rho a^3\right)^{1/2}\sim 10^{-3}$.
The result (\ref{eq:deplq})
also gives us the opportunity to recall that the number of non-condensed particles
should not be confused with the number of quasi-particles for
an interacting Bose gas:
one notes here that the second quantity vanishes at $T=0$ whereas the first
one does not.

At finite temperature there is, in addition to the quantum depletion,
a thermal depletion of the condensate:
\begin{equation}
\frac{\langle\delta \hat{N}\rangle}{N}(T)-
\frac{\langle\delta \hat{N}\rangle}{N}(0) = \frac{32}{\pi^{1/2}}
\int_0^{+\infty} dq\; \frac{q(q^2+1/2)}{\left(q^2+1\right)^{1/2}}
\left[\exp\left(2\beta\rho g q (q^2+1)^{1/2}\right)-1\right]^{-1}.
\end{equation}
The integral over $q$ depends only on the parameter $\rho g/(k_B T)$.
At low temperature $k_B T \ll \rho g$ this parameter is large 
so that the modes with a $q\sim 1$ have a very low occupation number:
one can neglect $q$ as compared to one and one-half
(which amounts to restricting to
the linear part of the Bogoliubov spectrum) and one obtains a 
small correction to the zero temperature case:
\begin{equation}
\frac{\langle\delta \hat{N}\rangle}{N}(T)=
\frac{\langle\delta \hat{N}\rangle}{N}(0)
\left[1+ \left(\frac{\pi k_B T}{2\rho g}\right)^2+\ldots\right].
\end{equation}
At high temperature $k_B T\gg \rho g$ the major fraction of the populated
Bogoliubov modes have a $q$ much larger than unity, so that we can now neglect
one and one-half as compared to $q^2$ (which amounts to restricting
to the quadratic part of the Bogoliubov spectrum). To this approximation
we recover the ideal Bose gas result
\begin{equation}
\frac{\langle\delta \hat{N}\rangle}{N}(T)\simeq
\frac{\zeta(3/2)}{\rho\lambda_{\mbox{\scriptsize dB}}^3}
\end{equation}
where $\lambda_{\mbox{\scriptsize dB}}$ is the thermal de Broglie wavelength
and $\zeta$ stands for the Riemann Zeta function (see \S\ref{Cap:ideal}).
For a given temperature indeed the density of non-condensed particles
in presence of a condensate
saturates to its maximal value
$\zeta(3/2)\lambda_{\mbox{\scriptsize dB}}^{-3}$ for the ideal Bose gas
in a box (see Eq.(\ref{eq:condfracidealbox}) and Eq.(\ref{eq:idealbox})).

In the high temperature regime $k_B T\gg \rho g$ our Bogoliubov approach
can therefore be valid only if $\rho\lambda_{\mbox{\scriptsize dB}}^3\gg \zeta(3/2)$.
Both inequalities on $\rho$ can be satisfied simultaneously only if 
$2\zeta(3/2) a/\lambda_{\mbox{\scriptsize dB}}\ll 1$.
Amusingly this condition is similar to the condition $a\Delta k\ll 1$ 
obtained in \S\ref{Cap:model} (see Eq.(\ref{eq:dk})) for the validity
condition of the Born approximation for the pseudo-potential.

\subsection{Fluctuations in the number of condensate particles}
The Bogoliubov theory that we have presented also allows a calculation of the fluctuations
of $N_0$ in the canonical ensemble. This has the advantage of removing the effect 
of fluctuations in the total number of particles present in the grand-canonical
ensemble, a \lq\lq trivial" contribution to the fluctuations of $N_0$.

As the total number of particles is fixed to $N$ the variance
of $N_0$ is exactly equal to the variance of $\delta \hat{N}$, number
of non-condensed particles. The mean value of $\delta \hat{N}$ has
been given in the previous section, we now have to calculate the mean
value of its square:
\begin{eqnarray}
\langle(\delta\hat{N})^2\rangle &=& 
\int d\vec{r}_1\int d\vec{r}_2\, \langle
\delta\hat{\psi}^\dagger(\vec{r}_1\,)\delta\hat{\psi}(\vec{r}_1\,)
\delta\hat{\psi}^\dagger(\vec{r}_2\,)\delta\hat{\psi}(\vec{r}_2\,)\rangle
\\
&=& 
\langle\delta\hat{N}\rangle+
\int d\vec{r}_1\int d\vec{r}_2\, \langle
\delta\hat{\psi}^\dagger(\vec{r}_1\,)\delta\hat{\psi}^\dagger(\vec{r}_2\,)
\delta\hat{\psi}(\vec{r}_1\,)\delta\hat{\psi}(\vec{r}_2\,)\rangle
\end{eqnarray}
where we have used the commutation relation Eq.(\ref{eq:commut})  and
the fact Eq.(\ref{eq:prop1}) that $\phi$ is orthogonal to $\delta\hat{\psi}$.
To lowest order in the Bogoliubov approximation we can replace
$\delta\hat{\psi}$ by $\hat{\Lambda}$ in the above expression and take the
approximation Eq.(\ref{eq:te}) for the equilibrium density operator. 
As ${\hat H}_{\mbox{\scriptsize quad}}$ is quadratic in the field $\hat{\Lambda}$,
Wick's theorem can be used. We finally find for
the variance of the number of non-condensed particles:
\begin{equation}
\mbox{Var}(\delta\hat{N}) \simeq
\langle\delta\hat{N}\rangle+
\int d\vec{r}_1\int d\vec{r}_2\, \left[
\left|\langle \hat{\Lambda}^\dagger(\vec{r}_1\,)\hat{\Lambda}(\vec{r}_2\,)\rangle\right|^2+
\left|\langle\hat{\Lambda}(\vec{r}_1\,)\hat{\Lambda}(\vec{r}_2\,)\rangle\right|^2
\right].
\label{eq:var}
\end{equation}

The variance in Eq.(\ref{eq:var}) is simple to calculate in the homogeneous case
of a gas in a cubic box of size $L$ with periodic boundary conditions.
The eigenmodes $(u,v)$ are simply plane waves Eq.(\ref{eq:planew})
with real coefficients $U_k,V_k$ given in Eq.(\ref{eq:u_et_v}) and depending only
on the modulus $k$ of the wavevector $\vec{k}$. We find for the two correlation
functions appearing in Eq.(\ref{eq:var}) the simple expression:
\begin{eqnarray}
\langle \hat{\Lambda}^\dagger(\vec{r}_1\,)\hat{\Lambda}(\vec{r}_2\,)\rangle
 &=& \frac{1}{L^3} \sum_{\vec{k}\neq\vec{0}}
\left[(U_k^2 + V_k^2) n_k + V_k^2\right]e^{i\vec{k}\cdot(\vec{r}_1-\vec{r}_2)} \\
\langle \hat{\Lambda}(\vec{r}_1\,)\hat{\Lambda}(\vec{r}_2\,)\rangle
 &=& \frac{1}{L^3} \sum_{\vec{k}\neq\vec{0}}
U_k V_k (1+2n_k)e^{i\vec{k}\cdot(\vec{r}_1-\vec{r}_2)} 
\end{eqnarray}
where we have introduced the mean occupation numbers $n_k = \langle \hat{b}_k^\dagger
\hat{b}_k\rangle$. After spatial integration of the square modulus of these
quantities we obtain
\be
\mbox{Var}(\delta\hat{N}) = \langle \delta\hat{N}\rangle +
\sum_{\vec{k}\neq\vec{0}} \left[\left(U_k^2+V_k^2\right)n_k+V_k^2\right]^2
+U_k^2 V_k^2 (1+2n_k)^2 
\label{eq:varex}
\ee
where the mean number of non-condensed particles $\langle \delta\hat{N}\rangle$
is already given in Eq.(\ref{eq:dnex}):
\be
\langle \delta\hat{N}\rangle = \sum_{\vec{k}\ne\vec{0}} \left( U_k^2 + V_k^2\right)
n_k + V_k^2.
\ee

We first apply formula (\ref{eq:varex}) to the limiting case of zero
temperature. All the occupation numbers $n_k$ vanish. For the ideal Bose gas 
($g=0$) all the $V_k$'s are zero and the variance of $\delta\hat{N}$ vanishes
as expected, since all the particles are in the ground state of the box.
For the interacting Bose gas we find
\be
\mbox{Var}(\delta\hat{N})(T=0)=
\frac{1}{8} \sum_{\vec{k}\neq\vec{0}}
\frac{1}{q^2\left(1+q^2\right)}
\ee
where $q$ is given as function of $k$ by Eq.(\ref{eq:cov}). In the thermodynamical
limit we replace the discrete sum by an integral and this leads to a
variance scaling as the number of particles for a fixed density:
\be
\frac{\mbox{Var}(\delta\hat{N})(T=0)}{N}= 2\pi^{1/2}\left(\rho a^3\right)^{1/2}.
\ee
In the regime of validity of the Bogoliubov approach one has $\rho a^3\ll 1$
so that the fluctuations of $N_0$ are sub-poissonian.

The situation can be totally different at finite temperature. Consider first the case
of the ideal Bose gas \cite{Var_parfait}:
\bea
\mbox{Var}(\delta\hat{N}) &=& \sum_{\vec{k}\neq\vec{0}} n_k (1+n_k) \nonumber\\
 &=& \sum_{\vec{k}\neq\vec{0}} \frac{1}{4\sinh^2(\beta\epsilon_k/2)}
\eea
with $\epsilon_k = \hbar^2 k^2/(2m)$.  In the thermodynamical limit one may be
tempted to replace in the usual manner the sum over $\vec{k}$ by an integral.
This leads however to an integral divergent in $k=0$: the integrand scales
as $1/k^4$, which is not compensated by
the Jacobian $k^2$ of three-dimensional integration
in spherical coordinates. In this case the contribution of the sum in the thermodynamical
limit is dominated by the terms close to $k=0$ where $\beta \epsilon_k\ll 1$
and the function $\sinh$ can be linearized. We then obtain
\be
\mbox{Var}(\delta\hat{N})_{g=0} \simeq \left(\frac{k_B T}{\Delta}\right)^2
\sum_{\vec{n}\neq\vec{0}} \frac{1}{n^4}.
\label{eq:var_ideal}
\ee
In this expression we have introduced the kinetic energy difference between the ground state
and the first excited state for a single particle in the box:
\be
\Delta = \frac{h^2}{2mL^2}
\ee
and the sum ranges over all the vectors $\vec{n}$ with integer components and a
non-vanishing norm $n$. By a numerical calculation we obtain
\be
\sum_{\vec{n}\neq\vec{0}} \frac{1}{n^4} = 16.53\ldots
\ee
A remarkable feature is that the variance of $N_0$ scales as $L^4$ that is as the volume
of the box to the power $4/3$, or equivalently 
as the number of particles to the power $4/3$
in the thermodynamical limit. This is much larger than $N$ in the thermodynamical
limit.

Do the fluctuations of $N_0$ remain large in presence of interactions~?
As the spectrum $\epsilon_k$ is linear at small $k$ the divergence of $n_k^2$ is only
as $1/k^2$, which has a finite integral in three dimensions. However the mode
functions $U_k,V_k$ are also diverging in $k=0$, each as $1/k^{1/2}$, so that one
recovers the $1/k^4$ dependence of the summand close to $k=0$. As in the ideal Bose
gas case we replace in the thermodynamical limit the summand by its
low $k$ approximation:
\bea
n_k &\sim& \frac{k_B T}{\hbar k c} \\
U_k &\sim& \frac{1}{2 q^{1/2}} \\
V_k &\sim& -\frac{1}{2 q^{1/2}}
\eea
and we keep in the summation the most diverging terms. We finally obtain 
\cite{Var_inter,Var_inter2}
\be
\mbox{Var}(\delta\hat{N})_{g>0} \simeq \frac{1}{2}\left(\frac{k_B T}{\Delta}\right)^2
\sum_{\vec{n}\neq\vec{0}} \frac{1}{n^4}.
\label{eq:var_inter}
\ee
Remarkably this result differs from the ideal Bose gas case Eq.(\ref{eq:var_ideal})
by a factor $1/2$ only: fluctuations of $N_0$ remain large.
The fact that Eq.(\ref{eq:var_inter}) does not depend on the strength $g$ of the
interaction is valid only at the thermodynamical limit and indicates
that the limit $g\rightarrow 0$ and the thermodynamical limit do not commute.

\subsection{A simple reformulation of thermodynamical stability condition}
The thermodynamical stability condition simply requires that the
Bogoliubov Hamiltonian $\hat{H}_{\mbox{\scriptsize quad}}$ is
the sum of a constant (function of $\hat{N}$) and of a {\sl positive} quadratic
operator in the field variables. In the diagonal form (\ref{eq:hbd})
positivity is clearly equivalent to the requirement $\epsilon_k$ positive for
all $k$ in the $+$ family. How to express this condition from
the non-diagonal form (\ref{eq:hquad})?
We simply have to rewrite Eq.(\ref{eq:hquad}) as
\be
\hat{H}_{\mbox{\scriptsize quad}} = f(\hat{N}) +
\frac{1}{2} \int d^3\vec{r}\, (\hat{\Lambda}^\dagger,\hat{\Lambda})
\eta{\cal L} 
\left( \begin{tabular}{c}
$\hat{\Lambda}$ \\ $\hat{\Lambda}^\dagger$
\end{tabular}
\right)
\ee
where $\eta$ is the operator
\be
\eta = \left(\begin{array}{rr}1 & 0 \\ 0 & -1\end{array}\right)
\ee
so that $\eta{\cal L}$ is an Hermitian operator.

The thermodynamical stability condition is therefore equivalent to
the requirement that $\eta {\cal L}$ is positive:
\be
\eta{\cal L} \ge 0.
\ee
More precisely, as $\hat{\Lambda}$ is orthogonal to ${\phi_0}$,
$\eta{\cal L}$ has to be strictly positive in the subspace orthogonal
to $(|{\phi_0}\rangle,0)$ and $(0,|{\phi_0}^*\rangle)$.

We can give a simple physical interpretation of this condition: 
${\phi_0}$ has to be a local minimum of the Gross-Pitaevskii energy
functional 
\be
\label{eq:gpef_encore}
E[\phi,\phi^*] = N \int d^3\vec{r} \left[ {\hbar^2\over 2m}|\vec{\mbox{grad}}\,\phi|
^2 +
U(\vec{r}\,)|\phi(\vec{r}\,)|^2 +{1\over 2}Ng|\phi(\vec{r}\,)|^4\right],
\ee
which is the expression of \S \ref{Cap:GPE} with the approximation
$N_0 \simeq N$.
Let us consider indeed the variation $\delta E$
of $E$ from $E_0\equiv E[{\phi_0},{\phi_0}^*]$
up to second order in a small deviation $\delta\phi$ of $\phi$ from
${\phi_0}$. 

The terms linear in $\delta\phi$ are given by:
\be
\delta E^{(1)}  = N\int d^3\vec{r}\, \left[\frac{\hbar^2}{2m}\vec{\mbox{grad}}\,
{\phi_0}^*\cdot\vec{\mbox{grad}}\,\delta\phi + U(\vec{r}\,){\phi_0}^*\delta\phi
+Ng{\phi_0}^{*2}{\phi_0}\delta\phi +\mbox{c.c.}\right].
\ee
By integration by parts and using the fact that ${\phi_0}$ solves the 
Gross-Pitaevskii equation Eq.(\ref{eq:zero}) we rewrite
this expression as
\be
\delta E^{(1)} = N\mu [\langle{\phi_0}|\delta\phi\rangle + \langle\delta\phi|
{\phi_0}\rangle].
\ee
As both $\phi$ and ${\phi_0}$ are normalized to unity $\delta\phi$ actually
fulfills the identity
\be
\langle{\phi_0}|\delta\phi\rangle+ \langle\delta\phi|{\phi_0}\rangle
= -\langle \delta\phi|\delta\phi\rangle.
\label{eq:1devient2}
\ee
The {\sl a priori}
first order energy change is {\sl a posteriori} of second order: 
\be
\delta E^{(1)} =
-N\mu \langle \delta\phi|\delta\phi\rangle!
\ee

The terms {\sl a priori} quadratic in $\delta\phi$ are given by:
\bea
\delta E^{(2)} = N\int d^3\vec{r}\, &&\left[\frac{\hbar^2}{2m}
\vec{\mbox{grad}}\,\delta\phi^*\cdot\vec{\mbox{grad}}\,\delta\phi+U(\vec{r}\,)
\delta\phi^*\delta\phi+2Ng|{\phi_0}|^2\delta\phi^*\delta\phi\right.\nonumber\\
&&
\left.+\frac{1}{2}Ng{\phi_0}^{*2}\delta\phi^2+\frac{1}{2}Ng{\phi_0}^{2}\delta\phi^{*2}\right].
\label{eq:complete}
\eea
We transform this expression by splitting $\delta\phi$ in a part parallel to ${\phi_0}$
and a part orthogonal to ${\phi_0}$:
\be
\delta\phi(\vec{r}\,) = \gamma{\phi_0}(\vec{r}\,) + \delta\phi_\perp(\vec{r}\,).
\ee
Using integration by parts, the Gross-Pitaevskii equation and the fact that
operator $\cal L$ contains projectors orthogonally to ${\phi_0},{\phi_0}^*$
we are able to write $\delta E^{(2)}$ as
\bea
\delta E^{(2)} &=& |\gamma|^2N\mu +\frac{1}{2}(\gamma+\gamma^*)^2N^2g
\int d^3\vec{r}\, |{\phi_0}|^4
+(\gamma+\gamma^*)N^2g\int d^3\vec{r}\,|{\phi_0}|^2
({\phi_0}^*\delta\phi_\perp+\mbox{c.c.}) \nonumber \\
&&+\frac{1}{2}N\int d^3\vec{r}\, (\delta\phi_\perp^*,\delta\phi_\perp)
(\eta{\cal L}+\mu\,\mbox{Id})
\left(\begin{array}{c}\delta\phi_\perp \\ \delta\phi_\perp^*\end{array}\right).
\label{eq:transit}
\eea
Note that we had to add $\mu$ times the identity matrix $\mbox{Id}$
to $\eta\cal{L}$ as Eq.(\ref{eq:complete}), contrarily to
$\eta{\cal L}$, does not contain any term proportional
to $\mu$.
From Eq.(\ref{eq:1devient2}) we see that $\gamma+\gamma^*$ is actually of second order
in $\delta\phi_\perp$ so that it can be set to zero in Eq.(\ref{eq:transit}).

Summing the {\sl a priori} first and second order energy changes we see that 
$\delta E^{(1)}$
exactly cancels the terms involving explicitly $\mu$ in Eq.(\ref{eq:transit}) so
that we arrive at
\be
\delta E \simeq  \frac{1}{2} N\int d^3\vec{r}\, (\delta\phi_\perp^*,\delta\phi_\perp)
\eta{\cal L} \left(\begin{array}{c}\delta\phi_\perp \\ \delta\phi_\perp^*
\end{array}\right).
\ee
Thermodynamical stability that is positivity of
$\eta {\cal L}$ is therefore equivalent to the Gross-Pitaevskii
energy functional having a local minimum in ${\phi_0}$.

\subsection{Thermodynamical stability implies dynamical stability}
As we show now the positivity of $\eta{\cal L}$ automatically
leads to a purely real spectrum for $\cal L$, that is
to dynamical stability. Consider
an eigenvector $(u,v)$ of $\cal L$ with the eigenvalue $\epsilon$.
Contracting the operator $\eta{\cal L}$
between
the ket $(|u\rangle,|v\rangle)$ and the bra $(\langle u|,\langle v|)$
we get
\begin{equation}
(\langle u|,\langle v|)\eta{\cal L} \left(
\begin{array}{c}
|u\rangle \\
|v\rangle
\end{array}
\right)=
\epsilon \left[\langle u|u\rangle-\langle v|v\rangle\right].
\label{eq:sand}
\end{equation}
The matrix element of $\eta{\cal L}$ is real positive as
$\eta{\cal L}$ is supposed to be a positive hermitian operator. We now face two
possible cases for the real quantity $\langle u|u\rangle-\langle v|v\rangle$:
\begin{itemize}
\item $\langle u|u\rangle-\langle v|v\rangle= 0$. In this case
$\eta{\cal L}$ has a vanishing expectation value in
$(|u\rangle,|v\rangle)$; as $\eta{\cal L}$ is positive,
$(|u\rangle,|v\rangle)$ has to be an eigenvector
of $\eta{\cal L}$ with the eigenvalue zero; as $\eta$
is invertible we find that $(|u\rangle,|v\rangle)$
is an eigenvalue of $\cal L$ with the eigenvalue 0, so that $\epsilon=0$
is a real number.
\item $\langle u|u\rangle-\langle v|v\rangle > 0$:
we get $\epsilon$ as the ratio
of two real numbers, so that $\epsilon$ is real.
\end{itemize}

\subsection{Examples of thermodynamical instability}
\subsubsection{Real condensate wavefunction with a node}
Let us assume that the solution of the Gross-Pitaevskii equation is a real
function $\phi_0(\vec{r}\,)$. To decide if this solution is thermodynamically
stable one has to check the positivity of the operator $\eta{\cal L}$.
Consider an eigenvector of $\eta{\cal L}$ with eigenvalue $\varepsilon$:
\be
\eta{\cal L}\left(\begin{array}{c} u \\ v\end{array}\right)
= \varepsilon \left(\begin{array}{c} u \\ v\end{array}\right).
\label{eq:eigen_eta}
\ee
This $\varepsilon$ should not be confused with the quasi-particle energies
as $\eta{\cal L}$ and $\cal L$ have different spectra.
By performing the sum and the difference of the two lines of 
Eq.(\ref{eq:eigen_eta}) we get decoupled equations for the sum $\psi_s=
u+v$ and the difference $\psi_d = u-v$:
\bea
\label{eq:premier}
\varepsilon |\psi_s\rangle &=& \left[\frac{\vec{p}\,^2}{2m} + U(\vec{r}\,) +
Ng\phi_0^2(\vec{r}\,)+2Ng Q\phi_0^2(\vec{r}\,)Q-\mu\right]|\psi_s\rangle \\
\varepsilon |\psi_d\rangle &=& \left[\frac{\vec{p}\,^2}{2m} + U(\vec{r}\,) +
Ng\phi_0^2(\vec{r}\,)-\mu\right]|\psi_d\rangle.
\label{eq:second}
\eea
Both operators involved in these equations have to be positive
to achieve thermodynamical stability. Note that for $g>0$ the positivity
of the second operator Eq.(\ref{eq:second}) implies the positivity of
the first one Eq.(\ref{eq:premier}) as $g Q\phi_0^2(\vec{r}\,)Q$
is positive.

We therefore concentrate on Eq.(\ref{eq:second}). It involves the Gross-Pitaevskii
Hamiltonian
\be
{\cal H}_{GP}= \frac{\vec{p}\,^2}{2m} + U(\vec{r}\,) +
Ng\phi_0^2(\vec{r}\,)-\mu.
\ee
An obvious  eigenvector of this Hamiltonian
is $\psi_d=\phi_0$ with eigenvalue $\varepsilon=0$, 
as $\phi_0$ solves the Gross-Pitaevskii equation! 
The condition of a positive $\varepsilon$  in Eq.(\ref{eq:second}) simply
means that $\phi_0$ should be the ground state of ${\cal H}_{GP}$!

We can then invoke a theorem claiming that the ground state of a potential
has no node \cite{Peierls}. 
If $\phi_0(\vec{r}\,)$ has a node it cannot be the ground state
of ${\cal H}_{GP}$. The ground state of ${\cal H}_{GP}$
has therefore an eigenenergy $\varepsilon$ lower than the one
of $\phi_0$, that is lower than zero, so that the operator
$\eta {\cal L}$ is not positive and there is no thermodynamical stability.

As an example consider in a harmonic trap with eigenaxis $z$,
a solution of the Gross-Pitaevskii equation even along $x$ and $y$
but odd along $z$, so that it vanishes in the plane
$z=0$. Such a solution exists, for $g>0$: within the class
of {\sl real} functions $\phi$ odd along $z$ and even along $x$,$y$,
the Gross-Pitaevskii energy $E[\phi,\phi]$, bounded from below,
has a minimum, reached in $\phi_0$, and this $\phi_0$ then solves
the Gross-Pitaevskii equation. This solution however is no longer
a local minimum of $E[\phi,\phi^*]$ when one includes all possible deviations
of $\phi$ from $\phi_0$ (complex and with no well defined parity along $z$).

\subsubsection{Condensate with a vortex}
Can we get a thermodynamically stable condensate wavefunction with a node?
To beat the results of the previous subsection we now assume
$\phi_0$ to be complex. 

A particular class of complex wavefunctions with a node
are condensate wavefunctions with vortices. A vortex is characterized
(i) by a nodal, not necessarily straight,
line in $\phi_0$ (the center of the so-called vortex core) and (ii) by the
fact that the phase of $\phi_0$ changes by $2q\pi$, $q$ non-zero
integer, along a
closed path around the vortex core ($q$ is the so-called charge of the vortex).
This second property means that the circulation of the local
velocity field (defined in \S\ref{subsubsec:hye}) around the vortex
core is $2\pi\hbar q/m$.

A condensate wavefunction can have several vortices; the change
of the phase of $\phi_0$ along a closed path is now 
$2\pi q_{\mbox{\scriptsize sum}}$ where $q_{\mbox{\scriptsize sum}}$
is the algebraic sum of the charges of the vortex lines enclosed by the path.

It has been shown that 
a condensate wavefunction with a vortex in a harmonic trap is not
thermodynamically stable \cite{Rokhsar}. In the limit
of vanishing interaction between the particles ($g=0$) this is clear indeed.
Suppose that the trap is cylindrically symmetric with respect to $z$.
$|\phi_0\rangle$ can be chosen as $(|n_x=1,n_y=0,n_z=0\rangle + i
|n_x=0,n_y=1,n_z=0\rangle)/\sqrt{2}$ where $|n_x,n_y,n_z\rangle$
is the eigenstate of the harmonic oscillator with quantum number
$n_\alpha$ along axis $\alpha$ ($\alpha=x,y,z$). The chemical
potential is simply $2\hbar\omega_{x,y}+\frac{1}{2}\hbar\omega_z$
where $\omega_\alpha$ is the atomic oscillation frequency along axis $\alpha$.
One then finds that $(|u\rangle=|n_x=0,n_y=0,n_z=0\rangle,|v\rangle=0)$
is an eigenvector of $\eta{\cal L}$ with the strictly negative energy
$\varepsilon=-\hbar\omega_{x,y}$.

What happens in the opposite Thomas-Fermi regime of strong interactions?
An intuitive answer can be obtained in a 2D model of the Gross-Pitaevskii
equation, assuming for simplicity a quasi-isotropic
trapping potential and restricting to the following class of condensate
wavefunctions:
\be
\phi_0(x,y)= \phi_{\mbox{\scriptsize slow}}(x,y)
\tanh[\kappa|\vec{r}-\vec{\alpha} R|]e^{i\theta_{\vec{\alpha}R}}.
\ee
In this ansatz $\phi_{\mbox{\scriptsize slow}}(x,y)$ is the usual square root
of inverted parabola Thomas-Fermi approximation for a condensate wavefunction
without vortex, with a radius $R$; the $\tanh[\;]$ 
represents the correction to the modulus of $\phi_0$ due to the vortex
core (of adjustable position $\vec{\alpha}R$ and inverse width $\kappa$);
$\theta_{\vec{\alpha}R}$ is the polar angle of a
system of Cartesian coordinates $(X,Y)$ centered on the vortex core,
and represents (approximately for $\vec{\alpha}\ne\vec{0}$)
the phase of the unit-charge vortex.

\begin{figure}[htb]
\centerline{ \epsfysize=7cm \epsfbox{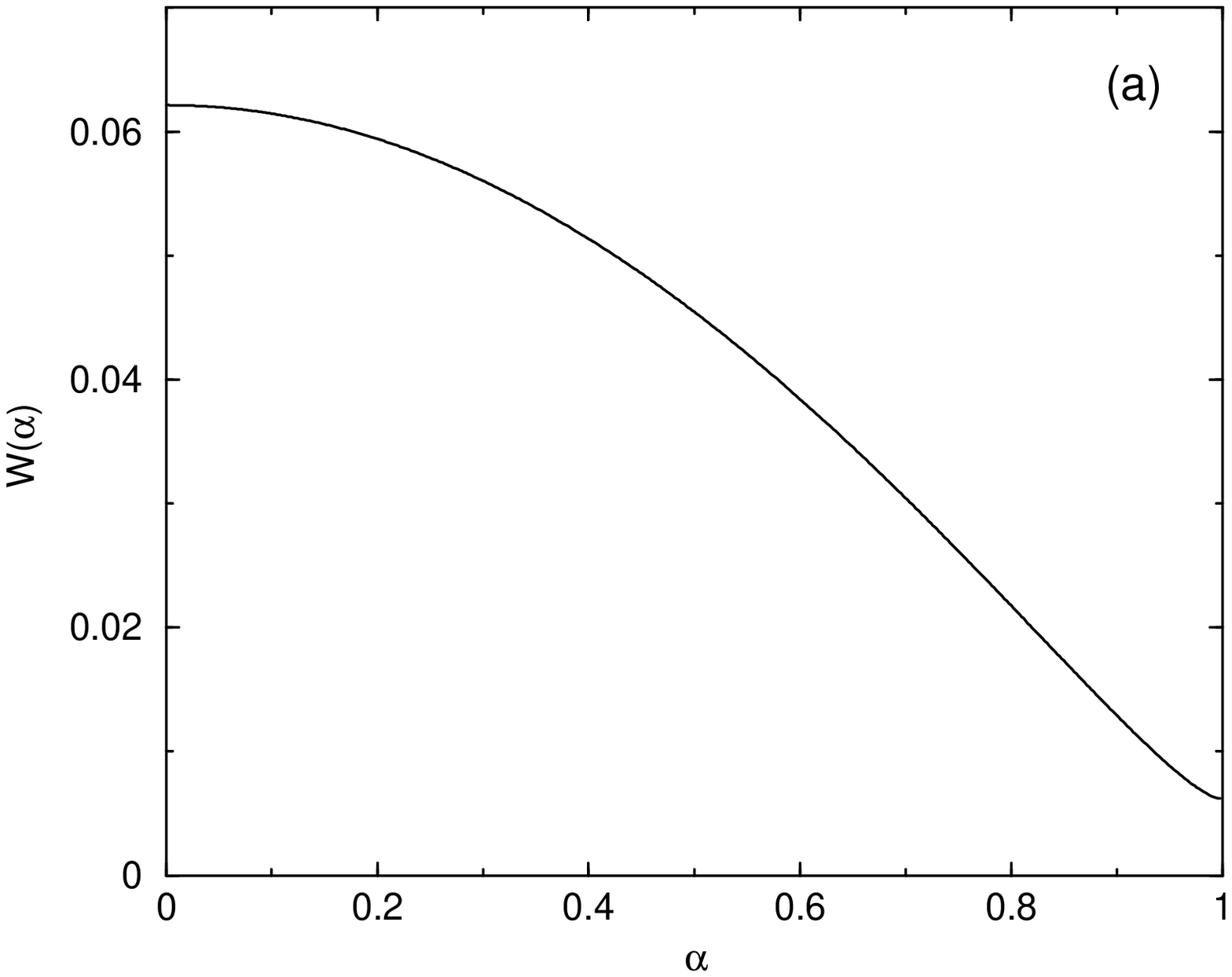} \ \ \
\epsfysize=7cm \epsfbox{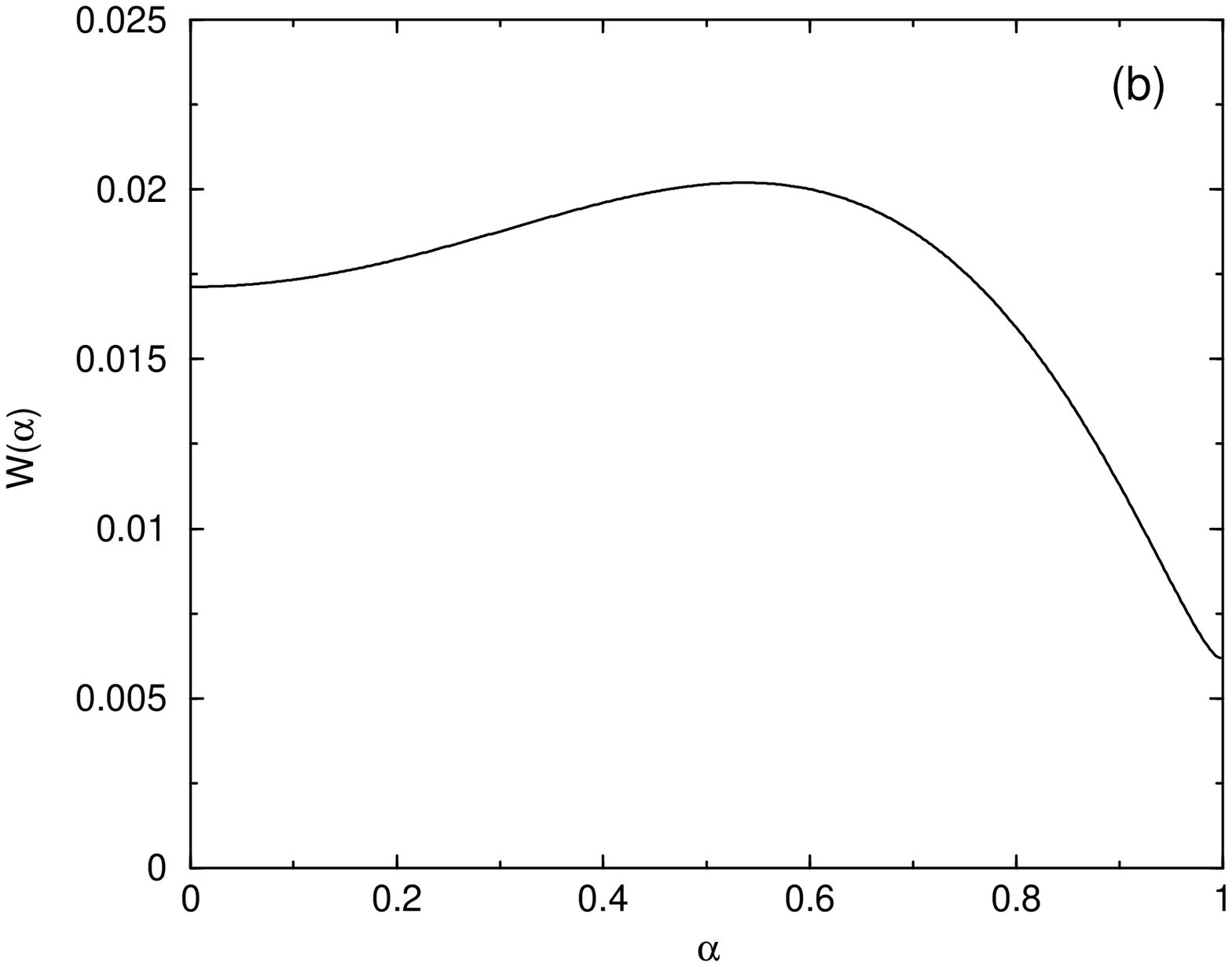}}
\caption{\small In a 2D model, 
effective potential energy $W$ of a vortex in a quasi-axisymmetric
harmonic trap as function of the distance $\alpha R$
of the core from the trap center, for $\mu_0= 80\hbar\omega$.
(a) $\Omega=0$ and (b) $\Omega=0.045\omega$.
The unit of energy is $N\hbar\omega$ where $\omega$ is the oscillation
frequency of the atoms in the trap.}\label{fig:W}
\end{figure}

One then calculates the mean energy of $\phi_0$, with the simplification
that $\phi_{\mbox{\scriptsize slow}}(x,y)$ varies very slow at the scale of
$\kappa^{-1}$, and one minimizes this energy over $\kappa$. This
leads to the inverse size of the vortex core on the order of the
local healing length of the condensate:
\begin{equation}
{\hbar ^2\kappa^2\over m} = 0.59\left[\mu-{1\over 2}m\omega^2
(\alpha R)^2\right].
\end{equation}
The mean energy of $\phi_0$ (\ref{eq:gpef_encore}) 
is now a function of the position
of the vortex core only,
\be
E = E_{\mbox{\scriptsize no vortex}} + W(\vec{\alpha})
\ee
where $E_{\mbox{\scriptsize no vortex}}$ is the energy of the condensate
with no vortex and 
\begin{equation}
W(\vec{\alpha}) = N\frac{(\hbar\omega)^2}{\mu_0}
\left\{ {1\over 2}+ (1-\alpha^2)\left[ {2\ln 2+1\over 3}+\ln
{\nu\mu_0\over\hbar\omega}+\ln(1-\alpha^2)
\right]
 \right\}
\label{eq:sans_rot}
\end{equation}
with $\nu=0.49312$ and $\mu_0$ the chemical potential in the absence of
vortex.
This function $W$ represents an effective potential seen by the vortex
core. As shown in Fig.\ref{fig:W}a this potential
is maximal at the center of the trap so that it
is actually an expelling potential for the vortex core: 
shifting the vortex core away from the center of the trap
lowers the condensate energy.

A method to stabilize the vortex is to rotate the harmonic trap around
$z$ at a frequency $\Omega$ (the trap is anisotropic in the $x-y$
plane otherwise rotation would have no effect).
Thermodynamical
equilibrium will now be obtained in the frame rotating at the frequency
$\Omega$, where the harmonic trap is time independent. As this frame
is non Galilean the Hamiltonian and therefore the Gross-Pitaevskii
energy functional have to be supplemented by the inertial energy
term $-\Omega L_z$ per atom, where $L_z$ is the angular momentum operator
along $z$. 
The effective potential $W(\vec{\alpha})$ gets an extra
term:
\be
W_{\Omega}(\vec{\alpha}) = W_{\Omega=0}(\vec{\alpha})
-N\hbar\Omega (1-\alpha^2)^2
\ee
where $W_{\Omega=0}$ is the result (\ref{eq:sans_rot})
in the absence of rotation.
As shown in Fig.\ref{fig:W}b this extra term can trap the
vortex core at the center of the harmonic trap if $\Omega$
is large enough.

What happens if $\Omega$ is increased significantly ? It becomes
favorable to put more vorticity in the condensate.
As the vortices with charge larger than one are unstable the way out is
to create several vortices with unit charge. This can be analyzed
along the previous lines by a generalized multi-vortex ansatz,
as discussed in \cite{Dum_vortex}. A condensate with vortices
has been recently obtained at the ENS in a rotating trap
\cite{Dalibard_vortex}.

Another philosophy was followed at JILA: rather than relying on thermal
equilibrium in a rotating trap to produce a vortex they used a
\lq\lq quantum engineering " technique \cite{Ignacio_vortex} 
to directly induce the vortex by giving 
angular momentum to the atoms through coupling to electromagnetic fields 
\cite{JILA_vortex}. It has also been suggested to imprint 
the phase of the vortex on the condensate through a lightshift induced
by a laser beam whose spatial intensity profile has been conveniently
tailored \cite{Maciek}. Such an imprinting technique has successfully
led to the observation of dark and gray solitons in atomic
condensates with repulsive interactions in Hannover \cite{Hannover}
and in the group of W.\ Phillips at NIST.
All these techniques illustrate again the powerfulness of atomic physics
in its ability to manipulate a condensate.

\section{Phase coherence properties of Bose-Einstein condensates}
\markright{Phase coherence}
\label{Cap:PC}

Consider two Bose-Einstein condensates prepared in spatially well separated
traps and that have \lq never seen each other' (e.g.\ one rubidium condensate at JILA
and one rubidium condensate at ENS). It is \lq natural'
to assume that these two  condensates do not have a well defined relative phase.
However the trend in the literature on Bose condensates is to assume
that the two condensates are in a {\sl coherent} state with a well defined
relative phase, the so-called \lq symmetry-breaking' point of view.
So imagine that one lets the two condensates spatially overlap. Will
interference fringes appear on the resulting atomic density or not~?

One of the goals of this chapter is to answer this question 
and to reconcile the symmetry breaking point of view 
with the \lq natural' point of view.

\subsection{Interference between two BECs}
At MIT a double well trapping potential was obtained by superimposing
a sharp barrier induced with laser light on top of the usual harmonic trap produced
with a magnetic field. In this way one can produce two Bose-Einstein condensates,
one on each side of the barrier. The height of the barrier can be made much larger than
the chemical potential of the gas so that
coupling between the two condensates {\sl via} tunneling through the wall 
is very small.
In this way one can consider the two condensates as independent.

\begin{figure}[htb]
\epsfysize=8cm \centerline{\epsfbox{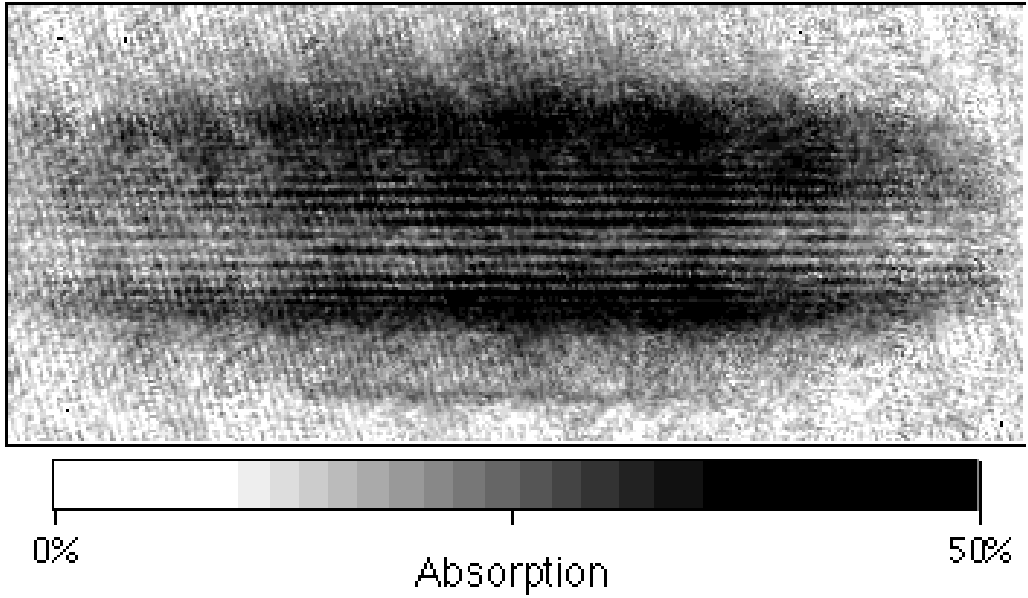}}
\caption{\small Interference fringes between two condensates observed at MIT
\cite{MIT_franges}.}\label{fig:franges}
\end{figure}

One can then switch off the barrier and magnetic trap, let the two condensates
ballistically expand and spatially overlap. One then measures the spatial density
of the cloud by absorption imaging. This spatial density exhibits clearly
fringes \cite{MIT_franges} (see figure \ref{fig:franges}). 
These fringes have to be interference fringes, as
hydrodynamic effects (such as sound waves) are excluded at the very low
densities of the ballistically expanded condensates.
We show  here on a simple model that we indeed expect to see interference fringes
in such an experiment, even if the two condensates have initially
no well defined relative phase.

\subsubsection{A very simple model}
In our simple modelization of an MIT-type interference experiment
we will concentrate on the positions of the particles
on an axis $x$ connecting the two condensates so that we use
a one-dimensional model  enclosed in a box of size $L$
with periodic boundary conditions. We assume that
the system is initially in the Fock state
\be
|\Psi\rangle =| \frac{N}{2}:k_a , \frac{N}{2}:k_b \rangle
\label{eq:Fock}
\ee
with $N/2$ particles in the plane wave of momentum $\hbar k_a$
and $N/2$ particles in the plane wave of momentum $\hbar k_b$:
\be
\langle x|k_{a,b} \rangle=\frac{1}{\sqrt{L}} \exp \left[{i k_{a,b}x} \right]
\,.
\ee
We assume that one detects the position of all the particles. What will be
the outcome~?
As the numbers of particles are exactly defined in the two modes $a$ and $b$
the relative phase between the atomic fields in the two modes is totally
undefined.

\subsubsection{A trap to avoid}\label{subsubsec:trap}
If we calculate the {\it mean density}
in the state given by (\ref{eq:Fock}) 
we find a uniform result
\be
\langle \hat{\psi}^\dagger(x) \hat{\psi}(x) \rangle=N/L
\label{eq:naive}
\ee
and we may be tempted to conclude that no interference fringes
will appear in the beating of two Fock state condensates.

Actually this naive statement is wrong. Interference fringes appeared
in a {\it single realization} of the experiment at MIT.
We have therefore to consider the probability of the 
outcome of a particular density profile in a single
realization of the measurement and not the
average of the density profile over many realizations of the experiment.
Indeed we will see that by interfering two independent Bose-Einstein condensates
we get interference fringes on the density profile
in each single realization of the experiment
but the position of the interference pattern is random
so that by averaging the density profile
over many realizations we wash out the fringes.

We wish to emphasize the following crucial point
of the quantum theory: Whatever single-time measurement is performed on the system
all the information about the outcomes of a single realization of the measurement 
procedure is contained in the $N-$body density matrix, here
\be
\hat{\rho}=|\Psi\rangle \langle \Psi|.
\ee
Indeed the only information we can get from quantum mechanics on
a single realization outcome is its probability $P$, which can
be obtained from $\hat{\rho}$ by
\be
P=\mbox{Tr}[\hat{O}\hat{\rho}]
\ee
where the operator $\hat{O}$ depends on the considered outcome. E.g.\
in our gedanken experiment $P$ is the
probability density of
finding the $N$ particles at positions $x_1,x_2,....x_N$
and the operator $\hat{O}$ is expressed in terms of the field operator as
\be
\hat{O}=\frac{1}{N!} \hat{\psi}^\dagger(x_1) .... \hat{\psi}^\dagger(x_N)
                   \hat{\psi}(x_N) .... \hat{\psi}(x_1).
\ee
In a first quantized picture this corresponds to the fact that
the probability density $P$
is equal to the modulus squared of the $N-$body wavefunction.

The complete calculation of the $N-$body distribution function $P(x_1,\ldots,x_N)$
for the state $|\Psi\rangle$ in Eq.(\ref{eq:Fock})
is involved and we will see in the coming subsections how to circumvent the
difficulty. But we can do a simple calculation of the pair distribution function
of the atoms in state $|\Psi\rangle$: 
\bea
\rho(x_1,x_2) &=& \langle \Psi|\hat{\psi}^\dagger(x_1)\hat{\psi}^\dagger(x_2)
\hat{\psi}(x_2)\hat{\psi}(x_1)|\Psi\rangle \\
&=& ||\hat{\psi}(x_2)\hat{\psi}(x_1)|\Psi\rangle||^2.
\eea
We expand the field operator on the two modes $\phi_{a,b}$ and on other arbitrary
orthogonal modes not relevant here as they are not populated in $|\Psi\rangle$:
\be
\hat{\psi}(x) = \hat{a}\langle x|k_a\rangle + \hat{b} \langle x|k_b\rangle+\ldots
\ee
where $\hat{a}$ and $\hat{b}$ annihilate a particle in state $k_a$ and $k_b$
respectively. We obtain
\bea
\hat{\psi}(x_2)\hat{\psi}(x_1)|\Psi\rangle &= &
\left[\frac{N}{2}\left(\frac{N}{2}-1\right)\right]^{1/2}
\langle x_2|k_a\rangle\langle x_1|k_a\rangle |\frac{N}{2}-2:k_a,\frac{N}{2}:k_b\rangle
\nonumber \\
&+&\left[\frac{N}{2}\left(\frac{N}{2}-1\right)\right]^{1/2}
\langle x_2|k_b\rangle\langle x_1|k_b\rangle |\frac{N}{2}:k_a,\frac{N}{2}-2:k_b\rangle
\nonumber \\
&+& \frac{N}{2} \Big[ \langle x_2|k_a\rangle\langle x_1|k_b\rangle+
\langle x_2|k_b\rangle\langle x_1|k_a\rangle
\Big]  |\frac{N}{2}-1:k_a,\frac{N}{2}-1:k_b\rangle.
\label{eq:pluss}
\eea
The last line of this expression exhibits an interference effect between two amplitudes,
that could not appear in the previous naive reasoning on the one-body density operator
Eq.(\ref{eq:naive})!
In the limit $N\gg 1$ and using the fact that the populated modes are plane waves
the pair distribution function simplifies to
\be
\rho(x_1,x_2)\simeq 
\left(\frac{N}{L}\right)^2\left\{ 1+ \frac{1}{2} \cos\Big[(k_a-k_b)(x_1-x_2)\Big]
\right\}.
\ee
This function exhibits oscillations around an average value equal to the square of
the mean density. The oscillations are due to the interference effect  in
Eq.(\ref{eq:pluss}): they favor detections of pairs of particles with a distance
$|x_1-x_2|$ equal to  $2n\pi/|k_a-k_b|$ ($n$ integer) and they rarefy detections of pairs of
particles with a distance $(2n+1)\pi/|k_a-k_b|$. We therefore see on the pair distribution
function a precursor of the interference fringes observed when the positions of all
the particles are measured!

\subsubsection{A Monte Carlo simulation}
\begin{figure}[htb]
\epsfysize=8cm \centerline{\epsfbox{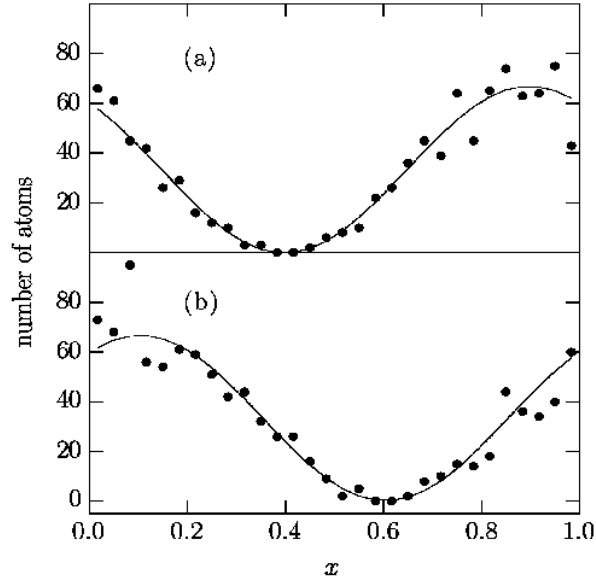}}
\caption{\small For two different Monte Carlo realizations (a) and (b) of the gedanken experiment,
histogram of the measured positions of $N=1000$ particles for an initial Fock
state with $N/2$ particles in plane wave $k_a$ and $N/2$ particles in plane wave
$k_b=-k_a$ \cite{Javanainen}. The positions of the particles are expressed in units of
$2\pi/(k_a-k_b)$ and are considered modulo $2\pi/(k_a-k_b)$.}\label{fig:java}
\end{figure}

By sampling the $N-$body distribution function $P$ with a
Monte Carlo technique, Javanainen and Sung Mi Yoo in \cite{Javanainen}
made a numerical experiment with $N=10^3$ particles and $k_b=-k_a$.
By distributing the measured positions in a given realization  $x_1,x_2,....x_N$
among 30 position bins they obtained histograms like the ones
in figure \ref{fig:java}.
It turns out that the density in the 
outcome of each realization of the numerical experiment
can be fitted by a cosine:
\be
\frac{N}{2L}\left|e^{i k_a x}e^{i \theta_a}+e^{i k_b x}e^{i \theta_b}\right|^2
\ee
where $\theta_a$ and $\theta_b$ are phases varying randomly from one
realization to the other. In other words one has the
impression that for each realization the system is in the state
\be
|\theta\rangle_N=\frac{1}{\sqrt{N!}}\left[ \frac{1}{\sqrt{2}}
     \left( a_{k_a}^\dagger e^{i\theta} + a_{k_b}^\dagger e^{-i\theta} \right)
     \right]^N |0\rangle
\label{eq:phase_state}
\ee
with the angle $\theta=(\theta_a-\theta_b)/2$ randomly distributed in
$[-\pi/2,\pi/2]$. Such a state, corresponding to a well defined phase between
the two modes $a$ and $b$, is called a phase state \cite{Leggett}.

\subsubsection{Analytical solution}
We wish to explain the result of the numerical experiment with an
analytical argument. This has been done with slightly different points
of view in \cite{mesure_franges2,mesure_franges1}.
We give here what we think is the simplest possible presentation.

Let us allow Poissonian fluctuations in the number
of particles $N_a$ and $N_b$, corresponding to the distribution probabilities:
\be
{\cal P}_{\eps}(N_\eps)=\frac{(\bar{N_\eps})^{N_\eps}}{N_\eps!}e^{-\bar{N_\eps}}
\hspace{1cm} \eps=a,b
\ee
with mean number of particles $\bar{N_a}=\bar{N_b}=\bar{N}/2$.
These fluctuations become very small as compared to $\bar{N}$
when the number of particles becomes large:
\be
\frac{\Delta N_\eps}{\bar{N_\eps}}=
        \frac{1}{\sqrt{\bar{N_\eps}}}\rightarrow 0
        \ \ \  \mbox{for} \ \ \ \bar{N_\eps}\rightarrow \infty.
\ee
The corresponding density operator is  a statistical mixture of Fock states:
\be
\hat{\rho}=\sum_{N_a,N_b=0}^{\infty} {\cal P}_a(N_a) {\cal P}_b(N_b)
        |N_a:k_a,N_b:k_b\rangle \langle N_a:k_a,N_b:k_b| .
\ee
From this form one can {\sl imagine} that a single realization of the experiment
is in a Fock state, provided that one keeps in mind that $N_a$ and $N_b$
vary in an impredictable way from one experimental realization to the other.
We known from the work \cite{Javanainen} that there will be interference 
fringes in each experimental realization, but this fact is not intuitive.

The same density operator can also be written in terms of a statistical mixture
of phase states:
\be
\hat{\rho}=\sum_{N=0}^{\infty} \frac{(\bar{N})^N}{N!}e^{-\bar{N}}
                \int_{-\pi/2}^{\pi/2} \frac{d\theta}{\pi}
                |\theta \rangle_N {}_N\langle \theta|.
\ee
From this form one can {\sl imagine} that a single realization of the experiment
is in a phase state, provided that one keeps in mind that the total number
of particles $N$ and the relative phase $\theta$ 
vary in an impredictable way from one realization to the other.
This last form leads to the following algorithm to generate the
positions of the particles according
to the correct probability distribution:
\begin{enumerate}             
\item generate an integer $N$ according to the Poisson distribution
of parameter $\bar{N}$
\item generate $\theta$ according to a uniform probability distribution
within $-\pi/2$ and $\pi/2$
\item generate the positions $x_1,....x_N$ {\it as if} the system was in the
state $|\theta\rangle_N$, in which case all the particles are in
the same single particle-state and
the probability density $P(x_1,....x_N)$ is factorized:
\be
P(x_1,....x_N)=\prod_{j=1}^{N} p(x_j)
\ee
where
\be
p(x)={1\over 2L}\left|e^{ik_a x}e^{i\theta}+e^{ik_b x}e^{-i\theta}\right|^2.
\ee
\end{enumerate}
One then obtains interference fringes in each experimental realization, 
in a very explicit way.

One could also use a third form of the same density operator
$\hat{\rho}$, that is
a statistical mixture of Glauber coherent states:
\be
\hat{\rho} = \int_0^{2\pi} {d\theta_a\over 2\pi}
\int_0^{2\pi} {d\theta_b\over 2\pi}
|\mbox{coh}:\bar{N_a}^{1/2}e^{i\theta_a},\mbox{coh}:\bar{N_b}^{1/2}e^{i\theta_b}\rangle
\langle \mbox{coh}:\bar{N_a}^{1/2}e^{i\theta_a},\mbox{coh}:\bar{N_b}^{1/2}e^{i\theta_b}|.
\ee
This mathematical form is at the origin of the popular belief that
condensates are in coherent states. 
From this form one can only {\sl imagine} that a single realization of the experiment
is in a coherent state, keeping in mind that the phases $\theta_a$
and $\theta_b$ 
vary in an impredictable way from one realization to the other.
In this representation the occurrence of interference fringes is straightforward.

There is an important difference between the coherent states and the Fock or
phase states: as the number of particles is a conserved quantity in the
non-relativistic Hamiltonian used to describe the experiments on atomic gases
it seems difficult to
produce a condensate in a coherent state in some mode $\psi$, that is with $\hat{\rho}$
being a pure state $|\mbox{coh}:\alpha\rangle\langle\mbox{coh}:\alpha|$ where $\alpha$
is a complex number.

On the contrary one could imagine producing a condensate in a Fock state 
by measuring the number of particles in the condensate. One could then
obtain a phase state by applying a $\pi/2$
Rabi pulse on the Fock state changing the internal atomic state $a$ to a 
superposition $(|a\rangle+|b\rangle)/\sqrt{2}$ where $b$ is another
atomic internal state; such a Rabi pulse has been demonstrated
at JILA and has allowed the measurement of the coherence time
of the relative phase between the $a$ and $b$ condensates \cite{JILA_phase}.

\subsubsection{Moral of the story}
\begin{itemize}
\item there is in general no
unique way of writing the density operator $\hat{\rho}$ as a statistical mixture.
The canonical form corresponding to the diagonalization
of $\hat{\rho}$ is always a possibility but not always the most convenient
one. E.g.\ in our simple model the eigenbasis (Fock states)
is less convenient than the non-orthogonal family of
phase states (symmetry breaking states).
\item no measurement or no set of measurements performed
{\it on the system} can distinguish between two different mathematical forms
of the same density matrix as a statistical mixture.
\item the symmetry breaking point of view consists in writing 
(usually in an approximate way) the $N-$body density operator
as a statistical mixture of Hartree-Fock states.
One can then imagine that
a given experimental realization of the system is a Hartree-Fock state,
whose physical properties are immediate to understand as all the particles
are in the same quantum state. 
\item If the system is not in a state that is as simple as a Hartree-Fock state
(e.g.\ in a Fock state for our simple model)
it is dangerous to make reasonings on the single particle density operator
(that is on the first order correlation function of the atomic field operator)
to predict outcomes of single measurements on the system: the relevant information
may be stored in higher order correlation functions of the field.
\end{itemize}

\subsection{What is the time evolution of an initial phase state ?}
\label{subsec:whatisphaseevol}

\subsubsection{Physical motivation}

Consider an interference experiment between two condensates $A$ and $B$ either
in spatially separated traps or in different internal states (JILA-type
configuration \cite{JILA_phase}). Assume that the two condensates
have been prepared initially in a state with a well defined
relative phase $\theta$; this has actually been achieved at JILA.
Let the system evolve freely for some time $t$. How long will the relative
phase remain well defined~?
This question is probably not an easy one to answer. We present here a simple
model including only two modes of the field. In real life the
other modes of the field are not negligible
(see for example \cite{Graham} for a discussion of finite temperature effects) 
and phenomena neglected here such as losses of particles from the
trap and fluctuations in the total number of particles
may be important in a real experiment \cite{resur,phase_mel}.

We assume that the state of the system at time $t=0$ is a phase
state.
More specifically, expanding the $N-$th power
in Eq.(\ref{eq:phase_state}) with the binomial formula, we take as initial
state:
\be
|\Psi(t=0)\rangle = 
2^{-N/2}\sum_{N_a=0}^{N}
\left(\frac{N!}{N_a!N_b!}\right)^{1/2}e^{i(N_a-N_b)\theta}|N_a:\phi_a,N_b:\phi_b\rangle
\label{eq:etat_in}
\ee
where $N_b=N-N_a$ and $\phi_{a,b}$ are the steady state condensate wavefunctions
with $N_{a,b}$ particles in condensates $A,B$ respectively. The time evolution
during $t$ is 
simple for each individual Fock states, as the system is then in a steady
state with total energy $E(N_a,N_b)$:
\be
|N_a:\phi_a,N_b:\phi_b\rangle \rightarrow e^{-iE(N_a,N_b)t/\hbar}|N_a:\phi_a,N_b:\phi_b\rangle.
\ee
The time evolution of the phase state Eq.(\ref{eq:etat_in}) is much more complicated:
the state vector $|\Psi(t)\rangle$ is a sum of many oscillating functions of
time. 

\subsubsection{A quadratic approximation for the energy}
The discussion can be greatly simplified if one uses the fact that the binomial
factor in Eq.(\ref{eq:etat_in}) for large $N$ is a function of $N_a$ and $N_b$ 
sharply peaked around $N_a=N_b=N/2$ with a width $\sqrt{N}$: 
from Stirling's formula $n!\simeq (n/e)^n\sqrt{2\pi n}$
we obtain indeed
\be
\frac{1}{2^N}\frac{N!}{N_a! N_b!} \simeq \frac{1}{\sqrt{2\pi}2^N}\left(\frac{N}{N_a N_b}\right)^{1/2}
e^{-N_a\log(N_a/N)-N_b\log(N_b/N)}\simeq \left(\frac{2}{\pi N}\right)^{1/2}
e^{-(N_a-N_b)^2/(2N)}.
\label{eq:gauss}
\ee
We therefore expand the energy $E$ in powers of $N_a-N/2$ and $N_b-N/2$ up to second order.
\bea
E(N_a,N_b) &=& E(N/2,N/2) + (N_a-N/2)\partial_{N_a}E + (N_b-N/2)\partial_{N_b}E
\nonumber \\
&&+\frac{1}{2}(N_a-N/2)^2\partial_{N_a}^2E
+\frac{1}{2}(N_b-N/2)^2\partial_{N_b}^2E \nonumber \\
&&+(N_a-N/2)(N_b-N/2)\partial_{N_a}\partial_{N_b}E+\ldots,
\eea
all the derivatives being taken in $(N_a,N_b)=(N/2,N/2)$. Note that the first derivatives
of the energy are the chemical potentials $\mu_{a,b}$ of the two condensates; as the
condensates are independent condensates (there is no mechanism locking
the relative phase of the condensates) one has in general $\mu_a\neq \mu_b$.
As we restrict to the set of occupation numbers such that $N_a+N_b=N$ we can rewrite the expansion
of the energy using $N_a-N/2= -(N_b-N/2)=(N_a-N_b)/2$:
\be
E(N_a,N_b) \simeq E(N/2,N/2) + \frac{1}{2}(N_a-N_b) (\mu_a-\mu_b)
+\frac{\hbar}{4}(N_a-N_b)^2\chi
\label{eq:energie}
\ee
where we have introduced the quantity
\be
\chi=\frac{1}{2\hbar}\left[\left(\partial_{N_a}-\partial_{N_b}\right)^2 E
\right]_{N_a=N_b=N/2}.
\label{eq:chi}
\ee

\subsubsection{State vector at time $t$}
If one uses the quadratic approximation of the energy
the system evolves from the initial state
Eq.(\ref{eq:etat_in}) to the state
\be
|\Psi(t)\rangle=
2^{-N/2}\sum_{N_a=0}^{N}
\left(\frac{N!}{N_a!N_b!}\right)^{1/2}e^{i(N_a-N_b)(\theta+vt)}
e^{-i(N_a-N_b)^2\chi t/4}
|N_a:\phi_a,N_b:\phi_b\rangle
\label{eq:au_temps_t}
\ee
The contribution of the term linear in $N_a-N_b$ in Eq.(\ref{eq:energie}) is contained
in the quantity 
\be
v= \frac{1}{2\hbar}(\mu_b-\mu_a).
\ee
The resulting effect on the time evolution is simply to shift the relative phase
between the condensate from $\theta$ to $\theta+vt$: this is a mere phase drift 
with a velocity $v$. This phase drift takes place only if the \lq frequencies'
$\mu_a/\hbar$ and $\mu_b/\hbar$ of the atomic fields in $A$ and in $B$ are different.

The effect of the quadratic term in Eq.(\ref{eq:energie}) 
is to spread the relative phase of the two condensates. This effect is formalized
in \cite{mesure_franges2}, we give here the intuitive result.
The spreading of the phase can 
be understood in analogy with the spreading of the wavepacket of a fictitious massive
particle, with the relative phase $\theta$ being the position $x$ of the particle and the
occupation number difference $N_b-N_a$ being the wavevector $k$ of the particle. 
The energy term proportional to $\chi$ plays the role of the kinetic energy of the particle responsible 
for the spreading in position. The effective mass of the fictitious particle
is $M$ such that
\be
\frac{1}{4}(N_a-N_b)^2\chi \longleftrightarrow \frac{\hbar k^2}{2M}
\ee
so that
\be
M = \frac{2\hbar}{|\chi|}.
\ee
Replacing the discrete sum in Eq.(\ref{eq:au_temps_t}) by an integral we formally
obtain the expansion of the time dependent state vector of the fictitious particle
over the plane waves in free space. 
In this case the variance of the position of the fictitious particle
spreads as
\be
\Delta x^2(t) = \Delta x^2(0) + \left(\frac{\hbar \Delta k}{M}\right)^2 t^2.
\ee
Within the approximation (\ref{eq:gauss})
the wavepacket of the fictitious particle is a Gaussian in momentum space,
with a standard deviation $\Delta k = (N/2)^{1/2}$. Initially the position
$x$ is well defined with a spread $\sim 1/\Delta k \ll 1$. The relative phase of the
condensates will start becoming undefined when the position spread $\Delta x$ of the
fictitious particle becomes on the order of unity. This happens after a time
\be
t_{\mbox{\scriptsize spread}} \sim \frac{M}{\hbar \Delta k} = \frac{2\sqrt{2}}{|\chi| N^{1/2}}.
\label{eq:tspread}
\ee

At times much longer than $t_{\mbox{\scriptsize spread}}$ it is not correct to
replace the discrete sum over $(N_a-N_b)/2$ by an integral.
The discreteness of $N_a-N_b$  leads to reconstructions of a phase state
(the so-called revivals) at times $t_q=q\pi/\chi$, $q$ integer: one
can check indeed from Eq.(\ref{eq:au_temps_t}) that a phase state is reconstructed
with a relative phase $\theta+v t_q + q\pi/2$ for $N$ even and $\theta+v t_q$ for $N$
odd. The observability of even the first revival at time $t_1$ is a non trivial 
question: the revivals are easily destroyed by decoherence phenomena such
as the loss of a few particles out of the condensate due to inelastic
atomic collisions \cite{resur}, and effects of the non-condensed fraction
also need to be investigated. This fragility of the revivals is not surprising
if one realizes that the state vector $|\Psi(t)\rangle$
in Eq.(\ref{eq:au_temps_t}) is a Schr\"odinger cat at time $t_1/2$, that is a coherent
superposition of the $N$ particles in some state $\phi_1$ and of the $N$ particles
in some state $\phi_2$ orthogonal to $\phi_1$: the revival at time $t_1$
is suppressed if the Schr\"odinger cat at time $t_1/2$ is transformed by
decoherence into a statistical mixture of the states $|N:\phi_{1,2}\rangle$, 
which is difficult to avoid for large values of $N$ (see the lecture notes
of Michel Brune in this volume).

\subsubsection{An indicator of phase coherence}
To characterize the degree of phase correlation between the two condensates
it is natural to consider the average of $\langle \hat{a}^\dagger \hat{b}\rangle$
where $\hat{a},\hat{b}$ annihilate a particle in condensates $A$ and $B$ respectively.
Consider indeed the average over many experimental realizations of some one-body
observable $\hat{O}$ sensitive to the relative phase of the two condensates.  
This observable
necessarily has a non-vanishing matrix element between the modes
$\phi_a$ and $\phi_b$ so that in second quantized form the 
part of $\langle\hat{O}\rangle$ sensitive
to the relative phase involves 
$\langle \hat{a}^\dagger \hat{b}\rangle$.
E.g.\ in the case of spatially separated condensates one can
beat on a 50$-$50 matter waves beam splitter atoms leaking out of the condensates
and detect the atoms in the output channels of the beam splitter \cite{mesure_franges2};
the number of counts in the $+$
output channel averaged of many experimental realizations
is proportional to the expectation value of
\be
\hat{O}= \frac{\hat{a}^\dagger + \hat{b}^\dagger}{\sqrt{2}}
 \frac{\hat{a} + \hat{b}}{\sqrt{2}}.
\label{eq:O}
\ee
Expanding this product of operators we get \lq diagonal' terms such as 
$\hat{a}^\dagger \hat{a}$ not sensitive to the relative phase, and crossed
terms (actually interference terms!) 
such as $\hat{a}^\dagger \hat{b}$ sensitive to the phase.
In the JILA-type configuration, where the condensates
$A$ and $B$ are in different internal
atomic states, an observable $\hat{O}$ similar to Eq.(\ref{eq:O})
has been achieved by mixing the internal states of the two condensates
by a $\pi/2$ electromagnetic pulse and by measuring the mean density of atoms
in $A$ and $B$ \cite{JILA_phase}.

From the Schwartz inequality $|\langle u|v\rangle|\le||u||\; ||v||$ and setting 
$|u\rangle = \hat{a}|\Psi\rangle$, $|v\rangle = \hat{b}|\Psi\rangle$ we obtain
an upper bound for the expectation value of $\hat{a}^\dagger \hat{b}$:
\be
|\langle\Psi| \hat{a}^\dagger \hat{b}|\Psi\rangle| \leq 
\langle\Psi| \hat{a}^\dagger\hat{a}|\Psi\rangle^{1/2} 
\langle \Psi| \hat{b}^\dagger\hat{b}|\Psi\rangle^{1/2}.
\ee
The case of a maximally well defined relative phase corresponds to an equality
in this inequality, obtained if $|u\rangle$ and $|v\rangle$
are proportional. 
In the present situation of equal mean numbers of particles $N/2$
in $A$ and in $B$ this corresponds to $|\Psi\rangle$ being a phase state.

For an initial phase state
it is possible to calculate the expectation value of $\hat{a}^\dagger \hat{b}$ 
as function of time from the expansion (\ref{eq:au_temps_t}). One obtains after
simple transformations the sum
\be
\langle \hat{a}^\dagger \hat{b}\rangle (t) =\frac{N}{2^N}e^{-2i(\theta+vt)}
\sum_{N_a=0}^{N-1} \frac{(N-1)!}{N_a! (N_b-1)!}e^{i\chi t[N_a-(N_b-1)]}
\ee
with $N_b=N-N_a$ as in Eq.(\ref{eq:au_temps_t}). After inspection one realizes
that this sum is the binomial expansion of a $(N-1)-$th power so that
the final result is \cite{Walls}:
\be
\langle \hat{a}^\dagger \hat{b}\rangle (t) = \frac{N}{2} e^{-2i(\theta+vt)}
\cos^{N-1}\chi t.
\label{eq:cosinus}
\ee

From this very simple expression one can calculate the time $t_c$ after which
the relative phase has experienced a significant spread. For short times 
$\chi t\ll 1$ one can expand the cosine function in Eq.(\ref{eq:cosinus})
to second order in $t$:
\be
\cos^N\chi t = e^{N\log\cos\chi t}\simeq e^{-N(\chi t)^2/2}.
\ee
One obtains a Gaussian decay of phase coherence with a collapse time
\be
t_c= \frac{1}{|\chi|N^{1/2}}
\label{eq:tc}
\ee
equivalent to the rougher estimate Eq.(\ref{eq:tspread}) up to a numerical factor.
One can also easily see the revivals (reconstruction of $|\Psi\rangle$
to a phase state) at times $t_q= q\pi /\chi$ when the cosine function
is equal to $\pm 1$ in Eq.(\ref{eq:cosinus}).

Formula (\ref{eq:tc}) can be used to calculate the coherence time of the
relative phase of the condensates in the present zero-temperature model.
As an interesting application of this formula
we now show that the spreading time of the relative
phase can be significantly different for mutually interacting and non-mutually
interacting condensates. Assume for simplicity that the two condensates
are stored in cubic 
boxes of identical size $L$ and with periodic boundary conditions.
In the MIT-type configuration the two boxes are spatially separated and the atoms
are in the same internal state; the energy of a configuration with $N_a,N_b$
atoms in the condensates $A,B$ is then
\be
E = \frac{g}{2 L^3} \left[N_a^2 + N_b^2\right].
\ee
From Eq.(\ref{eq:chi}) this form of $E$ leads for an initial phase state
to a collapse time of the relative phase
\be
t_c = N^{1/2} \frac{\hbar}{2\rho g}
\label{eq:cond_separ}
\ee
where $\rho=N/(2L^3)$ is the mean spatial density in each of the condensates.
In the JILA-type configuration the atoms are in the same
spatial box but in different internal
states; the energy of a configuration with $N_a,N_b$ atoms in the condensates
$A,B$ is given now by Eq.(\ref{eq:demix}) if the two internal states are subject
to spatial demixing, or by Eq.(\ref{eq:unif}) if there is no demixing instability.
The collapse time is then given by
\be
\label{eq:tc_demix}
t_c = N^{1/2} \frac{\hbar}{\rho [g_{aa} + g_{bb} -2 (g_{aa} g_{bb})^{1/2}]}
\ee
for a demixed condensates and by
\be
\label{eq:tc_unif}
t_c = N^{1/2} \frac{\hbar}{\rho [g_{aa} + g_{bb} -2 g_{ab}]}
\ee
for fully overlapping condensates.
When the coupling constants among the various internal states are close to each other
the denominator in Eqs.(\ref{eq:tc_demix},\ref{eq:tc_unif}) can become small,
which results in a relative phase coherence time 
$t_c$ much larger than in the MIT-configuration
Eq.(\ref{eq:cond_separ}). This fortunate feature of close coupling constants 
is present for rubidium in the JILA experiment \cite{JILA_phase}!

In real life the condensates are usually stored in harmonic traps; 
the simple formulas
obtained for a cubic box have to be revisited. This has been done 
analytically for spatially
separated condensates \cite{Maciek1,Maciek2} and numerically for mutually
interacting condensates \cite{phase_mel,Bigelow}.
\section{Symmetry breaking description of condensates}
\markright{Broken symmetry}
\label{Cap:BS}

We have already seen in chapter \ref{Cap:PC} that it is very convenient,
physically, to introduce phase states to understand the phenomenon of interference
between two Bose-Einstein condensates: rather than assuming that
two Bose-Einstein condensates that 
\lq\lq have never seen each other" are in Fock states,
one assumes that they are in a phase state with a relative phase varying
in an unpredictable way for any new experimental
realization. One can even suppose that the condensates are in coherent states
of the atomic field; this description is said to \lq break the symmetry',
here the $U(1)$ symmetry associated to the invariance of the Hamiltonian 
by a change of the phase of the atomic field operator.

In this chapter we consider other examples of symmetry breaking descriptions:
$SO(3)$ symmetry breaking (case of spinor condensates)
and spatial translational symmetry breaking (case of one dimensional condensates with
attractive interactions). In both cases the procedure is the same:
the ground state of the system is
symmetric, its mean-field approximation by Hartree-Fock states breaks the symmetry.
In both cases we will consider Gedanken experiments whose single outcomes
can be predicted from the exact ground state and from the Hartree-Fock state.
This will illustrate the ability of the mean-field approximation
to allow physical predictions in an easy and transparent way,
correct in the limit of a large number of particles.

\subsection{The ground state of spinor condensates}
The alkali atoms used in the Bose-Einstein condensation experiments have an
hyperfine structure in the ground state, each hyperfine level having several Zeeman
sublevels. We have up to now ignored this structure in the lecture, as we were
implicitly assuming that the atoms were polarized in a well defined Zeeman sublevel.

Consider for example ${}^{23}$Na atoms used at MIT in the group of Wolfgang Ketterle.
The ground state has an hyperfine splitting between the lower multiplicity 
of angular momentum $F=1$ and the higher multiplicity of angular momentum $F=2$.
All the three Zeeman sublevels $m_F =0,\pm 1$ of the lower multiplicity $F=1$
cannot be trapped in a magnetic
trap (if $m_F=-1$ is trapped than $m_F=+1$ which experiences an opposite Zeeman
shift is antitrapped). But they can all be trapped in an optical dipole trap,
produced with a far off-resonance laser beam, as
the Zeeman sublevels experience then all the same lightshift.
This optical trapping was performed at MIT \cite{MIT_optical_trap}, opening the way to a series of
interesting experiments with condensates of particles 
of spin one \cite{MIT_spin1}.

We concentrate here on a specific aspect, the ground state of the spinor
condensate, assuming for simplicity that 
the atoms are stored in a cubic box with periodic
boundary conditions.

\subsubsection{A model interaction potential}
We have to generalize the model 
scalar pseudo-potential of Eq.(\ref{eq:ppf}) to the case of particles having
a spin different from zero. As we want to keep the simplicity of a contact
interaction potential we choose the simple form
\be
V(1,2) \equiv
{\cal V}_{\mbox{\scriptsize spin}}(1,2)
\delta(\vec{r_1}-\vec{r_2}) \left[{\partial\over\partial{r_{12}}}\left(r_{12}
\ \cdot \ \right)\right]
\ee
that is the product of an operator acting only on the spin of the particles 1 and 2,
and of the usual regularized contact interaction acting only on the relative motion
of the two particles. 
The interaction potential $V(1,2)$ has to be invariant 
by a simultaneous rotation of the spin variables and of the position variables 
of the two particles. As the contact interaction is already rotationally invariant,
the spin part of the interaction ${\cal V}_{\mbox{\scriptsize spin}}(1,2)$ has to
be invariant by any simultaneous rotation of the two spins. 

This condition of rotational invariance of ${\cal V}_{\mbox{\scriptsize spin}}(1,2)$
is easy to express in the coupled basis obtained by the addition of the two spins
of particle 1 and particle 2: within each subspace of well defined total angular
momentum ${\cal V}_{\mbox{\scriptsize spin}}(1,2)$ has to be a scalar. Let us restrict
to the case studied at MIT, 
with spin one particles. By addition of $F=1$ and $F=1$
we obtain a total angular momentum $F_{\mbox{\scriptsize tot}}=2$, 1 or 0, so that
one can write
\be
{\cal V}_{\mbox{\scriptsize spin}}(1,2)= g_2 P_{F_{\mbox{\scriptsize tot}}=2}(1,2)
+g_1 P_{F_{\mbox{\scriptsize tot}}=1}(1,2) + g_0 P_{F_{\mbox{\scriptsize tot}}=0}(1,2)
\ee
where the $g$'s are coupling constants and
the $P(1,2)$'s are projectors on the subspace of particles 1 and 2 with a well
defined total angular momentum $F_{\mbox{\scriptsize tot}}$.
At this stage we can play a little trick, using the fact that the states of
$F_{\mbox{\scriptsize tot}}=1$ are antisymmetric by the exchange of particles 1 and 2 (whereas
the other subspaces are symmetric). The regularized contact interaction scatters
only in the $s$-wave, where the external wavefunction of atoms 1 and 2
is even by the exchange of the positions $\vec{r}_1$ and $\vec{r}_2$; as our atoms are
bosons, the spin part has also to be symmetric by exchange of the spins of atoms
1 and 2 so that the \lq fermionic' part of ${\cal V}_{\mbox{\scriptsize spin}}(1,2)$, that is
in the subspace $F_{\mbox{\scriptsize tot}}=1$, has no effect. We can therefore change
$g_1$ at will without affecting the interactions between bosons. The most convenient
choice is to set $g_1=g_2$ so that we obtain
\be
{\cal V}_{\mbox{\scriptsize spin}}(1,2)= g_2 \mbox{Id}(1,2) 
+(g_0-g_2) P_{F_{\mbox{\scriptsize tot}}=0}(1,2) 
\label{eq:Vs}
\ee
where $\mbox{Id}$ is the identity.
The subspace $F_{\mbox{\scriptsize tot}}=0$ is actually of dimension one, and it
is spanned by the vanishing total angular momentum state $|\psi_0(1,2)\rangle$. Using the
standard basis $|m=-1,0,+1\rangle$ of single particle
angular momentum with $z$ as quantization axis, one can write
\be
|\psi_0(1,2)\rangle = -\frac{1}{\sqrt{3}}\left[ |+1,-1\rangle+|-1,+1\rangle -|0,0\rangle\right].
\label{eq:spin0_s}
\ee
A more symmetric writing is obtained in the single particle angular momentum
basis $|x,y,z\rangle$ used in chemistry,  defined by
\bea
|+1\rangle &=& -\frac{1}{\sqrt{2}}\left( |x\rangle + i|y\rangle\right) \\
|-1\rangle &=& +\frac{1}{\sqrt{2}}\left( |x\rangle - i|y\rangle\right) \\
|0\rangle &=& |z\rangle.
\eea
The vector $|\alpha\rangle$ in this basis $(\alpha=x,y,z)$
is an eigenvector of angular momentum along axis
$\alpha$ with the eigenvalue zero. One then obtains
\be
|\psi_0(1,2)\rangle = \frac{1}{\sqrt{3}} \left[ |x,x\rangle +|y,y\rangle + |z,z\rangle
\right].
\label{eq:spin0_c}
\ee

To summarize the part of the Hamiltonian describing the interactions between
the particles can be written, if one forgets for simplicity the regularizing
operator in the pseudo-potential:
\bea
H_{\mbox{\scriptsize int}} &=& \frac{g_2}{2} \int d^3\vec{r}\,
\sum_{\alpha,\beta=x,y,z} \hat{\psi}_\alpha^\dagger \hat{\psi}_\beta^\dagger
\hat{\psi}_\beta \hat{\psi}_\alpha  \nonumber \\
&& + \frac{g_0-g_2}{6}\int d^3\vec{r}\, \sum_{\alpha,\beta=x,y,z}
\hat{\psi}_\alpha^\dagger\hat{\psi}_\alpha^\dagger
\hat{\psi}_\beta\hat{\psi}_\beta.
\label{eq:model_hamil}
\eea
where $\hat{\psi}_{\alpha}(\vec{r}\,)$ is the atomic field operator
for the spin state $|\alpha\rangle$.
This model Hamiltonian has also been proposed by \cite{Ho,Zhang,Machida}.

\subsubsection{Ground state in the Hartree-Fock approximation}
As we are mainly interested in the spin contribution to the energy we assume for
simplicity that the condensate is in a cubic box of size $L$ with periodic
boundary conditions. We assume that the interactions between
the atoms are repulsive ($g_2,g_0\geq 0$) and we suppose that there is no magnetic field
applied to the sample.

We now minimize the energy of the condensate within the Hartree-Fock trial 
statevectors $|N_0:\phi\rangle$ with the constraint that the number of particles
$N_0$ is fixed ($|\phi\rangle$ is normalized to unity)
but without any constraint on the total angular momentum of the spins.
The external part of the condensate wavefunction is simply the plane wave with momentum
$\vec{k}=\vec{0}$ whereas the spinor part of the wavefunction remains
to be determined:
\be
\langle \vec{r}\,|\phi\rangle = \frac{1}{L^{3/2}}\sum_{\alpha=x,y,z} c_\alpha |\alpha\rangle
\ \ \ \mbox{with}\ \ \ \sum_\alpha |c_\alpha|^2=1.
\label{eq:cond_norm_spin}
\ee
Using the model interaction Hamiltonian Eq.(\ref{eq:model_hamil}) 
we find for the mean energy
per particle in the condensate
\be
\frac{E}{N_0} = \frac{N_0-1}{2L^3} g_2 +\frac{N_0-1}{6L^3} (g_0-g_2) |A|^2
\label{eq:ener_spin}
\ee
where we have introduced the complex quantity
\be
A = \sum_{\alpha=x,y,z} c_\alpha^2=\vec{c}\,^2
\ee
where $\vec{c}$ is the vector of components $(c_x,c_y,c_z)$.
We have to minimize the mean energy over the state of the spinor.

\begin{itemize}
\item Case $g_2> g_0$
\end{itemize}
This is the case of sodium \cite{MIT_spin1}. As the coefficient $g_0-g_2$ is negative
in Eq.(\ref{eq:ener_spin}) we have to maximize the modulus of the complex quantity
$A$. As the modulus of a sum is less than the sum of the moduli we immediately
get the upper bound
\be
|A|\leq \sum_{\alpha=x,y,z} |c_\alpha|^2 =1
\ee
leading to the minimal energy per particle
\be
\frac{E}{N_0} = \frac{N_0-1}{2L^3}g_2  +\frac{N_0-1}{6L^3}(g_0-g_2).
\ee
The upper bound for $|A|$
is reached only if all complex numbers $c_\alpha^2$ have the same
phase modulo $2\pi$. This means that one can write 
\be
c_\alpha = e^{i\theta} n_\alpha
\label{eq:def_n}
\ee
where $\theta$ is a constant phase and $\vec{n}=(n_x,n_y,n_z)$ is any unit vector
with real components. Physically this corresponds to a spinor condensate wavefunction
being the zero angular momentum state for a quantization axis pointing in the direction
of $\vec{n}$. The direction of $\vec{n}$ is well defined in the Hartree-Fock
ansatz, but it is arbitrary as no spin direction is privileged by the Hamiltonian.
We are facing symmetry breaking, here a rotational $SO(3)$ symmetry breaking,
as we shall see.

\begin{itemize}
\item Case $g_2< g_0$
\end{itemize}
In this case we have to minimize $|A|$ to get the minimum of energy. The minimal value
of $|A|$ is simply zero, corresponding to spin configurations such that
\be
\vec{c}\,^2 \equiv \sum_{\alpha=x,y,z} c_\alpha^2 =0
\label{eq:carre_zero}
\ee
with an energy per condensate particle
\be
\frac{E}{N_0} = \frac{N_0-1}{2L^3} g_2.
\ee
To get more physical understanding we split the vector $\vec{c}$ as
\be
\vec{c} = \vec{R} + i\vec{I}
\ee
where the vectors $\vec{R}$ and $\vec{I}$ have purely real components. Expressing the 
fact that the real part and imaginary part of $\vec{c}\,^2$ vanish, and 
using the normalization condition 
$\vec{c}\cdot\vec{c}\,^*=1$ in Eq.(\ref{eq:cond_norm_spin}) we finally obtain
\bea
\vec{R}\cdot\vec{I} &=& 0 \\
\vec{R}\,^2 &=& \vec{I}\,^2 = \frac{1}{2}.
\eea
This means that the complex vector $\vec{c}$ is circularly polarized with respect
to the axis $Z$ orthogonal to $\vec{I}$ and $\vec{R}$. Physically this corresponds
to a spinor condensate wavefunction having an angular momentum $\pm \hbar$ along
the axis $Z$. The direction of axis $Z$ is well defined in the Hartree-Fock ansatz
but it is arbitrary.  

\subsubsection{Exact ground state of the spinor part of the problem}
Imagine that we perform some intermediate approximation, assuming that
the particles are all in the ground state $\vec{k}=\vec{0}$ of the box 
but not assuming that they are all in the same spin state. We then have to diagonalize
the model Hamiltonian 
\be
H_{\mbox{\scriptsize spin}} = \frac{g_2}{2L^3} \sum_{\alpha,\beta=x,y,z} 
\hat{a}_\alpha^\dagger \hat{a}_\beta^\dagger \hat{a}_\beta \hat{a}_\alpha 
+\frac{1}{6L^3}(g_0-g_2)\hat{A}^\dagger \hat{A}
\label{eq:hspin}
\ee
where $\hat{a}_\alpha$ annihilates a 
particle in state $|\vec{k}=0\rangle|\alpha\rangle$
($\alpha=x,y,z$) and where we have introduced
\be
\label{eq:definit_A}
\hat{A} = \hat{a}_x^2+\hat{a}_y^2+\hat{a}_z^2.
\ee
Up to a numerical factor 
$\hat{A}$ annihilates a pair of particles in the two-particle spin state 
$|\psi_0(1,2)\rangle$
of vanishing total angular momentum, as shown by Eq.(\ref{eq:spin0_c}).

The Hamiltonian  Eq.(\ref{eq:hspin}) can be diagonalized exactly \cite{edhs}. This
is not surprising as (i) it is rotationally invariant and (ii) the 
bosonic $N_0-$particle
states with a well defined total angular momentum $S_{N_0}$ can be calculated:
one finds that $S_{N_0}=N_0,N_0-2,\ldots$, leading to degenerate multiplicities
of $H_{\mbox{\scriptsize spin}}$ of degeneracy $2S_{N_0}+1$. 

In practice one
may use the following tricks: The double sum proportional to
$g_2$ in Eq.(\ref{eq:hspin}) can be expressed in terms of the operator number $\hat{N}_0$ 
of condensate particles only, 
\be
\hat{N}_0=\sum_\alpha \hat{a}_\alpha^\dagger \hat{a}_\alpha.
\ee
So diagonalizing $H_{\mbox{\scriptsize spin}}$ amounts to
diagonalizing $\hat{A}^\dagger \hat{A}$!

Second the total momentum operator $\hat{\vec{S}}$
of the $N_0$ spins, defined as the sum of all the spin operators of
the individual atoms in units of $\hbar$,
can be checked to satisfy the identity 
\be
\hat{\vec{S}}\cdot\hat{\vec{S}} + \hat{A}^\dagger \hat{A} = \hat{N_0}(\hat{N_0}+1) 
\label{eq:iden_avec_s}
\ee
so that the Hamiltonian for $N_0$ particles becomes a function of $\hat{\vec{S}}$
\cite{edhs}:
\be
H_{\mbox{\scriptsize spin}} = \frac{g_2}{2L^3} \hat{N_0}(\hat{N_0}-1)
+\frac{1}{6L^3}(g_0-g_2)\left[\hat{N_0}(\hat{N_0}+1) - \hat{\vec{S}}\cdot\hat{\vec{S}}\right].
\label{eq:hspin_simple}
\ee
We recall that $\hat{\vec{S}}\cdot\hat{\vec{S}}=S_{N_0}(S_{N_0}+1)$ within the subspace of total spin
$S_{N_0}$.

When $g_2<g_0$ the ground state of $H_{\mbox{\scriptsize spin}}$ corresponds to the
multiplicity $S_{N_0}=N_0$, containing {\sl e.g.} the state with all the
spins in the state $|+\rangle$. In this case the $N_0-$particle states obtained
with the Hartree-Fock approximation 
are exact eigenstates of $H_{\mbox{\scriptsize spin}}$.

When $g_2>g_0$ the ground state of $H_{\mbox{\scriptsize spin}}$ corresponds to the 
multiplicity of minimal total angular momentum, $S_{N_0}=1$ for $N_0$ odd
or $S_{N_0}=0$ for $N_0$ even. In this case the Hartree-Fock state is a symmetry
breaking approximation of the exact ground state of $H_{\mbox{\scriptsize spin}}$. 
The error on the energy per
particle tends to zero in the thermodynamical limit; for $N_0$ even one finds indeed
\be
\frac{\delta E}{N_0} = -\frac{1}{3L^3}(g_0-g_2).
\ee

But what happens if one restores the broken symmetry by summing up the Hartree-Fock
ansatz over the direction $\vec{n}$ defined in Eq.(\ref{eq:def_n})? Assume that
$N_0$ is even; one has then to reconstruct from the Hartree-Fock ansatz a rotationally
invariant state. This amounts
to considering the following normalized state for the $N_0$ spins:
\be
|\Psi\rangle =  \sqrt{N_0+1}
\int \frac{d^2\vec{n}}{4\pi}\, |N_0:\vec{n}\,\rangle
\label{eq:exact}
\ee
where $d^2\vec{n}$ indicates the integration over the unit sphere (that is over
all solid angles) and $|N_0:\vec{n}\,\rangle$ is the state with $N_0$ particles
in the single particle state
\be
|\vec{n}\,\rangle = n_x|x\rangle +n_y|y\rangle + n_z|z\rangle.
\ee
The state vector $|\Psi\rangle$, being non zero and having a vanishing 
total angular momentum,
is equal to the exact ground state of $H_{\mbox{\scriptsize spin}}$!

The expression (\ref{eq:exact}) can be used as a starting point to obtain
various forms of $|\Psi\rangle$. If one expresses the Hartree-Fock state as the 
$N_0$-th power of the creation operator $\sum_\alpha \hat{a}_{\alpha}^{\dagger} n_{\alpha}$
acting on the vacuum $|\mbox{vac}\rangle$,
and if one expands this power with the usual binomial formula, the integral
over $\vec{n}$ can be calculated explicitly term by term and one obtains:
\be
|\Psi\rangle = {\cal N} \left(\hat{A}^\dagger\right)^{N_0/2} |\mbox{vac}\rangle
\label{eq:cdp}
\ee
where $\cal N$ is a normalization factor and the operator
$\hat{A}$ is defined in Eq.(\ref{eq:definit_A}). 
Formula (\ref{eq:cdp}) indicates that $|\Psi\rangle$
is simply a \lq condensate' of pairs in the pair state $|\psi_0(1,2)\rangle$. 
It can be used
to expand $|\Psi\rangle$ over Fock states with a well 
defined number of particles in the
modes $m=0,m=\pm 1$, reproducing Eq.(13) of \cite{edhs}.

To be complete we mention another way of constructing the exact eigenvectors
and energy spectrum of $H_{\mbox{\scriptsize spin}}$. The idea is to diagonalize $\hat{A}^\dagger
\hat{A}$ using the fact that $\hat{A}$ obeys a commutation relation 
that is reminiscent of that
of an annihilation operator:
\be
[\hat{A},\hat{A}^\dagger ] = 4\hat{N}_0+6.
\ee
In this way $\hat{A}^\dagger$ acts as a raising operator: acting on an eigenstate 
of $\hat{A}^\dagger\hat{A}$ with eigenvalue $\lambda$ and $N_0$ particles,
it gives an eigenstate of $\hat{A}^\dagger\hat{A}$ with eigenvalue $\lambda+4 N_0+6$
and with $N_0+2$ particles. One can also check from the identity (\ref{eq:iden_avec_s})
that the action of $\hat{A}^\dagger$ does not change the total spin:
\be
[\hat{A}^\dagger, \hat{\vec{S}}\cdot \hat{\vec{S}} ]=0.
\ee
By repeated actions of $\hat{A}^\dagger$ starting from
the vacuum one arrives at Eq.(\ref{eq:cdp}), creating the eigenstates with $N_0$ even
and vanishing total spin $S=0$. By repeated actions of $\hat{A}^\dagger$ starting from
the eigenstates with $N_0=2$ and total spin $S=2$ (e.g.\ the state $|++\rangle$) one
obtains all the states with $N_0$ even and total spin $S=2$. 
More generally the eigenstate of $H_{\mbox{\scriptsize spin}}$ with total spin $S$, a spin
component $m=S$ along $z$ and $N_0$ particles is:
\be
||N_0,S,m=S\rangle \propto \left(\hat{A}^\dagger\right)^{(N_0-S)/2} 
|S:+1\rangle
\label{eq:form_gene}
\ee
where $|S:+1\rangle$ represents $S$ particles in the state $|+1\rangle$.
From Eq.(\ref{eq:form_gene})
one can generate the states with spin components $m=S-1,\ldots,-S$ 
by repeated actions of the spin-lowering operator $\hat{S}_-=\hat{S}_x-i\hat{S}_y$
in the usual way.
 We note that formula (\ref{eq:form_gene}) was derived
independently in \cite{Ho_preprint}.

\subsubsection{Advantage of a symmetry breaking description}
Imagine that we have prepared a condensate of sodium atoms ($g_2>g_0$)
in the collective ground spin state, and that we
let the atoms leak one by one out of the trap, in a way that does not
perturb their spin. We then measure the spin component along $z$ of the outgoing
atoms. Suppose that we have performed this measurement on $k$ atoms, with
$k\ll N_0$. We then raise
the simple question: what is the probability $p_k$ that all the $k$ detections give
a vanishing angular momentum along $z$?

Let us start with a naive reasoning based on the one-body density matrix
of the condensate (even if the reader has been warned already in \S\ref{subsubsec:trap}
on the dangers of such an approach!).
The mean occupation numbers of the single particle spin states $|m=-1\rangle$,
$|m=0\rangle$ and $|m=+1\rangle$ in the initial condensate are obviously all equal to $N_0/3$, 
as the condensate is initially in a rotationally symmetric state. The probability of detecting
the first leaking atom in $|m=0\rangle$ is therefore $1/3$. Naively we assume that since
$k\ll N_0$ the detections have a very weak effect on the state of the condensate
and the probability of detecting the $n$-th atom ($n\leq k$)
in the $m=0$ channel is nearly independent of the $n-1$ previous detection results.
The probability for $k$ detections in the $m=0$ channel should then be
\be
p_k^{\mbox{\scriptsize naive}} = \frac{1}{3^k}.
\label{eq:Naive}
\ee

Actually this naive reasoning is wrong (and by far) as soon as $k\geq 2$. The first detection
of an atom in the $m=0$ channel projects 
the spin state of the remaining atoms in
\be
|\Psi_1\rangle = {\cal N}_1 \hat{a}_0 |\Psi\rangle
\ee
where $\hat{a}_0$ annihilates an atom in spin state $m=0$, $|\Psi\rangle$ is the collective
spin ground state (\ref{eq:exact}) and ${\cal N}_1$ is a normalization factor.
The probability of detecting the second atom in $m=0$ (knowing that the first atom
was detected in $m=0$) is then given by
\be
\frac{p_2}{p_1} = \frac{\langle\Psi_1|\hat{a}_0^\dagger\hat{a}_0|\Psi_1\rangle}
{\langle\Psi_1|\sum_{m=-1}^{+1} \hat{a}_m^\dagger\hat{a}_m|\Psi_1\rangle}.
\ee
The denominator is simply equal to $N_0-1$ as $|\Psi_1\rangle$ is a state with
$N_0-1$ particles. Using the integral form (\ref{eq:exact}) and the simple
effect of an annihilation operator on a Hartree-Fock state, {\sl e.g.}
\be
\hat{a}_0^2|N_0:\vec{n}\rangle =\left[N_0(N_0-1)\right]^{1/2}n_z^2 |N_0-2:\vec{n}\rangle
\ee
we are able to express the probability in terms of integrals over solid angles:
\be
\frac{p_2}{p_1} = \frac{\displaystyle\int d^2\vec {n} \int d^2\vec{n}\,' \ 
n_z^2 n_z'^2(\vec{n}\cdot\vec{n}\,')^{N_0-2}}
{\displaystyle\int d^2\vec {n} \int d^2\vec{n}\,' \ n_z n_z'
(\vec{n}\cdot\vec{n}\,')^{N_0-1}}.
\ee
We suggest the following procedure to calculate these integrals. One first integrates
over $\vec{n}\,'$ for a fixed $\vec{n}$, using spherical coordinates relative to
the \lq vertical' axis directed along $\vec{n}$: the polar angle $\theta'$ is then the 
angle between $\vec{n}\,'$ and $\vec{n}$ so that one has simply
$\vec{n}\cdot\vec{n}\,'=\cos\theta'$. The integral over $\theta'$ and over the azimuthal
angle $\phi'$ can be performed, giving a result involving only $n_z$. The remaining integral
over $\vec{n}$ is performed with the spherical coordinates of vertical axis $z$.
This leads to
\be
\frac{p_2}{p_1} = \frac{3}{5} +\frac{2}{5(N_0-1)}.
\ee

The ratio $p_2/p_1$ is therefore different from the naive (and wrong!)
prediction (\ref{eq:Naive}). For $N_0=2$ one finds $p_2/p_1=1$ so that
the second atom is surely in $m=0$ if the first atom was detected in $m=0$.
As the two atoms were initially in the state with total angular momentum
zero, this result could be expected from the expression (\ref{eq:spin0_s}) of
the two-particle spin state. In the limit of large $N_0$ we find
that once the first atom has been detected in the $m=0$ channel, the probability
for detecting the second atom in the same channel $m=0$ is $3/5$. This somehow
counter-intuitive result shows that the successive detection probabilities
are strongly correlated in the case of the spin state (\ref{eq:exact}).

The exact calculation of the ratio
\be
\frac{p_{k+1}}{p_k} = \frac{\displaystyle\int d^2\vec {n} \int 
d^2\vec{n}\,' \ 
n_z^{k+1} n_z'^{k+1}(\vec{n}\cdot\vec{n}\,')^{N_0-(k+1)}}
{\displaystyle \int d^2\vec {n} \int d^2\vec{n}\,' \ n_z^k n_z'^k
(\vec{n}\cdot\vec{n}\,')^{N_0-k}}
\ee
is getting more difficult when $k$ increases. The large $N_0$ limit
for a fixed $k$ is easier to obtain: in the integral over $\vec{n}\,'$
the function $(\vec{n}\cdot\vec{n}\,')^{N_0-(k+1)}$
is extremely peaked around $\vec{n}'=\vec{n}$ so that we can replace $n_z'^{k+1}$
by $n_z^{k+1}$. This leads to
\be
\lim_{N_0\rightarrow+\infty} \frac{p_{k+1}}{p_k} = \frac{2k+1}{2k+3}.
\label{eq:exact_ratio}
\ee

We now give the reasoning in the symmetry breaking point of view, which
assumes that a single experimental realization of the condensate corresponds
to a Hartree-Fock state $|N_0:\vec{n}\rangle$ with the direction
$\vec{n}$ being an impredictable random variable with uniform distribution
over the sphere. If the system is initially in the spin state $|N_0:\vec{n}\rangle$
there is no correlation between the spins, and the probability of having $k$ detections
in the channel $m=0$ is simply $(n_z^2)^{k}$. One has to average over the unknown 
direction $\vec{n}$ to obtain
\be
p_k^{\mbox{\scriptsize sb}} = \int \frac{d^2\vec{n}}{4\pi} n_z^{2k} = \frac{1}{2k+1}.
\label{eq:sb}
\ee
One recovers in an easy calculation the large $N_0$ limit of the exact result,
Eq.(\ref{eq:exact_ratio})!
We note that the result (\ref{eq:sb}) is much larger than the
naive (and wrong) result (\ref{eq:Naive}) as soon as $k\gg 1$.

\subsection{Solitonic condensates}
We consider in this section a Bose-Einstein condensate 
with effective attractive interactions
subject to a strong confinement in the $x-y$ plane so that it constitutes
an approximate one-dimensional interacting Bose gas along $z$. 
Such a situation is interesting
physically as it gives rise in free space
to the formation of \lq bright' solitons well
known in optics but not yet observed with atoms.
Also the model of a one-dimensional Bose gas with a $\delta$ interaction
potential has known exact solutions in free space, that can be used to test
the translational symmetry breaking Hartree-Fock approximation. 

\subsubsection{How to make a solitonic condensate ?}
\label{subsubsec:sol_mf}
Consider a steady state condensate with effective attractive interactions
in a three dimensional harmonic trap. The confinement in the $x-y$ plane is
such that the transverse quanta of oscillation  $\hbar\omega_{x,y}$ are much larger
than the typical mean field energy per particle $N_0 |g| |\phi|^2$, where $\phi$ is
the condensate wavefunction with $N_0$ particles. This confinement prevents
the occurrence of a spatial collapse of the condensate (see \S\ref{subsubsec:tic}). 
The confinement is however
not strong enough to violate the validity condition of the Born approximation
for the pseudo-potential,
$k |a|\ll 1$ with $k\simeq (m\omega_{x,y}/\hbar)^{1/2}$.

In this case we face a quasi
one-dimensional situation, 
where the condensate wavefunction is approximately factorized
as
\be
\phi(x,y,z) = \psi(z)\chi_x(x)\chi_y(y)
\label{eq:fact}
\ee
where $\chi_x$ and $\chi_y$ are the normalized ground states of the harmonic oscillator
along $x$ and along $y$ respectively. By inserting the factorized form (\ref{eq:fact})
in the Gross-Pitaevskii energy functional Eq.(\ref{eq:GPef}) and by integrating over
the directions $x$ and $y$ we obtain an energy functional for $\psi$:
\be
E[\psi,\psi^*]= N_0 \int dz \left[ {\hbar^2\over 2m}\left|\frac{d\psi}{dz}\right|^2 +
\frac{1}{2}m\omega_z^2 z^2|\psi(z)|^2 +{1\over 2}N_0g_{1d}|\psi(z)|^4\right]
\label{eq:GPef1d}
\ee
where we have dropped the zero-point energy of the transverse motion 
and we have called $g_{1d}$ the quantity
\be
g_{1d} = g\int dx\int dy\; |\chi_x(x)|^4 |\chi_y(y)|^4
=g\frac{m(\omega_x\omega_y)^{1/2}}{2\pi\hbar}.
\ee
The corresponding time independent Gross-Pitaevskii equation for
$\psi$ is
\be
\mu \psi(z) =-\frac{\hbar^2}{2m}\frac{d^2\psi}{dz^2} + \left[
\frac{1}{2}m\omega_z^2 z^2 + N_0 g_{1d}  |\psi(z)|^2\right] \psi(z).
\label{eq:gpe1d}
\ee
The energy functional Eq.(\ref{eq:GPef1d}) corresponds to
a one-dimensional interacting Bose gas
with an effective coupling constant between the atoms equal to $g_{1d}$, that is
one can imagine that the particles have a binary contact interaction 
\be
V(z_1,z_2) = g_{1d} \delta(z_1 - z_2).
\label{eq:pot1d}
\ee
Note that such a Dirac interaction potential leads to a perfectly well defined 
scattering problem in one dimension, contrarily to the three dimensional case.

Imagine now that we slowly decrease the
trap frequency along $z$ while keeping intact the transverse trap frequencies,
until $\omega_z$ vanishes. What will happen then?
If $g$ was positive the cloud would simply expand without limit along $z$.
With attractive interaction the situation is dramatically different:
due to the slow evolution of $\omega_z$ 
the condensate wavefunction will follow adiabatically the minimal energy
solution of the Gross-Pitaevskii equation. For $\omega_z=0$ this minimal energy
solution is the so-called bright soliton, well known in non-linear optics.
We recall the analytic form of the solitonic wavefunction:
\be
\psi(z) = \frac{1}{(2l)^{1/2}} \; \frac{1}{\cosh(z/l)}
\label{eq:soli}
\ee
where $l$ is the spatial radius of the soliton:
\be
l= -\frac{2\hbar^2}{N_0 m g_{1d}}.
\ee
Note that this size $l$ results of a compromise between minimization of
kinetic energy by an increase of the size and minimization of interaction energy by 
a decrease of the size, so that the typical kinetic energy per particle
$\hbar^2/(m l^2)$ is roughly opposite to the interaction energy per
particle  $N_0 g_{1d}/l$.
We also give the corresponding chemical potential:
\be
\mu = -\frac{1}{8} N_0^2 \frac{m g_{1d}^2}{\hbar^2}.
\ee

We briefly address the validity of the Gross-Pitaevskii solution 
(\ref{eq:soli}).
As we have pointed out in the three dimensional case  
(see for example \S\ref{subsubsec:why_not})
we wish that the Born approximation for the interaction potential
be valid.
In one dimension the $\delta$ interaction potential can be treated in the Born
approximation only if the relative wavevector of the colliding particles is
{\bf high} enough (in contrast to the three-dimensional case): 
\be
\left|\frac{\hbar^2k}{m g_{1d}}\right| \gg 1.
\ee
This condition can be obtained of course from a direct calculation, but also from
a dimensionality argument ($m g_{1d}/\hbar^2$ is the {\bf inverse} of a length)
and from the fact that the Born approximation should apply in the
limit $g_{1d}\rightarrow 0$ for a fixed $k$.
If we use the estimate $k \simeq 1/l$ we obtain the condition
\be
-\frac{\hbar^2}{m g_{1d}l} \simeq N_0 \gg 1,
\label{eq:condi}
\ee
implicitly valid here as we started from a condensate!

Another phenomenon neglected in the prediction (\ref{eq:soli}) is the spreading
of the center of mass coordinate during the switch-off of the trapping
potential along $z$. Whereas Eq.(\ref{eq:soli})
assumes that the abscissa of the center of the soliton $z_0$ is exactly $0$
the spreading of the center of mass leads in real life to
a finite width probability distribution for $z_0$.
This spreading can be calculated simply for an almost
pure condensate $N_0\simeq N$, using the fact that the 
center of mass coordinate operator $\hat{Z}$ and the total momentum operator
$\hat{P}$ of the gas along 
$z$ axis
are decoupled from the relative
coordinates of the particles in a harmonic
potential, in presence of interactions depending only on the relative
coordinates. To prove this assertion one expresses the operators
$\hat{Z}$ and $\hat{P}$ 
in terms of the position and momentum operators of each particle $i$ of the
gas:
\bea
\label{eq:def_hat_Z}
\hat{Z} &=& \frac{1}{N} \sum_{i=1}^{N} z_i \\
\hat{P} &=& \sum_{i=1}^{N} p_i
\label{eq:def_hat_P}
\eea
and one derives the following equations of motion in Heisenberg point of
view:
\bea
\frac{d\hat{Z}}{dt} &=& \frac{\hat{P}}{Nm} \\
\frac{d\hat{P}}{dt} &=& -Nm\omega_z^2(t) \hat{Z}.
\eea
The spreading acquired by $\hat{Z}$ is not negligible when it becomes comparable
to the size $l$ of the soliton. 

The spreading of $\hat{Z}$ is interesting to calculate in the absence 
of harmonic confinement along $z$, $\omega_z\equiv 0$,
with the simple assumption that all the particles
of the gas are at time $t=0$
in the soliton state $|\psi\rangle $ of Eq.(\ref{eq:soli}).
As $\hat{P}$ is a constant of motion for $\omega_z=0$ one has simply 
\be
\hat{Z}(t) = \hat{Z}(0) + \frac{\hat{P}t}{Nm}
\ee
so that the variance of the center of mass coordinate at time $t$
is
\be
\mbox{Var}(\hat{Z})(t) = \mbox{Var}(\hat{Z})(0)+\frac{t}{Nm}
\langle \hat{Z}(0)\hat{P}+\hat{P}\hat{Z}(0)\rangle +
\frac{t^2}{N^2 m^2} \mbox{Var}(\hat{P}).
\ee
One then replaces $\hat{Z}(0)$ and $\hat{P}$ by the sums 
(\ref{eq:def_hat_Z},
\ref{eq:def_hat_P}). As the single particle wavefunction $\psi$ has vanishing
mean position and mean momentum all the `crossed terms' 
expectation values involving two different
particles vanish. As $\psi(z)$ is a real wavefunction one finds
also $\langle \psi|zp+pz|\psi\rangle=0$ so that the contribution linear
in time vanishes. One is left with
\be
\mbox{Var}(\hat{Z})(t)=\frac{1}{N}\langle\psi|z^2|\psi\rangle +
\frac{t^2}{Nm^2}\langle\psi|p^2|\psi\rangle.
\ee
The variance of $\hat{Z}$, initially $N$ times smaller than the single
particle variance $\langle\psi|z^2|\psi\rangle$, becomes equal to
the single particle variance after a time
\be
t_c = \left(\frac{Nm^2\langle\psi|z^2|\psi\rangle}
{\langle\psi|p^2|\psi\rangle}\right)^{1/2}= N^{1/2} \frac{\pi m l^2}{2\hbar}
\label{eq:tcsol}
\ee
where we used the explicit expressions
\bea
\langle \psi |z^2 |\psi\rangle &=& \frac{\pi^2 l^2}{12} \\
\langle \psi |p^2 |\psi\rangle &=& \frac{\hbar^2}{3 l^2}.
\label{eq:p2_soliton}
\eea
The spreading phenomenon of the position of the soliton
is formally equivalent to the spreading of
the relative phase of two condensates initially prepared in a phase
state (see \S\ref{subsec:whatisphaseevol}). The critical time $t_c$
in (\ref{eq:tcsol}) scales as $N^{1/2}\hbar/|\mu|$ as in
Eq.(\ref{eq:cond_separ}).

\subsubsection{Ground state of the one-dimensional attractive Bose gas}
We consider here the model of the one-dimensional gas of $N$ bosonic
particles interacting 
with the contact potential Eq.(\ref{eq:pot1d}) and in the
absence of any confining potential. 

It turns out that in this model with $g_{1d}>0$
one can calculate exactly the eigenenergies and eigenstates of
the Hamiltonian for $N$ particles using the Bethe
ansatz \cite{Gaudin}. We consider here the less studied attractive case $g_{1d}<0$, 
where several exact results are also available.
In particular the exact expression for the ground state energy
is known \cite{Guire}:
\be
E_0(N) = -\frac{1}{24} \frac{m g_{1d}^2}{\hbar^2} N (N^2-1)
\label{eq:exe}
\ee
and the corresponding $N-$particle wavefunction
of the ground state is \cite{Herzog}:
\be
\Psi(z_1,\ldots,z_N) = {\cal N}
\exp\left[\frac{m g_{1d}}{2\hbar^2}\sum_{1\leq i < j\leq N} |z_i-z_j|\right].
\label{eq:psi_ex}
\ee
To determine the normalization factor $\cal N$ we enclose the gas in a 
fictitious box of size $L$ tending to $+\infty$:
\footnote{
The center of mass of the gas corresponds to a fictitious particle of wavevector
$K$, where $\hbar K$ is the total momentum of the gas, and of position $Z$,
where $Z$ is the centroid of the gas. In the ground state $|\Psi\rangle$
the center of mass is completely delocalized with $K=0$. The
factor $1/L$ in $|{\cal N}|^2$ originates
from the normalization of the fictitious
particle plane wave in the fictitious box of size $L$,
$\langle Z |K\rangle = e^{i K Z}/\sqrt{L}$.
The more correct mathematical way (not used here) is to normalize in free
space (no box) using the closure relation
$\int dK |K\rangle\langle K| = \mbox{Id}$, which amounts to replace
$L$ by $2\pi$.
}
\be
|{\cal N}|^2 = \frac{(N-1)!}{NL}\left(\frac{m|g_{1d}|}{\hbar^2}\right)^{N-1}.
\ee

To what extent can we recover these results using a Hartree-Fock ansatz
$|N:\psi\rangle$
for the ground state wavefunction? As discussed around Eq.(\ref{eq:GPef_bis})
we get a mean energy for the Hartree-Fock state very similar to Eq.(\ref{eq:GPef1d}):
\be
E[\psi,\psi^*]= N \int dz \left[ {\hbar^2\over 2m}\left|\frac{d\psi}{dz}\right|^2 +
{1\over 2}(N-1)g_{1d}|\psi(z)|^4\right].
\ee
We minimize this functional using the results of \S\ref{subsubsec:sol_mf}, replacing
$N_0$ by $N-1$, and we obtain
\be
E_0^{\mbox{\scriptsize hf}}(N) = -\frac{1}{24} \frac{m g_{1d}^2}{\hbar^2} N (N-1)^2.
\label{eq:hfe}
\ee
The deviation of the Hartree-Fock result from the exact result is a fraction $1/N$
of the energy and is small indeed in the large $N$ limit, as expected
from the validity condition (\ref{eq:condi})!

There is a notable difference of translational properties however.
Whereas the exact ground state (\ref{eq:psi_ex}) is invariant by a global translation
of the positions of the particles, as it should be, the Hartree-Fock ansatz
leads to condensate wavefunctions $\psi$ localized within the length $l$ around some
arbitrary point $z_0$ (around $z_0=0$ in Eq.(\ref{eq:soli})):
\be
\psi_{z_0} (z) =\frac{1}{(2l)^{1/2}}\frac{1}{\cosh[(z-z_0)/l]}
\label{eq:swfz0}
\ee
with a spatial radius
\be
l= -\frac{2\hbar^2}{(N-1) m g_{1d}}.
\label{eq:size}
\ee
The Hartree-Fock ansatz
$|N:\psi\rangle$ therefore breaks the translational symmetry of the system.

Breaking a symmetry of the system costs energy, and this can be checked
for the present translational symmetry breaking.
As the center of mass coordinates $Z,P$ of the $N$
particles are decoupled from the relative coordinates of the particles
we can write the total energy of the gas as the
sum of the kinetic energy of the center of mass 
and an \lq internal' energy including
the kinetic energy of the relative motion of the particles and the interaction
energy. 
Whereas the exact ground state wavefunction has a vanishing
center of mass kinetic energy, the symmetry breaking ansatz $|N:\psi\rangle$
contains a center of mass kinetic energy:
\be
E_{\mbox{\scriptsize c.o.m.}} =  
\langle N:\psi|\frac{\hat{P}^2}{2mN}| N:\psi\rangle
\ee
where $mN$ is the total mass of the gas and $\hat{P}$ is the 
total momentum operator. Using the definition (\ref{eq:def_hat_P}),
expanding 
the square of $\hat{P}$,
and using the fact that the soliton wavefunction $\psi$ has 
a vanishing mean momentum we obtain
\bea
E_{\mbox{\scriptsize c.o.m.}} &=& \langle \psi |\frac{p^2}{2m}|\psi\rangle \\
&=& \frac{1}{24} \frac{mg_{1d}^2}{\hbar^2}(N-1)^2.
\label{eq:ekin}
\eea
We see that $E_{\mbox{\scriptsize c.o.m.}}$ accounts for half the energy
difference between the exact ground state energy (\ref{eq:exe})
and the Hartree-Fock energy (\ref{eq:hfe}). 

\subsubsection{Physical advantage of the symmetry breaking description}
We now raise the question: is there a Bose-Einstein condensate in the one-dimensional
free Bose gas with attractive interaction? To make things simple we assume that the
gas is at zero temperature so that the $N-$particle wavefunction is known
exactly, see Eq.(\ref{eq:psi_ex}).

We start with a reasoning in terms of the one-body density operator (even if we know
from the previous physical examples that this may be dangerous). Paraphrasing the
usual three dimensional definition of a Bose-Einstein condensate in free space
we put the one-dimensional gas in a fictitious box of size $L$ 
and we calculate the mean number of particles $n_0$ in the plane wave with vanishing momentum
$p=0$ in the limit $L\rightarrow +\infty$.

The calculation with the exact ground state wavefunction has been done
\cite{Herzog}. One finds that $n_0$ is going to zero as $1/L$:
\be
\lim_{L\rightarrow+\infty}
n_0 L = C(N) \frac{2\hbar^2}{m|g_{1d}|}.
\label{eq:exact_n0}
\ee
The factor $C(N)$ is given by 
\be
C(N) = \sum_{i=1}^{N}\sum_{j=i}^{N}
\frac{(j-1)!}{(i-1)!}\frac{(N-i)!}{(N-j)!}
\prod_{k=i}^{j}\left[k(N+1-k)-\frac{1}{2}(N+1)\right]^{-1}
\ee
and converges to $\pi^2/2$ in the large $N$ limit, so that $n_0$ no longer depends
on $N$ in this limit. There is therefore no macroscopic population in the $p=0$
momentum state.
One may then be tempted to conclude 
that there is no Bose-Einstein condensate,
even at zero temperature, in the one-dimensional Bose gas with attractive contact
interactions. However we have learned that a reasoning based on the one-body
density matrix may miss crucial correlations between the particles, and
that the symmetry breaking point of view may be illuminating in this
respect.

The translational symmetry breaking point of view approximates the state of the gas by
the $N$-body density operator:
\be
\hat{\rho}^{\mbox{\scriptsize sb}} = \lim_{L\rightarrow +\infty}
\int_{-L/2}^{L/2} \frac{dz_0}{L} |N:\psi_{z_0}\rangle \langle N:\psi_{z_0}|.
\label{eq:sba}
\ee
In the large $N$ limit we expect this prescription to be valid
for few-body observables. Of course for a
$N-$body observable such as the kinetic energy of the center of mass of the gas,
the results will be different, Eq.(\ref{eq:ekin}) for the symmetry breaking
point of view vs.\ a vanishing value for the exact result.

Let us test this expectation by calculating in the Hartree-Fock 
approximation 
the mean 
number of particles in the plane wave $\langle z|k\rangle=\exp(ikz)/L^{1/2}$.
Using the following action of the annihilation operator $\hat{a}_k$ of a particle
with wavevector $k$ on the Hartree-Fock state:
\be
\hat{a}_k |N:\psi_{z_0}\rangle = N^{1/2} \langle k|\psi_{z_0}\rangle
|N-1:\psi_{z_0}\rangle
\ee
we obtain
\be
n_k^{\mbox{\scriptsize hf}}  = N |\langle k|\psi\rangle|^2.
\label{eq:mom_dist}
\ee
The momentum distribution of the particles in the gas in this approximation
is simply proportional to the momentum distribution of a single particle
in the solitonic wavefunction $\psi$! 
It turns out that the Fourier transform of the $1/\cosh$ function
can be calculated exactly, and it is also a $1/\cosh$ function.
We finally obtain:
\be
n_k^{\mbox{\scriptsize hf}}  \simeq \frac{1}{L} \frac{\pi^2\hbar^2}{m |g_{1d}|}
\frac{1}{\cosh^{2}\left(\frac{\pi k l}{2}\right)}
\label{eq:nks}
\ee
where $l$ is the soliton size given in Eq.(\ref{eq:size}).
For $k=0$ one recovers 
the large $N$ limit of the exact result (\ref{eq:exact_n0}).

In more physical terms, one can imagine from Eq.(\ref{eq:sba}) that 
a given experimental realization of the Bose gas corresponds to  
a condensate of $N$ particles in the solitonic wavefunction
(\ref{eq:swfz0}), with a central position $z_0$ being a random variable varying
in an unpredictable way for any new realization of the experiment.  
There is therefore
a Bose-Einstein condensate in the one-dimensional attractive Bose gas!

An illustrative gedanken experiment would be to measure the positions along $z$
of all the particles of the gas.
In the symmetry breaking point of view the positions $z_1,\ldots, z_N$ obtained in
a single measurement are randomly distributed according to the density $|\psi_{z_0}^2|
(z)=|\psi(z-z_0)|^2$
where $z_0$ varies from shot to shot as the relative
phase of the two condensates did in the MIT interference experiment. 
As we know the exact ground
state (\ref{eq:psi_ex}) we also know the exact $N-$body distribution function,
$|\Psi(z_1,\ldots,z_N)|^2$. This is however not so easy to use! 

So we suggest instead to consider 
the mean spatial density of the particles {\sl knowing} that the center
of mass of the cloud has a position $Z$. In the exact formalism this gives \cite{Herzog}:
\bea
\rho(z|Z) &=& \int dz_1\ldots\int dz_N |\Psi(z_1,\ldots,z_N)|^2
\left(\sum_{j=1}^{N}\delta(z-z_j)\right)
L\delta\left(Z-\frac{1}{N}\sum_{n=1}^{N}z_n\right)
\label{eq:def_rho}\\
&=&   \frac{2 N}{l} \sum_{k=0}^{N-2}
\frac{(N-2)!}{(N-k-2)!}  \frac{N!}{(N+k)!}(-1)^k(k+1)
\exp\left[-(k+1)\frac{2N}{N-1}\frac{|z-Z|}{l}\right]
\nonumber
\eea
where $l$ is the $N$-dependent length of the soliton (\ref{eq:size}),
the integrals are taken in the range $[-L/2,L/2]$ and $L\rightarrow
+\infty$; the factor $L$, compensating the one in the normalization
factor of $\Psi$, ensures that the integral of $\rho(z|Z)$
over $z$ is equal to $N$.

In the symmetry breaking point of view the definition of $\rho(z|Z)$ 
is similar to Eq.(\ref{eq:def_rho}); 
the factor $L$ cancels with the $1/L$ factor of Eq.(\ref{eq:sba}).
This leads to
\bea
\rho^{\mbox{\scriptsize sb}}(z|Z) &=& \int\! dz_0 \int\! dz_1\ldots\int\! dz_N \left(\prod_{k=1}^{N}|\psi(z_k-z_0)|^2\right)
\left(\sum_{j=1}^{N}\delta(z-z_j)\right)
\delta\left(Z-\frac{1}{N}\sum_{n=1}^{N}z_n\right)
\nonumber
\\
&=& N\int dz_1\ldots\int dz_N \left(\prod_{k=1}^{N}|\psi(z_k)|^2\right)
\delta\left(Z-z+z_1-\frac{1}{N}\sum_{n=1}^{N}z_n\right)
\label{eq:complique}
\eea
where we have made the change of variables $z_k\rightarrow z_k+z_0$ (which allows
to integrate over $z_0$) and we have replaced the sum over the indiscernible particles
$j$ by $N$ times the contribution of particle $j=1$.
The multiple integral over the positions $z_1,\ldots,z_N$
can be turned into a single integral over a wavevector $q$ by using the identity
$\delta(X)= \int dq/(2\pi) \exp(iqX)$, allowing a numerical calculation of
$\rho^{\mbox{\scriptsize sb}}(z|Z)$. 

\begin{figure}[htb]
\centerline{ 
\epsfysize=7cm \epsfbox{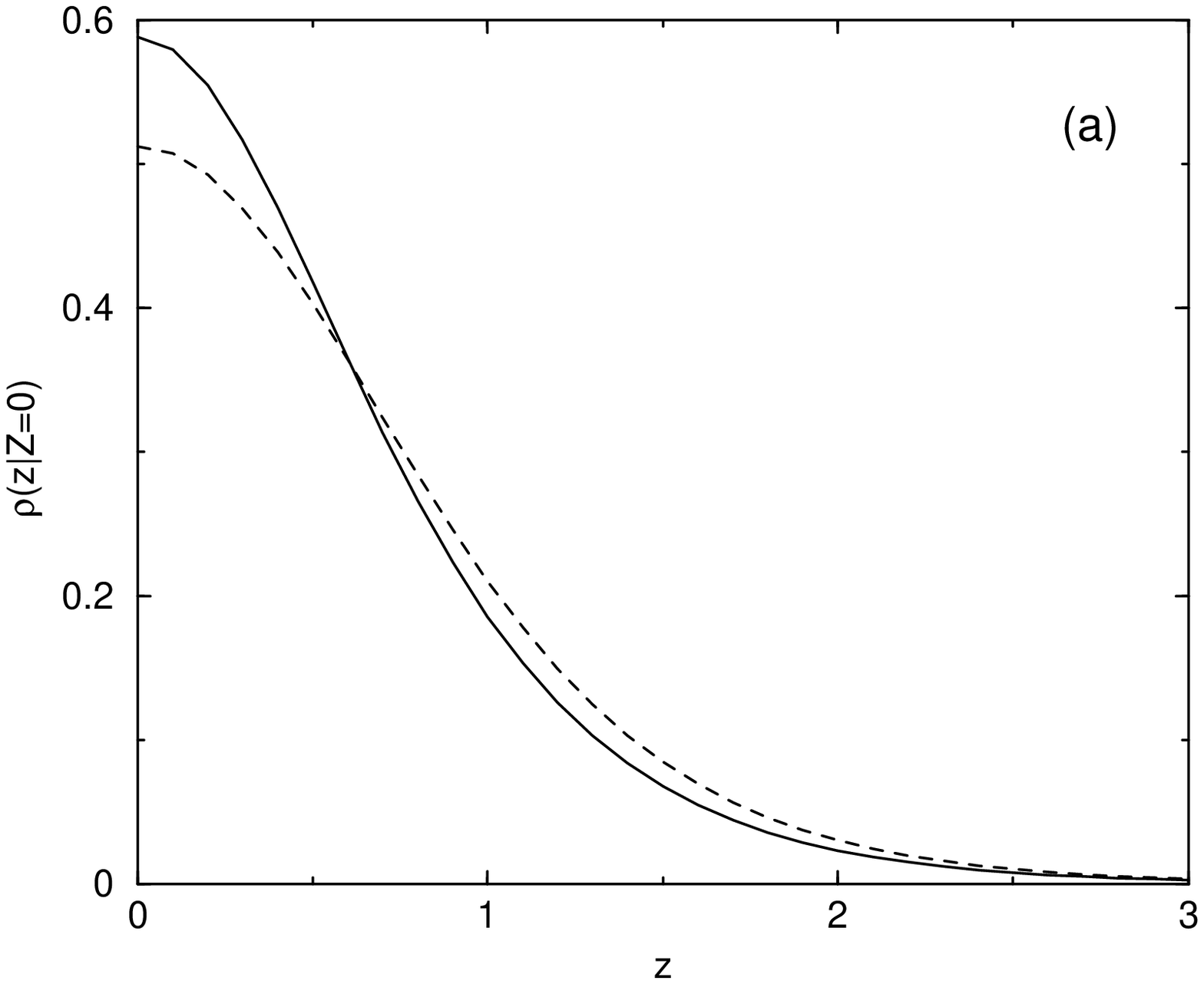} \ \ \
\epsfysize=7cm \epsfbox{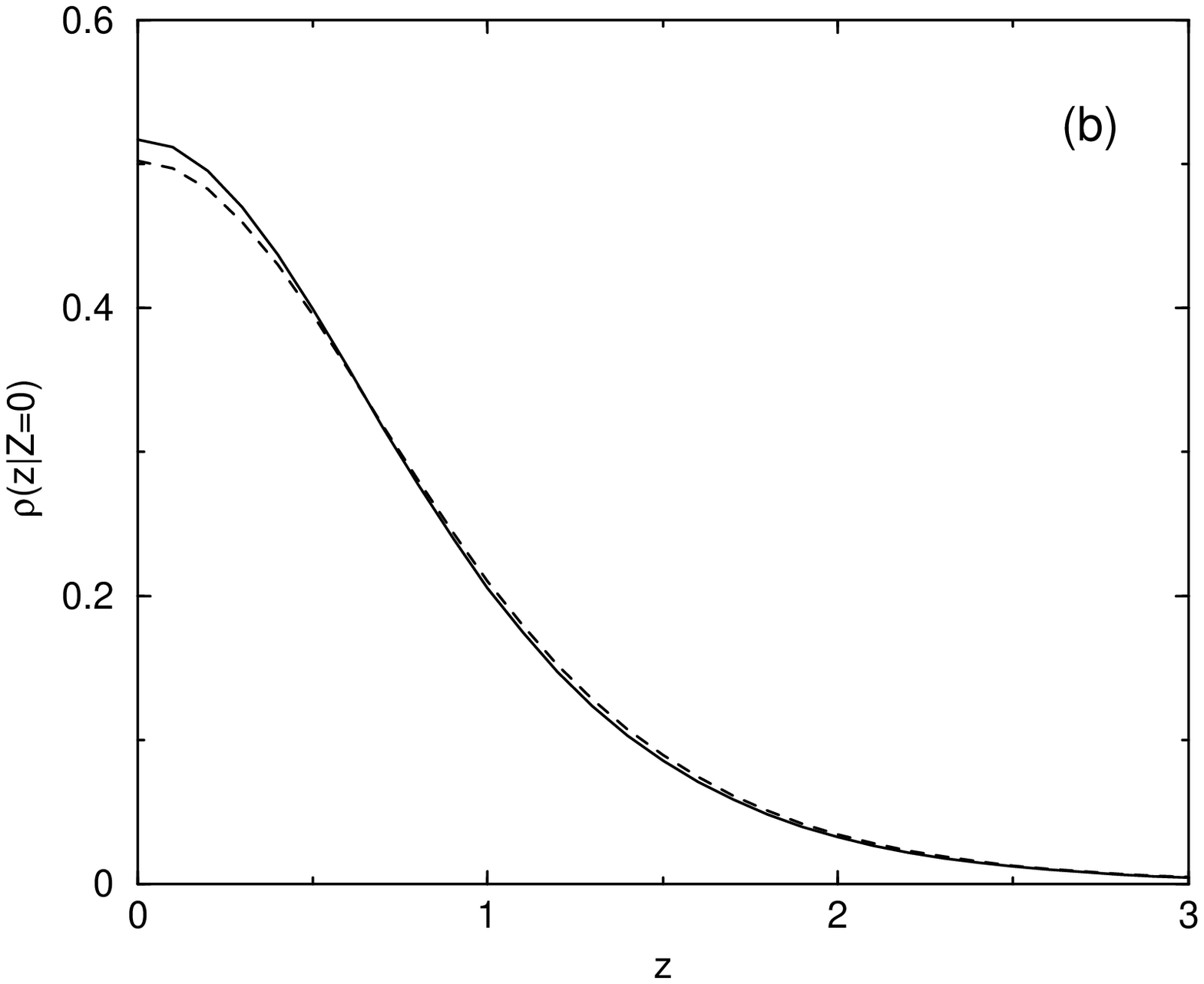}}
\caption{\small For the ground state of the one-dimensional attractive Bose gas,
position dependence of the
mean density of particles knowing that the center-of-mass of the gas is
in $Z=0$. Solid line: exact result $\rho(z|Z=0)$. Dashed line: mean-field approximation
$\rho^{\mbox{\scriptsize sb}}(z|Z=0)$. The position $z$ is expressed in units
of the \lq soliton' radius $l$ given in Eq.(\ref{eq:size}), and the linear
density in units
of $N/l$. The number of particles is (a) $N=10$ and (b) $N=45$. 
\label{fig:soli}}
\end{figure}

Does the approximate result (\ref{eq:complique}) 
get close to the exact result
for large $N$?  We compare numerically in figure~\ref{fig:soli} 
the exact density $\rho(z|Z)$ to the symmetry breaking mean-field prediction
$\rho^{\mbox{\scriptsize sb}}(z|Z)$:
modestly large values of $N$ give already good agreement between the two densities.
This validates the symmetry breaking approach for the considered gedanken
experiment.

What happens in the large $N$ limit?
In Eq.(\ref{eq:complique}) each variable $z_k$ explores an interval of size
$\sim l$ so that the quantity $(z_1+\ldots+z_N)/N$ has a standard
deviation $\sim l/\sqrt{N}$ much smaller than $l$ and can be neglected as compared
to $z_1$ inside the $\delta$ distribution. This leads to
\be
\rho^{\mbox{\scriptsize sb}}(z|Z) 
\simeq N|\psi_{z_0=Z}(z)|^2\ \ \ \mbox{for}\ \ \ \sqrt{N}\gg 1
\label{eq:proper}
\ee
where the solitonic wavefunction $\psi_{z_0=Z}$ is given in Eq.(\ref{eq:swfz0}).
Numerical calculation of $\rho^{\mbox{\scriptsize sb}}(z|Z)$ 
shows that Eq.(\ref{eq:proper})
is a good approximation over the range $|z-Z|\simeq l$ for $N=10$
already!

\noindent {\bf Acknowledgments}

\noindent I warmly thank all my colleagues of the \'Ecole normale sup\'erieure
and all my coworkers
for their contribution to my understanding and knowledge
of Bose-Einstein condensates; in particular the lectures of Claude
Cohen-Tannoudji at the Coll\`ege de France have been an illuminating example
and the notes taken by David Gu\'ery-Odelin at an early version of this
course  have been useful.
I am very indebted to Franck Lalo\"e for useful comments
on the manuscript.
I am immensely grateful to my wife Alice for her invaluable help
in the production of these lecture notes; her careful notes taken in
Les Houches and her help in typing the manuscript have been crucial.
\addcontentsline{toc}{section}{{\bf Bibliography } }
\vfill\eject

\renewcommand{\baselinestretch} {1}

\end{document}